\documentclass[superscriptaddress, nofootinbib, amsfonts, amsmath, amssymb, aps, prd, notitlepage, floatfix]{revtex4-2}
\usepackage[colorlinks=true, linkcolor=blue, citecolor=blue, urlcolor=blue]{hyperref}
\usepackage{xcolor}
\usepackage{graphicx}
\usepackage{subfigure}
\usepackage{multirow}
\usepackage{slashed}
\usepackage{lineno}
\usepackage[percent]{overpic}
\usepackage{ulem}
\allowdisplaybreaks
\newcommand{\Tr}[1]{\rm{Tr}\!\left[#1\right]}
\newcommand{\na}{\notag\\} 
\newcommand{\bb}{\mathbb}
\newcommand{\dhd}{{\textstyle d} \lower.03ex\hbox{\kern-0.38em$^{\scriptstyle-}$}\kern-0.05em{}}
\usepackage{amsmath,amssymb}
\usepackage{booktabs}
\usepackage{tabularx}
\usepackage{multirow}
\usepackage{enumitem}
\usepackage{array}
\usepackage{todonotes}

\newcolumntype{Y}{>{\raggedright\arraybackslash}X}

\begin{document}
%
\title{Back-to-back dijet production in DIS at arbitrary Bjorken-x:\\ TMD gluon distributions  to twist-3 accuracy}
%
%
\author{Swagato~Mukherjee}
\affiliation{Physics Department, Brookhaven National Laboratory, Upton, New York 11973, USA}
\author{Vladimir~V.~Skokov}
\affiliation{Department of Physics and Astronomy, North Carolina State University, Raleigh, NC 27695, USA}
\author{Andrey~Tarasov}
\affiliation{Department of Physics and Astronomy, North Carolina State University, Raleigh, NC 27695, USA}
\affiliation{Center for Frontiers in Nuclear Science (CFNS) at Stony Brook University, Stony Brook, NY 11794, USA}
\author{Shaswat~Tiwari}
\email{sstiwari@ncsu.edu}
\affiliation{Physics Department, Brookhaven National Laboratory, Upton, New York 11973, USA}
\affiliation{Department of Physics and Astronomy, North Carolina State University, Raleigh, NC 27695, USA}
\author{Fei~Yao}
\email{fyao@bnl.gov}
\affiliation{Physics Department, Brookhaven National Laboratory, Upton, New York 11973, USA}
\begin{abstract}
We derive the gluon transverse-momentum-dependent (TMD)  operator structure of back-to-back quark–antiquark dijet production in deep inelastic scattering at arbitrary Bjorken-$x$ to twist-3 accuracy. Working at leading order in the strong coupling and in the kinematic regime where the transverse momentum imbalance of the jets is much smaller than their individual transverse momenta, we perform a systematic gradient expansion of the quark propagator in a background gluon field. This expansion organizes multiple interactions with the target in terms of longitudinal Wilson lines and gauge-invariant field-strength insertions, yielding a TMD description valid beyond the strict high-energy eikonal ($x \to 0$) approximation. We obtain explicit cross sections for longitudinally and transversely polarized virtual photons, identifying all contributing gluon TMD operators up to twist-3, including structures involving $F^{+-}$, $F^{ij}$, and three-gluon correlators. The full longitudinal phase $e^{ixP^+z^-}$ associated with Bjorken-$x$ is retained throughout. In the small-$x$ limit, our results reproduce the known sub-eikonal expressions obtained in the Color Glass Condensate framework, establishing a direct connection between the general-$x$ TMD expansion and high-energy factorization. We further reduce the operator basis using equations of motion, minimizing the number of independent nonperturbative matrix elements entering the cross section. This work provides a systematic foundation for extending TMD analyses of dijet production beyond leading twist, establishing a unified operator framework valid at arbitrary Bjorken-x that smoothly interpolates between moderate- and small-x descriptions of gluon TMDs.

\end{abstract}

\date{\today}
\maketitle


\section{Introduction}

Accessing gluon dynamics in nucleons and nuclei, particularly at moderate and small Bjorken-$x$, is a central scientific motivation of the future Electron--Ion Collider (EIC)~\cite{Page:2019gbf}. Achieving this objective requires identifying processes that are directly sensitive to the transverse-momentum-dependent (TMD) structure of gluons across a wide kinematic range. Quark--antiquark dijet production in deep inelastic scattering (DIS) has emerged as one of the most promising channels for accessing gluon TMDs 
at the EIC~\cite{Metz:2011wb,Dumitru:2015gaa,Dumitru:2018kuw,Mantysaari:2019hkq,Caucal:2023fsf}. In this process, a virtual photon with virtuality $Q^2 \equiv -q^2$ splits into a quark--antiquark pair that subsequently interacts with the gluon field of the target nucleon or nucleus. The sensitivity of the dijet cross section to the transverse momentum imbalance of the final-state jets provides direct access to the intrinsic transverse structure of  gluons~\cite{Dumitru:2018kuw}, making this observable central to the broader program of nucleon and nuclear tomography~\cite{Boer:2011fh,Meissner:2009ww,Lorce:2013pza}.

In general kinematics, dijet production in DIS probes gluon TMD distributions that formally receive contributions from all twists. 
A key simplification occurs in the back-to-back limit, where the transverse momentum imbalance of the dijet system $|k_{1\perp}+k_{2\perp}|$ is much smaller than the individual jet momenta $k_{1,2\perp}$. In this regime, the dijet cross section admits a TMD-factorized description, and at leading power { in $1/k_{1\perp}^2\sim 1/k_{2\perp}^2$} provides access to the Weizs\"acker--Williams gluon TMD~\cite{Mulders:2000sh,Dominguez:2011wm,Metz:2011wb,delCastillo:2020omr,Dumitru:2015gaa,Dumitru:2016jku,Dumitru:2018kuw}. The back-to-back limit has consequently been studied extensively as a theoretically clean setting for extracting gluon TMDs from dijet observables.

Building on earlier studies of dijet and heavy quarkonium production~\cite{Fleming:2003gt,Becher:2010tm}, TMD factorization for dijet production in DIS at large-$x$ was performed~\cite{delCastillo:2020omr} within the soft-collinear effective theory (SCET) framework at next-to-leading order and leading-twist approximations. Within high-energy (small-$x$) factorization, dijet production in DIS was analyzed based on the Color Glass Condensate (CGC) effective theory~\cite{McLerran:1993ni,McLerran:1993ka,Iancu:2003xm,Gelis:2010nm,Kovner:2005pe}. Back-to-back dijet production in DIS, including leading-order (LO) as well as next-to-leading-order (NLO) quantum corrections, was calculated in the strict eikonal limit corresponding to $x=0$ in Refs.~\cite{Dominguez:2011wm,Boussarie:2021ybe,Caucal:2021ent,Caucal:2023fsf,Caucal:2023nci}.

For realistic EIC kinematics, however, the accessible values of $x$ are typically $\gtrsim 10^{-2}$ at perturbatively {large $k_{1\perp}^2 \sim k_{2\perp}^2$} (see, e.g., Ref.~\cite{Dumitru:2018kuw}). In this regime, sub-eikonal corrections associated with nonzero $x$ can become significant. This raises the question of how NLO quantum corrections compare in importance to finite-$x$ power corrections in the phenomenologically relevant kinematic region.

Leading sub-eikonal corrections to dijet production in DIS at LO in quantum fluctuations were investigated within the CGC framework~\cite{Altinoluk:2022jkk}. In the back-to-back limit,  these corrections are found to be suppressed by {$1/k_{1,2 \perp  }^{2}$ }and involve both kinematic power corrections and genuine twist-3 gluon operators~\cite{Altinoluk:2024zom}. 
While these studies provide important insights, they rely on the high-energy expansion inherent to the CGC framework and are therefore restricted to the small-$x$ regime.

To firmly establish the structure of higher-twist contributions and assess their relevance for EIC phenomenology, it is desirable to obtain these effects within a factorization framework valid for general Bjorken-$x$. Such an approach would permit transparent identification of operator structures, retains the full longitudinal phase factors appearing in standard TMD definitions, and, simultaneously, enable controlled comparison with the small-$x$ limit.

In this work, we address this issue using the TMD factorization framework recently proposed in Refs.~\cite{Mukherjee:2023snp,Mukherjee:2025aiw}, hereafter referred to as the MSTT framework. This framework is based on the background field method~\cite{Abbott:1980hw}, in which QCD degrees of freedom are separated into quantum and background modes. Importantly, unlike TMD factorization, the background gluon field in the MSTT framework carries both longitudinal and transverse momentum components without assuming any a priori ordering between them. Depending on the kinematic regime, the background gluons can be nearly on-shell (the partonic limit), highly off-shell (the small-$x$ regime), or in intermediate configurations. Consequently, the formalism retains the full $x$-dependent longitudinal phase factor $e^{ixP^{+}z^{-}}$, where $P^{+}$ is the large light-cone momentum of the target and $z^{-}$ is the light-cone separation between the gluon fields. By contrast, sub-eikonal corrections in CGC calculations rely on expansions in inverse powers of the collision energy and hence on expanding the phase factor $e^{ixP^{+}z^{-}}$ in powers of $x$~\cite{Kovchegov:2015pbl,Kovchegov:2016zex}.

We apply the MSTT framework to back-to-back quark--antiquark dijet production in DIS at leading order in $\alpha_s$. Going beyond leading twist, we incorporate all kinematic and dynamic twist-3 gluon operators contributing to the process. Since the framework is valid for general $x$, our results enable a direct and systematic comparison with eikonal and sub-eikonal expressions obtained from CGC framework. In the appropriate limit, we find full agreement with existing sub-eikonal CGC results~\cite{Altinoluk:2022jkk,Altinoluk:2024zom}, while providing a unified description valid for an arbitrary Bjorken-$x$. Furthermore, we improve upon Ref.~\cite{Altinoluk:2024zom} by expressing results in terms of a reduced set of independent twist-3 operators related through equations of motion, facilitating future phenomenological applications.

The remainder of this paper is organized as follows. In Sec.~\ref{sec:prelim}, we introduce the dijet production process in DIS, establish our conventions, and identify the quark propagator in a background color field as the central object of interest. In Sec.~\ref{sec:background}, we evaluate this propagator using the background field method in the TMD kinematic limit, systematically organizing multiple interactions with the target field. In Sec.~\ref{sec:ampl}, we construct the scattering amplitude from these results. In Sec.~\ref{sec:xsect}, we compute the dijet cross sections for longitudinally and transversely polarized virtual photons, compare with existing leading-twist calculations, and discuss the extension to higher twist. Finally, Sec.~\ref{sec:outl} summarizes our findings and outlines directions for future work.

\section{General Structure of the Dijet Amplitude and Cross Section}
\label{sec:prelim}

\begin{figure}[h]
\begin{center}
\begin{overpic}[width=0.4\textwidth]{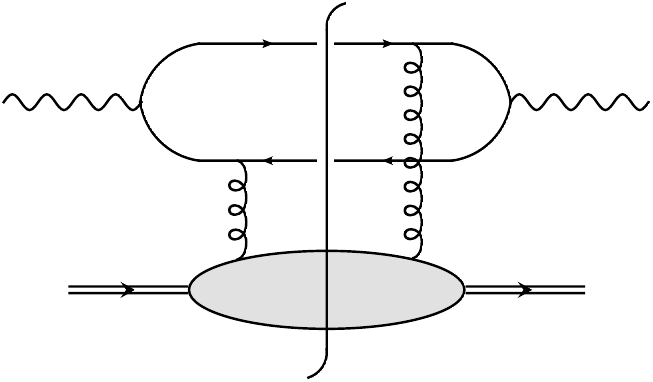}
    \put(5,47){ $\gamma^*\,(q)$} 
    \put(41,54){ $k_1$} 
    \put(41,36){ $k_2$} 
    \put(16,17){ $P$}
\end{overpic}
\end{center}
\caption{Diagrammatic representation of a virtual photon $\gamma^*$ splitting into a quark--antiquark pair, which interacts with the target hadron through multiple gluon exchanges. These gluon lines represent the background-field expansion, with each line corresponding to one insertion of the background gluon field in our calculation. The vertical cut separates the amplitude from its complex conjugate; the dijet momenta are identical on the two sides of the cut.}
\label{fig:dijet_diagram}
\end{figure}
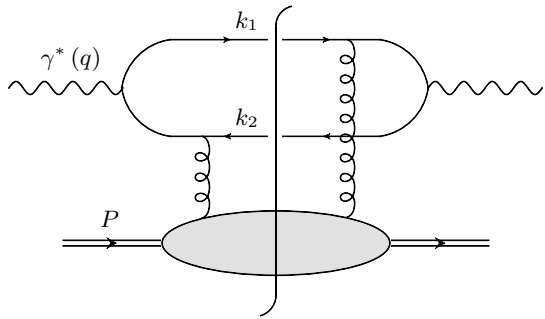

In the dijet process illustrated in Fig.~\ref{fig:dijet_diagram}, the virtual photon splits into a quark-antiquark pair that interacts with the background gluon field of the target hadron. Within the background-field approach, this interaction is represented by multiple gluon insertions on the quark and antiquark lines, resummed into the background-field quark propagator $i/\slashed{\rm{P}}$ with ${\rm P}_\mu = p_\mu + A_\mu$, where $A_\mu(x) = t^a A_\mu^a(x)$ is the background gauge field and $t^a$ are the generators in the fundamental representation.\footnote{The corresponding field-strength tensor is $F_{\mu\nu}(x) = t^a F_{\mu\nu}^a(x)$.} We perform the calculation using Schwinger notation $|x)$ and $|p)$, which provides a compact notation for background-field calculations (see App.~\ref{app:schwinger} for details). In this section, we establish our conventions, define the relevant kinematic variables, and outline the structure of the scattering amplitude and cross section.

The amplitude for producing a quark-antiquark pair with momenta $k_1$ and $k_2$ from a virtual photon of momentum $q$ and polarization $\epsilon_\rho(q)$ is
\begin{align} 
i \mathcal{M} = - i e \,\epsilon_\rho(q) 
\:  \lim_{k_{1}^2, k_2^2 \to 0}\,
\int d^4y \, e^{-iq\cdot y} \, \bar{u}(k_1)\, (k_1 | \slashed{\rm p}\frac{i}{\slashed{\rm P }} | y ) \gamma^\rho ( y | \frac{i}{\slashed{\rm{P}}} \slashed{\rm{p}}| -k_2 ) v(k_2)\, ,
\end{align}
where $\bar{u}(k_1)$ and $v(k_2)$ are the quark and antiquark spinors, and $i/\slashed{\rm{P}}$ is the quark propagator in the background gluon field that resums all interactions with the target to all orders.

To facilitate the subsequent gradient expansion, we rewrite the amplitude in terms of the squared Dirac operator in the background field. Using the identity
\begin{align}
\bar{u}(k_1)\,
(
  k_1 | \slashed{\rm{p}}\,\frac{i}{\slashed{\rm{P}}} | y
)
=
\bar{u}(k_1)\,
(
  k_1 | \slashed{\rm{p}}\slashed{\rm{P}}\,
  \frac{i}{ \slashed{\rm{P}} \slashed{\rm{P}}  } | y
)
=
\bar{u}(k_1) 
k_1^2\,
(
  k_1 |
  \frac{i}{{\rm P}^2+\frac{1}{2}\sigma^{\mu\nu}F_{\mu\nu}}
  | y
)\,,
\label{eq:leg_identity}
\end{align}
where the last step uses on-shell reduction for the external leg, with the factor $k_1^2$ implementing the standard LSZ reduction of the external leg in the on-shell limit~\cite{Lehmann:1954rq} \footnote{The term $ \bar{u}(k_1) (k_1 | \slashed{p} \slashed{A} \frac{i}{  \slashed{\rm{P}} ^2} |y)$ vanishes due to the equations of motion $\bar{u}(k) \slashed{k} = 0$. The pole $i/p^2$ does not go on-shell in this case due to the presence of the background field $\slashed{A}$ between $(k_1 |$  and $\frac{i}{  \slashed{\rm{P}} ^2}$.}. 
The condition of the zero background field in the asymptotic past is assumed.   
The second equality in Eq.~\eqref{eq:leg_identity} follows from the squared Dirac operator identity~\cite{Schwinger:1951nm}
\begin{align}
\slashed{\rm P}\slashed{\rm P}
&= {\rm P}_\mu {\rm P}_\nu\left(g^{\mu\nu}-i\sigma^{\mu\nu}\right)
= {\rm P}^2+\frac{1}{2}\sigma^{\mu\nu}F_{\mu\nu}\,,
\label{eq:PP_identity}
\end{align}
where $\sigma^{\mu\nu}=\frac{i}{2}\,[\gamma^\mu,\gamma^\nu]$ and we  used the commutator relation given in Eq.~\eqref{eq:PP_DP_com}. The additional term $\frac{1}{2}\sigma^{\mu\nu}F_{\mu\nu}$ (or in shorthand notation, $\frac{1}{2}\sigma F$) represents the standard Pauli-type spin-field coupling. Applying the same manipulation to the antiquark leg, the amplitude becomes
\begin{align}\label{eq:AmplitudeM}
  i \mathcal{M}=
  &\; i e \,\epsilon_\rho(q)
  \int d^4y \, e^{-iq\cdot y}
  \lim_{k_{1}^2, k_2^2 \to 0}
  \bar{u}(k_1)\,k_1^2\,
 (k_1|
  \frac{1}{{\rm P}^2+\frac{1}{2}\sigma^{\mu\nu}F_{\mu\nu}}
  |y )
  \gamma^\rho
  (y|
  \frac{1}{{\rm P}^2+\frac{1}{2}\sigma^{\mu\nu}F_{\mu\nu}}
 |-k_2)\,
  k_2^2\,v(k_2)\,.
\end{align}
This form makes the propagator structure explicit in terms of the covariant kinetic operator $({\rm P}^2+\frac{1}{2}\sigma^{\mu\nu}F_{\mu\nu})^{-1}$ and serves as the starting point for the systematic gradient expansion in powers of the field strength tensor $F_{\mu\nu}$.

To compute observables from the amplitude in Eq.~\eqref{eq:AmplitudeM}, we express the differential cross section in terms of the light-cone momenta of the final-state jets. In light-cone variables, the Lorentz-invariant phase-space measure for particle $i$ is
\[
\frac{d k_i^-\, d^2 k_{i\perp}}{2 k_i^- (2\pi)^3 },
\]
so that the two-particle phase space contributes a factor $(2k_1^-)^{-1}(2k_2^-)^{-1}$. The differential cross section is then
\begin{align}
\,k_1^- k_2^-\,
\frac{d \sigma}{\dhd k_1^-\, \dhd^2 k_{1\perp}\, \dhd k_2^-\, \dhd^2 k_{2\perp}}
=
 \pi\, q^-\,
\delta(k_1^- + k_2^- - q^-)\,
|i\mathcal{M}|^2 ,
\label{eq:diff_xsec_def}
\end{align}
where the $\delta$ function enforces minus-momentum conservation, and factors of $2\pi$ from the phase space integration were absorbed into the definition of $\dhd$ (see also Eq.~\eqref{eq:BalitNot}) in the differential cross section on the left-hand side.

{In this work, we retain operator structures up to dynamical twist-3 in the cross section. This includes the leading twist-2 gluon operators involving the field strengths $F_{-i}F_{-j}$, as well as genuine twist-3 components. The latter comprise three-gluon (three-body) operators, terms containing the transverse field-strength tensor $F_{ij}F_{-j}$,  and dynamical target effects encoded in $F_{+-}F_{-j}$.

In addition to this operator classification, we now specify the kinematic setup and power counting employed in our calculation. We work in a frame where the target hadron carries a large light-cone momentum $P^+$, 
while the incoming virtual photon has no transverse momentum, $q_\perp = 0$, so that $2\,q^+ q^- = -Q^2$. We introduce the relative transverse momentum $P_\perp$ of the quark--antiquark pair, defined in App.~\ref{App:Kinematic_setup}, which sets the hard scale of the back-to-back configuration. With this definition, the plus and minus components of the quark and antiquark momenta scale parametrically as $k_{1,2}^- \sim q^- \sim P_\perp$  and $k_{1,2}^+ \sim q^+ \sim P_\perp$. The individual transverse momenta are hard, $k_{1\perp} \sim k_{2\perp} \sim P_\perp$, while their imbalance is parametrically small, $\Delta_\perp \equiv k_{1\perp}+k_{2\perp} \ll P_\perp$.
This hierarchy allows for a systematic expansion in the small parameter $\Delta_\perp/P_\perp$ associated with the back-to-back limit. For dynamical twist-2 operators, this expansion generates both kinematic twist-2 and kinematic twist-3 contributions. In other words, the power correction linear in $\Delta_\perp/P_\perp$  originating from twist-2 operators is identified as kinematic twist-3.

Throughout this work, we consistently retain contributions 
up to twist-three in both the dynamical and kinematic sense, 
while neglecting terms that are simultaneously subleading 
in both power countings.
 }

\section{Background field propagators}
\label{sec:background}

As discussed in Sec.~\ref{sec:prelim}, the key object in our calculation is the quark propagator in the background field, expressed in mixed momentum-position representation as $(k | ({\rm P}^2 + \tfrac{1}{2}\sigma^{\mu\nu}F_{\mu\nu} + i\epsilon)^{-1} | y )$, evaluated in the on-shell limit which is for massless quarks reduces to  $k^2 \to 0$. In this section, we develop a systematic approach to compute this propagator using a gradient expansion in the background field. First, we compute the scalar part of the propagator $(k | ({\rm P}^2 + i\epsilon)^{-1} | y )$. We then systematically include the spin-field coupling term $\tfrac{1}{2}\sigma^{\mu\nu}F_{\mu\nu}$. Both contributions include the full longitudinal phase structure and capture contributions from the field strength tensor through the gradient expansion.
Finally, we impose the on-shell condition $k^2 \to 0$, which implements the Lehmann-Symanzik-Zimmermann (LSZ) reduction for external quarks and yields the amputated propagator entering the production amplitude in Eq.~\eqref{eq:AmplitudeM}.

\subsection{Scalar propagator}

The main technical challenge is evaluating the propagator $(k | ({\rm P}^2 + i\epsilon)^{-1} | y)$ in the presence of a non-trivial background field.  Working in a frame where the target carries momentum dominated by $P^+$, with $P^+ \gg P^-$, we approximate the background field as independent of light-cone time, $A_\mu(y) = A_\mu(y^-, y_\perp)$. This approximation is justified by the factorization between different kinematic modes in the high-energy limit and allows us to treat the background field as static in $y^+$. The independence of the background field on the plus coordinate implies that the minus--momentum operator commutes with \({\rm P}^\mu\), namely \([p^-, {\rm P}^\mu] = 0\). Taking this into account, and imposing the gauge condition \(A_+ = 0\) (the so-called ``wrong'' light-cone gauge~\cite{Roy:2018jxq} $A^- =0 $ for a hadron moving with large plus momentum $P^+$), 
we obtain
\begin{align}
\label{eq:dijetSCprop}
 (k \,|\, \frac{1}{{\rm P}^2 + i\epsilon} \,|\, y)
 &=
\frac{1}{2k^-}
 \int d^4 x \,
 (k \,|\, x)\,
 (x\, | \frac{1}{{\rm P}^+ - \tfrac{{\rm P}_\perp^2}{2k^-} + i\epsilon }  |\, y)\,
\end{align}
where $k^-$ denotes the minus-component of the outgoing quark momentum  ($k^- > 0$). 

We now focus on evaluating the operator $(x| \left({\rm P}^+ - \tfrac{{\rm P}_\perp^2}{2k^-} + i\epsilon \right)^{-1} |y)$. Expanding it formally in powers of \({\rm P}_\perp^2\), we obtain
\begin{align}
\label{eq:expansionPperp}
  (x|  \frac{1}{{\rm P}^+ - \frac{ {\rm P}_\perp^2 }{2 k^-} + i \epsilon}  |y) &=
  (x| \Big[ \frac{1}{{\rm P}^+ + i\epsilon} + \frac{1}{{\rm P}^+ + i\epsilon } \frac{ {\rm P}_\perp^2 }{2 k^-} \frac{1}{{\rm P}^+ + i\epsilon} 
   +   \frac{1}{{\rm P}^+ + i\epsilon } \frac{ {\rm P}_\perp^2 }{2 k^-} \frac{1}{{\rm P}^+ + i\epsilon } \frac{ {\rm P}_\perp^2 }{2 k^-} \frac{1}{{\rm P}^+ + i\epsilon}
  + \cdots \Big] |y).
\end{align}
To evaluate these terms, we insert complete sets of states in the minus coordinate between each operator. Using the resolution of identity $\int dz^- |z^-)(z^-| = 1$, the expansion can be rewritten as
\begin{align}
  (x|  \frac{1}{{\rm P}^+ - \frac{ {\rm P}_\perp^2 }{2 k^-} + i \epsilon}  |y) 
  =& 
  (x|  \frac{1}{{\rm P}^+ + i\epsilon} |y) 
  + \int d z^- (x| \frac{1}{{\rm P}^+ + i\epsilon } |z^-)(z^-| \frac{ {\rm P}_\perp^2 }{2 k^-} \frac{1}{{\rm P}^+ + i\epsilon} |y)
  \na
  & +  \int dz_1^- dz_2^- (x| \frac{1}{{\rm P}^+ + i\epsilon }|z_1^-)(z_1^-| \frac{ {\rm P}_\perp^2 }{2 k^-} \frac{1}{{\rm P}^+ + i\epsilon } |z_2^-)(z_2^-|\frac{ {\rm P}_\perp^2 }{2 k^-} \frac{1}{{\rm P}^+ + i\epsilon} |y) + \cdots \,.
\end{align}

Since the minus momentum operator does not enter into the right-hand side, we get an overall factor of $(x^+|y^+) = \delta(x^+ - y^+)$.  The minus-coordinate integrations are performed in App.~\ref{sec:WL} and lead to the string of Wilson lines which resum the interactions with $A_-$ background field to all orders. Thus, the above expression simplifies to 
\begin{align}
\label{eq:expansionPperp}
  (x | \frac{1}{{\rm P}^+ - \tfrac{{\rm P}_\perp^2}{2k^-} + i\epsilon }  | y)
 =&\,
 \delta(x^+ - y^+)\,
 \theta(x^- - y^-)\,(x_\perp \,|\,
 \Big[
   -i\,[x^-,y^-] 
   - \int_{y^-}^{x^-} dz^- \,
     [x^-,z^-]\,
     \frac{{\rm P}_\perp^2(z^-)}{2k^-}\,
     [z^-,y^-]
 \notag\\
 &
   + i \int_{y^-}^{x^-} dz_1^- \,
     [x^-,z_1^-]\,
     \frac{{\rm P}_\perp^2(z_1^-)}{2k^-}
     \int_{y^-}^{z_1^-} dz_2^- \,
     [z_1^-,z_2^-]\,
     \frac{{\rm P}_\perp^2(z_2^-)}{2k^-}\,
     [z_2^-,y^-]
   + \cdots
 \Big]|\, y_\perp)
 \, ,
\end{align}
where the square bracket denotes the Wilson line
\begin{align}
\label{eq:def-Wl-noxp}
[x^-,y^-]_{y_\perp}
=
\mathcal{P}\exp\!\left[
i \int_{y^-}^{x^-} dz^-\,A_-(z^-,y_\perp)
\right] .
\end{align}
 Throughout this work, we consider zero background gauge potential at the asymptotic future $A(x^-\to+\infty) \to 0$. In the back-to-back limit, where no finite transverse separation is involved, transverse Wilson lines at infinity simplify to unity and therefore do not appear explicitly in our final expressions.

The expansion in Eq.~\eqref{eq:expansionPperp} admits, in principle, an all-order resummation in insertions of ${P}_\perp^2$, as discussed in Ref.~\cite{Kar:2026vzk}. Such a resummation, however, lies beyond the scope of the present work. Our objective is to systematically isolate leading- and subleading-twist contributions to the cross section, including both kinematic corrections and genuinely dynamic higher-twist effects. Within consistent twist power counting, higher-order insertions of ${P}_\perp^2$ correspond to operators of higher twist; retaining them would introduce contributions beyond the accuracy of our analysis. As demonstrated explicitly in Sec.~\ref{sec:xsect}, it therefore suffices to keep only the terms shown in Eq.~\eqref{eq:expansionPperp} and neglect the higher-order contributions represented by the ellipsis. We proceed by systematically commuting all insertions of ${P}_\perp^2$ in Eq.~\eqref{eq:expansionPperp} to the left, rewriting the resulting structures in terms of gauge-invariant field-strength tensors and their covariant derivatives. This reorganization allows us to clearly identify the leading- and subleading-twist operator structures entering the cross section. We begin by considering
\begin{align}
  {\cal D}_1(x^-, y^-)
  &= \int_{y^-}^{x^-} dz^- [x^-, z^-] \frac{ {\rm P}_\perp^2 }{2 k^-}  [z^-, y^-]
  %
  \na & = \int_{y^-}^{x^-} dz^- \Bigg( {\rm {P}_i}(x^-)  [x^-, z^-] \frac{  {\rm P}_i(z^-) }{2 k^-}  [z^-, y^-] - \int_{z^-}^{x^-}\:dz_1^-\: [x^-, z_1^-]F_{-i}(z_1^-)[z_1^-,z^-] \frac{ {\rm P}_i(z^-) }{2 k^-}  [z^-, y^-] \Bigg), 
\end{align}
where we  commuted the covariant momentum operator ${\rm P}_i$ through the Wilson line $[x^-, z^-]$ using Eq.~\eqref{eq:wilson_comm}. Repeated application of this identity allows us to move all factors of ${\rm P}_i$ to the leftmost coordinate $x^-$, yielding
\begin{align}
   {\cal D}_1(x^-, y^-)
  =& (x^- - y^-)\frac{ {\rm P}_\perp^2(x^-) }{2 k^-}[x^-, y^-] -  \int_{y^-}^{x^-}\:dz_1^- \:(z_1^- - y^-)\frac{{2{\rm P}_i(x^-)}}{2k^-} [x^-, z_1^-]F_{-i}(z_1^-)[z_1^-, y^-] \na &  +   \frac{1}{k^-}  \int_{y^-}^{x^-}\:dz_1^-\:\int_{y^-}^{z_1^-} dz_2^-(z_2^- - y^-) [x^-, z_1^-]F_{-i}(z_1^-)[z_1^-,z_2^-]  F_{-i}(z_2^-)[z_2^-, y^-] \na &  + \frac{1}{2k^-}  \int_{y^-}^{x^-}\:dz_1^-\:(z_1^- - y^-)\: [x^-, z_1^-]i D_iF_{-i}(z_1^-)[z_1^-, y^-] \,, 
\end{align}
where the difference in the minus coordinate (e.g. $z_1^- - y^-$) arise from performing the $z^-$ integration.

\begin{figure}[t]
  \centering
  \includegraphics[width=\linewidth]{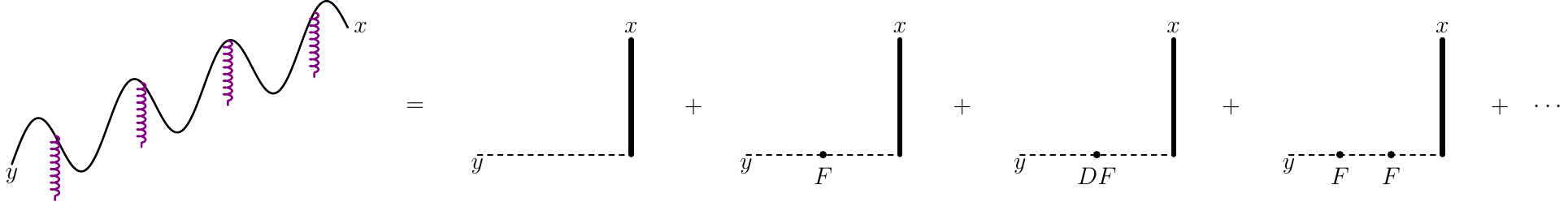}
  \caption{Diagrammatic representation of  Eq.~\eqref{eq:scpropresumfinal}. Horizontal dotted lines denote  Wilson lines along the light cone minus direction, while the thick vertical lines represent the transverse propagation given by $(x_\perp| \exp[-i P_\perp^2(x^-)(x^- - y^-)/(2 k^-)]|y_\perp)$ and its transverse derivatives. We do not further decompose the transverse exponential into Wilson links and field-strength tensors, as such transformation becomes irrelevant after taking the on-shell limit in Sec.~\ref{sec: on-shell}.}
  \label{fig:prop}
\end{figure}

We apply the same strategy to the double ${\rm P}_\perp^2$ insertion. The resulting expression can be written conveniently as a convolution of the single-insertion result,
\begin{align}
{\cal D}_2(x^-, y^-)
  =& \int_{y^-}^{x^-} dz_1^-  [x^-, z_1^-] \frac{ {\rm P}_\perp^2(z_1^-) }{2 k^-} \int_{y^-}^{z_1^-} dz_2^-   [z_1^-, z_2^-] \frac{ {\rm P}_\perp^2(z_2^-) }{2 k^-} [z_2^-, y^-]  
  \na  =&  \int_{y^-}^{x^-} dz_2^-  {\cal D}_1(x^-, z_2^-) \:\frac{ {\rm P}_\perp^2(z_2^-) }{2 k^-} [z_2^-, y^-].
\end{align}
Substituting the explicit expression for $\mathcal{D}_1$ and commuting the ${\rm P}_\perp^2$ insertions to the leftmost coordinate $x^-$ using Eqs.~\eqref{eq:comm_DF} and~\eqref{eq:wilson_comm}, we isolate the operator structures relevant at twist-3 accuracy. As shown in Secs.~\ref{sec:ampl} and~\ref{sec:xsect}, only specific operator combinations in $\mathcal{D}_2(x^-, y^-)$ contribute to the cross section at this order. Retaining only these terms, we obtain\footnote{For the double-insertion contributions, we neglect terms containing covariant derivatives acting on the field strength in mixed $DF\, F$ structures, as they are twist-suppressed. We also discard terms involving $D^2 F$, which correspond to higher-order kinematic twist.}
\begin{align}
{\cal D}_2(x^-, y^-) =&   \,\frac{(x^- - y^-)^2}{2}\frac{ {\rm P}_\perp^4(x^-) }{(2 k^-)^2}[x^-, y^-] 
- \frac{2 {{\rm P}_\perp^2(x^-)}{{\rm P}_i(x^-)} }{ (2k^-)^2} \int_{y^-}^{x^-}\:dz_1^- (x^- - y^-) (z_1^- - y^-) [x^-, z_1^-]F_{-i}(z_1^-)[z_1^-, y^-]  \na & 
+ \frac{{ 4{\rm P}_i(x^-)\: {\rm P}_j(x^-)}}{(2k^-)^2} \int_{y^-}^{x^-} dz_1^-\int_{y^-}^{z_1^-}\:dz_2^- \:(z_1^- - y^-)\:(z_2^- - y^-)\: [x^-, z_1^-]F_{-j}(z_1^-)[z_1^-, z_2^-]F_{-i}(z_2^-)[z_2^-, y^-]  \na & + \frac{ 2{\rm P}_\perp^2(x^-) }{ (2k^-)^2}   \int_{y^-}^{x^-}\:dz_1^-\:\int_{y^-}^{z_1^-} dz_2^-\:(x^- - y^-) (z_2^- - y^-) [x^-, z_1^-]F_{-i}(z_1^-)[z_1^-,z_2^-]  F_{-i}(z_2^-)[z_2^-, y^-] \na &  + \frac{  {\rm P}_\perp^2(x^-) }{(2 k^-)^2} \int_{y^-}^{x^-}\:dz_1^-\:(x^- - y^-)\:(z_1^- - y^-) [x^-, z_1^-]i D_iF_{-i}(z_1^-)[z_1^-, y^-]  \na&
+ \frac{ 2{\rm P}_i(x^-) {\rm P}_j(x^-)}{(2k^-)^2} \int_{y^-}^{x^-}\:dz_1^- \:(z_1^- -y^-)^2 [x^-, z_1^-]iD_jF_{-i}(z_1^-)[z_1^-, y^-] .  
\end{align}
Collecting the contributions from the single- and double-insertion terms $\mathcal{D}_1(x^-,y^-)$ and $\mathcal{D}_2(x^-,y^-)$ in Eq.~\eqref{eq:expansionPperp}, we observe a notable simplification: the local surface terms proportional to ${\rm P}_\perp^2(x^-)$ combine order by order into an overall exponential factor. This exponentiation reflects the geometric-series structure of the underlying ${\rm P}_\perp^2$ expansion. The scalar propagator can thus be cast into the compact form
\begin{align}
\label{eq:scpropresumfinal}
  &(x | \frac{1}{{\rm P}^+-\tfrac{{\rm P}_\perp^2}{2k^-} + i\epsilon} | y) =
\delta(x^+-y^+) \theta(x^-  - y^-)  (x_\perp| e^{ -i (x^--y^-) \frac{ {\rm P}_\perp^2 (x^-) }{2 k^-}}  \bigg[ -i \,[x^-, y^-]\,+ \frac{1}{2k^-}  \int_{y^-}^{x^-} dz^- (z^- - y^-) \, \times
   \na & \bigg(2 {\rm P}_k (x^-) \left[x^{-}, z^{-} \right] F_{-k}(z^-) \! - \!i \left[x^{-}, z^{-}\right] D_k F_{-k} (z^-)
     \! - \!2 \int^{x^-}_{z^{-}} d \zeta^{-} \left[x^{-}, \zeta^{-}\right] F_{-k} (\zeta^{-}) \left[\zeta^{-}, z^-\right] F_{-k}\left(z^{-}\right)\bigg) \left[z^{-}, y^{-}\right]  
   \na & +  \frac{4i}{(2k^-)^2} \int_{y^-}^{x^-} dz_1^- \int_{y^-}^{z_1^-} dz_2^-   (z_2^- - y^-) (z_1^- - y^-) \, {\rm P}_k(x^-)\,{\rm P}_l(x^-) \,[x^-, z_1^-] \, F_{-k} (z_1^-) \, [z_1^-, z_2^-] \,F_{-l} (z_2^-)\, [z_2^-, y^-] 
  \na &  -  \frac{2}{(2k^-)^2}  \int_{y^-}^{x^-} dz_1^- (z_1^- - y^-)^2  {\rm P}_l(x^-) {\rm P}_k(x^-)  [x^-, z_1^-]   D_l F_{-k}  (z_1^-)  [z_1^-, y^-] + \ldots \bigg]   |y_\perp)\,,
\end{align}
as can be verified by computing higher orders. 
The structure of Eq.~\eqref{eq:scpropresumfinal} admits a useful diagrammatic interpretation. The propagator consists of three elements: (i) a kinematic exponential factor encoding transverse momentum propagation, (ii) light-cone Wilson lines connecting different longitudinal positions, and (iii) field-strength insertions at intermediate points capturing dynamic twist corrections. This structure is illustrated in Fig.~\ref{fig:prop}, which displays the organization of longitudinal and transverse separations, Wilson line connections, and field-strength insertions. The thick vertical lines represent exponential propagation in the transverse plane, while the horizontal structure encodes the longitudinal Wilson line resummation.

\subsection{ Quark propagator}
Having derived the scalar propagator, we now generalize to the quark propagator.\footnote{By quark propagator we mean $(x| \frac{1}{{\rm P}^2 + \frac{1}{2}\sigma F + i\epsilon}|y)$ appearing in Eq.~\eqref{eq:AmplitudeM}.} This generalization amounts to replacing ${\rm P}^2 \to {\rm P}^2 + \frac{1}{2}\sigma F$ in Eq.~\eqref{eq:dijetSCprop} and repeating the calculation. The derivation simplifies considerably because only a single insertion of $\sigma F$ is required at the accuracy we work, yielding a perturbative correction to the scalar propagator. Explicitly, the quark propagator takes the form
\begin{align}
\label{eq:startingexpression}
(x| {1 \over {\rm P}^+ - {{\rm P}_\perp^2 \over 2 k^-} + {\sigma F \over 4 k^-} + i \epsilon}|y)  =& (x|  \frac{1}{{\rm P}^+ - \frac{ {\rm P}_\perp^2 }{2 k^-} + i \epsilon}  |y)  - (x|{1 \over {\rm P}^+ + i \epsilon}\:{{\sigma F} \over 4 k^-}\:{1 \over {\rm P}^+ + i \epsilon}\:+{1 \over {\rm P}^+ +i \epsilon  }\:{{{\rm P}_\perp^2} \over 2 k^-}\:{1 \over {\rm P}^+ + i \epsilon}\:{{\sigma F} \over 4 k^-}\:{1 \over {\rm P}^+ + i \epsilon}\na&\quad
    + {1 \over {\rm P}^+ + i \epsilon }\:{{\sigma F} \over 4 k^-}\:{1 \over  {\rm P}^+ + i \epsilon}\:{{{\rm P}_{\perp}^2} \over 2 k^-}\:{1 \over  {\rm P}^+ + i \epsilon} - {1 \over {\rm P}^+ + i \epsilon}\frac{\sigma F}{4 k^-}{1 \over {\rm P}^+ + i \epsilon}\frac{\sigma F}{4 k^-} {1 \over {\rm P}^+ + i \epsilon} +... |y)\,.
\end{align}
The simplification proceeds in close parallel to the scalar case. Using the results of App.~\ref{sec:WL}, we rewrite the right-hand side of Eq.~\eqref{eq:startingexpression} as
\begin{align}
\label{eq:wilsonlinequarks}
&(x| {1 \over {\rm P}^+ - {{\rm P}_\perp^2 \over 2 k^-} + {\sigma F \over 4 k^-} + i \epsilon}|y)\na
&  
 =  (x|  \frac{1}{{\rm P}^+ - \frac{ {\rm P}_\perp^2 }{2 k^-} + i \epsilon}  |y)- \delta(x^+ - y^+) \theta(x^- - y^-) (x_\perp|-{1 \over 4k^-}\int_{y^-}^{x^-} dz^-\:[x^-,z^-]\:{\sigma F}(z^-) \:[z^-,y^-]\:\na
&\quad{ +}\:{i \over 2(2k^-)^2}\:\int_{y^-}^{x^-}\:dz_1^-\:\int_{y^-}^{z_1^-}\:dz_2^-\:[x^-,z_1^-]\:{\rm P}_\perp^2(z_1^-)\:[z_1^-, z_2^-]\:{\sigma F}(z_2^-)\:[z_2^-, y^-]\na&\quad
{  +}\:{i \over 2(2k^-)^2}\:\int_{y^-}^{x^-}\:dz_1^-\:\int_{y^-}^{z_1^-}\:dz_2^-\:[x^-,z_1^-]\:{\sigma F}(z_1^-)\:[z_1^-, z_2^-]\:{\rm P}_\perp^2(z_2^-)\:[z_2^-, y^-] \na&\quad
{ -}\:{i \over 4(2k^-)^2}\:\int_{y^-}^{x^-}\:dz_1^-\:\int_{y^-}^{z_1^-}\:dz_2^-\:[x^-,z_1^-]\:{\sigma F}(z_1^-)\:[z_1^-, z_2^-]\:{\sigma F}\:[z_2^-, y^-] \,|y_\perp).
\end{align}

We follow the same strategy as for the scalar propagator: terms involving ${\rm P}_\perp^2$ enclosed between Wilson lines are rewritten in terms of insertions of $\mathcal{D}_1$, and the resulting covariant momenta are commuted to the leftmost coordinate using Eqs.~\eqref{eq:comm_DF} and~\eqref{eq:wilson_comm}. Retaining only operators at twist-2 accuracy, the quark propagator takes the form
\begin{align}
\label{eq:quarkpropfinal}
&(x| {1 \over {\rm P}^+ - {{\rm P}_\perp^2 \over 2 k^-} + {\sigma F \over 4 k^-} + i \epsilon}|y)\:\na 
&= \: (x|  \frac{1}{{\rm P}^+ - \frac{ {\rm P}_\perp^2 }{2 k^-} + i \epsilon}  |y)\:+\:\delta(x^+ - y^+)\:\theta(x^- - y^-)\: (x_\perp| e^{ -i (x^--y^-) \frac{ {\rm P}_\perp^2 (x^-) }{2 k^-}}\bigg[ {1 \over 4k^-}\int_{y^-}^{x^-}\:dz^- [x^-,z^-]\:\sigma F(z^-)\:[z^-,y^-]\:\na & \quad
-\:{1 \over (2k^-)^2}\int_{y^-}^{x^-}\:dz^-\:(z^- - y^-)\: {\rm P}_k(x^-) [x^-, z^-]\: D_k(\sigma F(z^-))\:[z^-, y^-]\:\na& \quad { +}\, {i \over (2k^-)^2} \int_{y^-}^{x^-}\:dz_1^-\:\int_{y^-}^{z_1^-}\:dz_2^-\:(z_1^- - y^-)\:{\rm P}_k(x^-)\:[x^-, z_1^-] \:F_{-k}(z_1^-)\:[z^-, z_2^-]\:\sigma F(z_2^-)[z_2^-,y^-]
\na & \quad {  +} \,{i \over (2k^-)^2} \int_{y^-}^{x^-}\:dz_1^-\:\int_{y^-}^{z_1^-}\:dz_2^-\:(z_2^- - y^-)\:{\rm P}_k(x^-)\:[x^-, z_1^-] \:\sigma F(z_1^-)\:[z_1^-, z_2^-]\:F_{-k}(z_2^-)[z_2^-,y^-] \na &\quad {  +} \, {i \over 4(2k^-)^2} \int_{y^-}^{x^-}\:dz_1^-\:\int_{y^-}^{z_1^-}\:dz_2^-\:[x^-, z_1^-] \:\sigma F(z_1^-)\:[z_1^-, z_2^-]\:\sigma F(z_2^-)[z_2^-,y^-] 
\bigg] |y_\perp).
\end{align}

\subsection{On-shell limit}
\label{sec: on-shell}

The results in Eqs.~\eqref{eq:scpropresumfinal} and~\eqref{eq:quarkpropfinal} can be combined into the compact form
\begin{align}
    (x| \frac{1} { {\rm P}^+ - {{\rm P}_\perp^2 \over 2 k^-} + {{\sigma} F \over 4 k^-} + i \epsilon}|y)
     = -i \delta(x^+ - y^+)\:\theta(x^- - y^-)\: (x_\perp| e^{ -i (x^--y^-) \frac{ {\rm P}_\perp^2 (x^-) }{2 k^-}} 
     { O}(x^-, y^-) |y_\perp)\,. 
\end{align}
where the operator ${O}(x^-, y^-)$ satisfies
$${  O}(x^-, x^-)  = 1 \, . $$
To incorporate this propagator into the dijet amplitude, we evaluate it in the on-shell limit. In this limit, the operator $\mathcal{O}(x^-, y^-)$ organizes the surviving twist-2 contributions, while subleading terms are either power-suppressed or vanish identically. The detailed derivation is presented in App.~\ref{sec:LC_LSZ_simplify}; here we quote only the final result:
\begin{align}
\lim_{k^2\to 0}  k^2 (k| \frac{1}{{\rm P}^2 + \frac{1}{2} \sigma F + i \epsilon } |y)
      &= \lim_{k^2\to 0}  \frac{k^2}{2k^-}\int d^4 x (k|x) (x|  \frac{1}{{\rm P}^+ - \frac{ {\rm P}_\perp^2 }{2 k^-} + {{\sigma} F \over 4 k^-} + i \epsilon }  |y)\na [4pt] &
=  \left.  e^{i (k^+ y^- + k^- y^+ )}   (k_\perp|  { O}(\infty, y^-) |y_\perp) \right|_{k^+ = \frac{k^2_\perp}{2k^-}}
     \,,
     \label{eq:lsz}
\end{align}
where we  used the asymptotic behavior ${\rm P}_\perp^2(x^- \to \infty) \to k_\perp^2$. Substituting the explicit form of $\mathcal{O}(\infty, y^-)$ completes the resummation of background-field effects and yields the amputated quark propagator through twist-2:
\begin{align}
\label{eq:mixedprop}
     &\lim_{k^2\to 0}  k^2 (k| \frac{1}{P^2 + \frac{1}{2} \sigma F + i \epsilon } |y) \na &=
      \!i  e^{i k y} \! \bigg\{\! \!
 -\!i [\infty, y^-]_{y_\perp} \!+\!  \!\int_{y^-}^{\infty} dz^-  [\infty, z^-]_{y_\perp}  \bigg[ \frac{k_k}{k^-} (z^- \!-\! y^-)   F_{-k} (z^-, y_\perp)\!+\! {1 \over 4k^-}\sigma F(z^-, {y_\perp})\!-\!  { k_k \over (2k^-)^2}(z^- \!-\! y^-) D_k(\sigma F(z^-, {y_\perp}))
  \na [4pt] & 
   - \frac{i}{2k^-}  (z^-\! -\! y^-) \left( \delta_{lk} \!-\! i (z^--y^-) \frac{k_l k_k}{k^-} \right)  D_l  F_{-k}(z^-,y_\perp)   +\,{i \over  2k^-}
F_{+-}(z^-,y_\perp)   
+(z^- - y^-){k_\perp^2 \over 2 (k^-)^2}F_{+-}(z^-,y_\perp)\bigg][z^-,y^-]_{y_\perp}
   \na [4pt] & - \frac{1}{k^-} \int_{y^-}^{\infty} dz_1^-  \left(\delta_{kl} - i\, (z_1^--y^-)\, \frac{k_l k_k}{k^-}  \right) \int_{y^-}^{z_1^-} dz_2^-   \, (z_2^- - y^-) \,  [\infty, z_1^-]_{y_\perp} F_{-k}(z_1^-, y_\perp) \, [z_1^-, z_2^-]_{y_\perp}  F_{-l} (z_2^-, y_\perp)  \,[z_2^-, y^-]_{y_\perp} 
     \na [4pt]&
     { +} \,{i \over (2k^-)^2} \int_{y^-}^{\infty}\:dz_1^-\:\int_{y^-}^{z_1^-}\:dz_2^-\:(z_1^- - y^-)\:k_k\:[\infty, z_1^-]_{y_\perp}  \:F_{-k}(z_1^-,y_\perp)\:[z_1^-, z_2^-]_{y_\perp} \:\sigma F(z_2^-,y_\perp)[z_2^-,y^-]_{y_\perp} 
\na [4pt] & { +}\, {i \over (2k^-)^2} \int_{y^-}^{\infty}\:dz_1^-\:\int_{y^-}^{z_1^-}\:dz_2^-\:(z_2^- - y^-)\:k_k\:[\infty, z_1^-]_{y_\perp}  \:\sigma F(z_1^-,y_\perp)\:[z_1^-, z_2^-]_{y_\perp} \:F_{-k}(z_2^-,y_\perp)[z_2^-,y^-]_{y_\perp}  \na [4pt]& + {i \over 4(2k^-)^2} \int_{y^-}^{\infty}\:dz_1^-\:\int_{y^-}^{z_1^-}\:dz_2^-\:[\infty, z_1^-]_{y_\perp}  \:\sigma F(z_1^-,y_\perp)\:[z_1^-, z_2^-]_{y_\perp} \:\sigma F(z_2^-,y_\perp)[z_2^-,y^-]_{y_\perp}  \bigg\}\,,
\end{align}
where we included the twist-2 effects of a dynamic target by incorporating the $F_{+-}$ insertion arising from the $z^+$ dependence. The details of this calculation are given in App.~\ref{sec:dynintar}. We also changed the order of integration:
$$
\int_{y^-}^\infty dz^- \int^\infty_{z^-} d\zeta^- \ldots = \int_{y^-}^\infty d\zeta^- \int_{y^-}^{\zeta^-} dz^- \ldots\,,
$$
followed by renaming variables to obtain a more symmetric expression. In the amplitude above, we ordered the terms by increasing the leading twist. It is worth emphasizing that the last term, to twist-2 accuracy, can only involve $\sigma^{-i} F_{-i}$ combinations. However, such a term vanished in the dijet production cross section.

For completeness, we quote the corresponding antiquark propagator. In the on-shell limit, the antiquark contribution
\begin{equation}
\lim_{k^2 \to 0} k^2 \, (y| \frac{1}{{\rm P}^2 + \frac{1}{2}\sigma F + i\epsilon} | -k)
\end{equation}
is obtained from the quark result by charge conjugation, which amounts to replacing $k^\mu \to -k^\mu$ and $i \to -i$. The twist expansion proceeds analogously to the quark case.

\section{Dijet production}\label{x_section}
Having established the structure of the background-field propagators, we now apply these results to dijet production. The quark and scalar propagators derived in the previous section provide the building blocks for constructing the amplitude. We then evaluate the corresponding cross section, enabling direct comparison with existing results in the literature.
\subsection{Amplitude}
\label{sec:ampl}

Using the quark propagator in the background field given in Eq.~\eqref{eq:mixedprop}, together with the corresponding antiquark propagator, we compute the dijet production amplitude defined in Eq.~\eqref{eq:AmplitudeM}. All remaining integrations over the lightcone minus coordinates can be carried out analytically; the detailed calculations are collected in App.~\ref{App:integrals}. The amplitude then reduces to a single integration over the spacetime coordinate $x$.
Moreover, since the background field is assumed to be independent of the lightcone time $x^+$, the integrand carries no $x^+$ dependence beyond the exponential phase. The $x^+$ integration therefore yields the overall minus-momentum-conserving delta function, and the phase factor reduces to
\begin{equation}
\int d^4x\, e^{i k_g \cdot x}
=
2\pi\,\delta(k_1^- + k_2^- - q^-)
\int dx^-\, d^2x_\perp\,
e^{i\bigl(k_g^+ x^- - {k}_{g\perp} {x}_\perp\bigr)}\,,
\label{eq:xplus_delta_normalization}
\end{equation}
where $k_g^\mu \equiv k_1^\mu + k_2^\mu - q^\mu$ denotes the momentum transferred from the background field. The factor $2\pi\,\delta(k_1^- + k_2^- - q^-)$ is a direct consequence of $x^+$-translation invariance and sets the overall normalization of the scattering amplitude. It cancels against the corresponding factor arising from the standard normalization of external states and phase space; we therefore absorb it into the normalization and suppress it henceforth, with the understanding that it is restored in the final cross section.

The resulting amplitude can be expressed in terms of field-strength insertions dressed by fundamental Wilson lines along the same lightlike path. For compactness, we introduce the parallel-transported field-strength tensor,
\begin{equation}
\overline{F}_{\mu\nu}(x^-,x_\perp)
\equiv
[\infty,x^-]_{x_\perp}\,F_{\mu\nu}(x^-,x_\perp)\,[x^-,\infty]_{x_\perp}\,,
\end{equation}
which appears throughout the following results. It is also convenient to package the double field-strength insertions into the tensors $\bb{D}_{kl}$:
\begin{align}
\label{eq:def_Dkl}
\bb{D}_{kl}(\,l^+, x^-,x_\perp)
&=  i\int \frac{\dhd k^+}{k^+-l^+ +i\epsilon}
\int dx^{\prime -} e^{i\,k^+( x^--x'^-)} [\infty,x^{\prime -}]_{x_\perp}\,
   F_{-k}(x^{\prime -},x_\perp)\,
 [x^{\prime -},x^-]_{x_\perp}\,
   F_{-l}(x^-, x_\perp)\,
 [x^-, \infty]_{x_\perp}.
\end{align}
This representation naturally reproduces the phase structure appearing in the definition of triple-gluon correlators in the cross section~\cite{Rodini:2022wki,Beppu:2010qn}. In our calculation, however, the double field-strength insertions enter in an equivalent form, obtained by evaluating the $k^+$ integral to produce a step function:
\begin{align}
\label{eq:def:theta_Dkl}
  i\int \frac{\dhd k^+}{k^+ - l^+ + i\epsilon}
  e^{i k^+(x^- - x'^-)}
  =
  e^{i l^+(x^- - x'^-)}\,
  \theta(x'^- - x^-)\,.
\end{align}

Retaining the contributions relevant for the cross section at twist-3 accuracy, we arrive at the following expressions for the amplitude. For clarity, we organize these results according to their distinct Dirac and tensor structures:\footnote{Here we also include the twist-2 contributions arising from the product of
twist-1 terms in the quark and antiquark propagators in
Eq.~\eqref{eq:mixedprop}.}

\begin{align}
 i\mathcal{M} =&- e\,\epsilon_\rho(q)\! \int \frac{dx^-  d^2 x_\perp}{2q^-}
e^{i\left(k_g^+ x^-- \Delta_{\perp}{x}_\perp\right)}    \bar{u}(k_1)  \bigg\{
  \bb{C}_1^i \gamma^\rho  
  \overline{F}_{-i}(x^-\!,{x_\perp}  )
 \! +\!
 i\left( \bb{S}_1 \sigma^{\mu\nu}\gamma^\rho
       \! + \bb{S}_2\gamma^\rho\sigma^{\mu\nu}
   \right)   \overline{F}_{\mu\nu}(x^-\!,{x_\perp} ) 
   \na
   &+\bb{C}_2^+ \,\gamma^\rho  \,\overline{F}_{+-}(x^-\!,{x_\perp} ) 
+
 \Big[ \left(  \bb{T}_1^{kl} \,\gamma^\rho+  i\,\bb{C}_3^l\,\sigma^{-k}\,\gamma^\rho + i\,\bb{C}_4^k\, \sigma^{-l}\gamma^\rho+i\,\bb{C}_5^k\,\gamma^\rho\,\sigma^{-l}\right) 
     \bb{D}_{kl}(\,l^+, x^-,x_\perp)\na &
     + \left( \bb{T}_2^{kl} \,\gamma^\rho +i\, \bb{C}_6^l\, \sigma^{-k}\,\gamma^\rho + i\,\bb{C}_7^l\,\gamma^\rho\,\sigma^{-k}  + i\,\bb{C}_8^k\,\gamma^\rho\,\sigma^{-l}\right) 
      \bb{D}^\dagger_{kl}(\,-l^+, x^-,x_\perp)\Big]_{l^+ = 0}\bigg\} v(k_2)
 \label{Eq:Mq}
\end{align}
where $\bb{C}^{\mu}$ ($\mu = +, i$), $\bb{T}^{kl}$ with $i,k,l = 1,2$,
and $\bb{S}$ multiply the corresponding Dirac structures, and we included an overall normalization factor of $2q^-$.
The ``$DF$'' structures appearing in the propagators of Eq.~\eqref{eq:mixedprop} were expressed in the basis adopted in this work (single $F$ and double $F$ insertions); for the details see App.~\ref{sec:DF_simplify}. 

Below, we provide the coefficients ($\bb{C}^{\mu}$, $\bb{T}^{kl}$ and  $\bb{S}_{1,d}$) for the generic kinematics and also near the back-to-back (correlation) limit ($\Delta_\perp / P_\perp \ll 1$) detailed in App.~\ref{App:Kinematic_setup}. Near the back-to-back kinematics, the dijet pair is characterized by the total transverse momentum ${P}_\perp = z  {k}_{1\perp} - \bar{z} {k}_{2\perp}$ and the momentum imbalance $\Delta_\perp = k_{1\perp} + k_{2\perp}$. It is also convenient to define the longitudinal momentum fractions 
$z = k_1^-/q^-$ and $\bar z = k_2^-/q^- = 1-z$
for the quark and antiquark, respectively. 
We also express our results in terms of the commonly used combination  $\epsilon_f^2 =z\bar{z}Q^2$ and adopt the shorthand notation $\Delta \cdot P \equiv \Delta_\perp \cdot P_\perp$ for the transverse momentum scalar product.
For the expansion near the back-to-back limit, we retain twist-1 operator structures up to $\mathcal{O}(\Delta_\perp / P_\perp)$ while keeping twist-2 structures at $\mathcal{O}(1)$; see Eq.~\eqref{Eq:Mq}: 
\begin{align}\label{eq:coeff_expand}
    {\bb{C}_1^i}&= \frac{(k_{ 1}^i\!-\!k_{2}^i) }{2 (k_g^+)^2}  \left(\frac{1 }{k_1^-} \!+\!\frac{1 }{k_2^-}  \right) \!+\! \frac{(k_{1}^l\!+\!k_{2}^l)}{\left(k_g^+\right)^3 } \left(\frac{k_{1}^l k_1^i}{ (k_1^-)^2 } \!-\!\frac{k_{2}^l k_2^i}{ (k_2^-)^2 }  \right) \!\approx\!
\frac{2z \bar{z}q^-}{(P_\perp^2 + \epsilon_f^2)^2} \left( {2  P^i}\!-\!{   \Delta^i ( \bar z \!-\! {z})} 
\!+\!\frac{4 (\Delta \cdot P)  P^i (\bar z \!-\! {z})}{(P_\perp^2 + \epsilon_f^2)}\right) 
, \notag\\[2pt]
    \bb{C}_2^+&=\frac{1}{2 k_g^+  }\left(\frac{1}{k_1^-}{{-}} \frac{1}{k_2^-} \right)-\frac{ 1 }{{2} (k_g^+ )^2}\left(\frac{k_{1\perp}^2}{(k_1^-)^2} {{-}} \frac{k_{2\perp}^2}{(k_2^-)^2} \right) \approx
     (z - \bar z) \frac{(P_\perp^2-\epsilon_f^2)}{(P_\perp^2 + \epsilon_f^2)^2} 
     , \notag\\[2pt]
     \bb{S}_1&=-\frac{1}{4 k_g^+ } \frac{1}{k_1^-} - \frac{ (k_{1l}+k_{2l})}{4(k_g^+ )^2}\frac{k_{1l} }{(k_1^-)^2}\approx- \frac{1}{2(P_\perp^2+\epsilon_f^2)} \left( \bar{z} +\frac{2 \bar{z}^2 (\Delta\cdot P)}{P_\perp^2+\epsilon_f^2 }\right) , \notag\\[2pt]
     \bb{S}_2&= -\frac{1}{4 k_g^+ }\frac{1}{k_2^-}  -\frac{ (k_{1l}+k_{2l})}{4(k_g^+ )^2}\frac{k_{2l} }{(k_2^-)^2}\approx  -\frac{1}{2(P_\perp^2+\epsilon_f^2)} \left( z -\frac{2 z^2 (\Delta\cdot P)}{P_\perp^2+\epsilon_f^2 }\right) ,
    \notag\\[2pt]
T^{kl}
&=
\frac{1}{(k_g^+)^3}\!
\Bigg[
\frac{k_g^+\delta^{kl}}{2}\!
\left(\frac{1}{k_1^-}\!+\!\frac{1}{k_2^-}\right)
-\left(
\frac{k_1^k k_1^l}{(k_1^-)^2}
+\frac{k_2^k k_2^l}{(k_2^-)^2}
\right)
+\left(2 - k_g^+\frac{\partial}{\partial l^+}\right)
\frac{k_1^k k_2^l}{k_1^- k_2^-}
\Bigg],
\notag\\[2pt]
\bb{T}_1^{kl}
&=T^{kl}
+\frac{i\,\delta x^-}{(k_g^+)^2}
\frac{k_1^k k_1^l}{(k_1^-)^2}
\approx \frac{2 z \bar{z}q^-
}{{(P_\perp^2+\epsilon_f^2)^2}} \left( {\delta_{kl}} -\frac{4 P^k P^l}{(P_\perp^2+\epsilon_f^2)} +  \frac{ 2 P^k P^l}{z\,q^-}\frac{\partial}{\partial l^+} \right) ,\notag\\
\bb{T}_2^{kl}&=T^{kl}
+\frac{i\,\delta x^-}{(k_g^+)^2}
\frac{k_2^k k_2^l}{(k_2^-)^2}\approx \frac{2 z \bar{z}q^-
}{{(P_\perp^2+\epsilon_f^2)^2}} \left( {\delta_{kl}} -\frac{4 P^k P^l}{(P_\perp^2+\epsilon_f^2)} + \frac{2P^k P^l}{\bar{z}\,q^-}\frac{\partial}{\partial l^+} \right)  ,
\notag\\[2pt]
\bb{C}_3^{\,l}
&= 
\frac{1}{2 k_1^- (k_g^+)^2}
\left(
\frac{k_1^{\,l}}{k_1^-}
-
\frac{k_2^{\,l}}{k_2^-}
\right) \approx  \frac{2 \bar{z} P^l}{(P_\perp^2+\epsilon_f^2)^2}  , \qquad
\bb{C}_4^{\,k}
=
-\frac{k_{1}^{\,k}}{2(k_1^-)^2}
\frac{ 1}{k_g^+} \frac{\partial}{\partial l^+} 
 \approx -\frac{ P^k}{(P_\perp^2+\epsilon_f^2)} \frac{ \bar{z} }{ zq^-}  \frac{\partial}{\partial l^+} ,\notag\\
\bb{C}_5^{\,k}
&=
 \frac{1}{2k_2^- (k_g^+)^2}\left(\frac{k_1^{\,k}}{k_1^-}-\frac{k_2^{\,k}}{k_2^-}\right) 
-\frac{ k_1^{\,k}}{2k_1^- k_2^-}\frac{1}{k_g^+} \frac{\partial}{\partial l^+}
\approx \frac{2 P^k}{(P_\perp^2+\epsilon_f^2)}\left( \frac{z}{(P_\perp^2+\epsilon_f^2)}-\frac{1}{2q^-}\frac{\partial}{\partial l^+}\right)  ,\na
\bb{C}_6^{\,l}&
=\frac{1}{2 k_1^- (k_g^+)^2}
\left(
\frac{k_1^{\,l}}{k_1^-}
-
\frac{k_2^{\,l}}{k_2^-}
\right)+\frac{ k_2^{\,l}}{2k_1^- k_2^-}\frac{1}{k_g^+} \frac{\partial}{\partial l^+}\approx \frac{2 P^l}{(P_\perp^2+\epsilon_f^2)}\left( \frac{\bar z}{(P_\perp^2+\epsilon_f^2)}-\frac{1}{2q^-}\frac{\partial}{\partial l^+}\right)
 ,
 \notag\\
 \bb{C}_7^{\,l}
&=\frac{k_{2}^{\,l}}{2(k_2^-)^2}
\frac{ 1}{k_g^+} \frac{\partial}{\partial l^+} 
 \approx -\frac{ P^l}{(P_\perp^2+\epsilon_f^2)} \frac{ z }{ \bar{z}q^-}  \frac{\partial}{\partial l^+}, \qquad
\bb{C}_8^{\,k} =
 \frac{1}{2k_2^- (k_g^+)^2}\left(\frac{k_1^{\,k}}{k_1^-}-\frac{k_2^{\,k}}{k_2^-}\right) 
\approx \frac{2 z P^k}{(P_\perp^2+\epsilon_f^2)^2}.
\end{align}

\subsection{Cross section}
\label{sec:xsect}

To obtain the differential cross section, we square the amplitude in Eq.~\eqref{Eq:Mq} and perform the Dirac traces for each independent $\gamma$-matrix structure. We present results separately for longitudinally and transversely polarized virtual photons. For longitudinal polarization, we adopt the convention\footnote{Any choice of longitudinal polarization vector $\tilde{\epsilon}_\mu^L$ satisfying $\tilde{\epsilon}_\mu^L = \epsilon_\mu^L + q_\mu$ yields the same result after contraction with the leptonic tensor.}
\begin{align}
\epsilon_\mu^L(q)
    = \frac{Q }{ q^-} { g_{\mu+}}\, ,
\end{align}
while for transverse polarization, summing over the two physical states $\lambda = \pm 1$ yields the projector
\begin{align}
{\frac{1}{2}
\sum_{\lambda=\pm 1} \epsilon_k^\lambda(q)\,\epsilon_l^{\lambda *}(q)
    = -\frac{1}{2}\, g_{kl}  } \, ,
\end{align}
with $k, l = 1, 2$ labeling the transverse directions.

\subsubsection{Longitudinally polarized photon}

For the longitudinal polarization of the virtual photon, the squared amplitude takes the following form,
\begin{align}
|i\mathcal{M}|^2_L
= & e^2\int \frac{dx^- d^2x_\perp dy^- d^2y_\perp}{(2q^-)^2} \,{e^{i {k}_g^+  (x^- - y^-)-i \Delta_{\perp}  (x_\perp - y_\perp) }}\,{8}k_1^- k_2^- \,{\rm{Tr}}\bigg[{ \bb{C}_1^i \bb{C}_1^j   }\overline{F}_{-i}(x^-,x_\perp)\overline{F}_{-j}(y^-,y_\perp) \na
&+{ \bb{C}_1^i \left(\bb{C}_2^+ +{2} \bb{S}_1-{2} \bb{S}_2\right)} \left(\overline{F}_{-i}(x^-,x_\perp)\overline{F}_{+-}(y^-,y_\perp)+\overline{F}_{+-}(x^-,x_\perp)\overline{F}_{-i}(y^-,y_\perp)\right) \na [4pt]
    &  + { \bb{C}_1^i }\left(\bb{T}_1^{kl} \,\overline{F}_{-i}(x^-,x_\perp)\,{\bb{D}_{kl}^\dagger}(\,l^+,y^-,y_\perp) +  \bb{T}_1^{kl}\,\bb{D}_{kl}(\,l^+,x^-,x_\perp)\,\overline{F}_{-i}(y^-,y_\perp) \right)_{l^+ = 0} \na
& + { \bb{C}_1^i} \left(\bb{T}_2^{kl} \,\overline{F}_{-i}(x^-,x_\perp)\,{\bb{D}_{kl}}(\,-l^+,y^-,y_\perp) + \bb{T}_2^{kl}\,{\bb{D}_{kl}^\dagger}(\,-l^+,x^-,x_\perp) \,\overline{F}_{-i}(y^-,y_\perp)\right)_{l^+ = 0}   \bigg]\,\frac{Q^2}{(q^-)^2 },
\end{align}
where the trace over fundamental indices can be rewritten as a sum over adjoint indices. Indeed, using the standard relation between Wilson lines in the fundamental and adjoint representations, we get 
\begin{equation}
[\infty,x^-]_{x_\perp} t^a\, F^a_{-i}(x^-,x_\perp) \;[x^-,\infty]_{x_\perp}
= t^b \,[x^-,\infty]^{ba}_{x_\perp}\, F^a_{-i}(x^-,x_\perp) \,
\end{equation}
where $[x^-,\infty]^{ba}_{x_\perp}$ denotes the Wilson line in the adjoint representation.
Consequently, the color trace appearing in the cross section produces color factors involving ${\rm Tr}[t^a\,t^b]$ for double-F operators and ${\rm Tr}[t^a\,t^b\,t^c]$ for triple-F operators.
We prefer to work in the fundamental representation.

Performing the expansion near the back-to-back limit  $\Delta_\perp / P_\perp$ and retaining all relevant contribution to the twist-3 precision we get 
\begin{align}\label{eq:lg_btb}
|i\mathcal{M}|^2_L
\approx\;&
e^2 \int  {dx^- d^2x_\perp dy^- d^2y_\perp} \,{e^{i {k}_g^+  (x^- - y^-)-i \Delta_{\perp}  (x_\perp - y_\perp) }}
\frac{16\,z^2\bar z^2\,\epsilon_f^2}{(P_\perp^2+\epsilon_f^2)^4}
\,{\rm Tr}\Big[
 {O}_1
+ {O}_2
+ {O}_3
+ {O}_4
\Big]
+\mathcal{O}\left(\frac{\Delta_\perp^2}{P_\perp^2}\right),
\end{align}
where the operator structures are classified according to their field content. Specifically,
\begin{align}\label{eq:coeff_btb_logi}
O_1
=&\;
\Big[
2P^iP^j
-(\bar z-z)(P^i\Delta^j+P^j\Delta^i)
+8(\bar z-z)\frac{(\Delta\!\cdot\!P)P^iP^j}{P_\perp^2+\epsilon_f^2}
\Big]
\overline{F}_{-i}(x^-,x_\perp)\,\overline{F}_{-j}(y^-,\, y_\perp),
\na [1pt]
O_2
=&
\frac{(z-\bar z)}{z\bar z\,q^-}\,
P^i P_\perp^2
\Big[
\overline{F}_{-i}(x^-,x_\perp)\,\overline{F}_{+-}(y^-,\, y_\perp)
+\overline{F}_{+-}(x^-,x_\perp)\,\overline{F}_{-i}(y^-,\, y_\perp)
\Big],
\na[4pt]
O_3
=&\;
P^i\, \Big[\mathcal{H}_{kl}(z)\,
\overline{F}_{-i}(x^-,x_\perp)\,{\bb{D}_{kl}}^\dagger(\,l^+,y^-,y_\perp) 
+
\mathcal{H}_{kl}(z)\,{\bb{D}_{kl}}(\,l^+,x^-,x_\perp) \,\overline{F}_{-i}(y^-,\, y_\perp)
\Big]_{l^+ = 0},
\na[4pt]
O_4 
=&\;
P^i\,\Big[\mathcal{H}_{kl}(\bar z)\,
\overline{F}_{-i}(x^-,x_\perp)\,{\bb{D}_{kl} }(\,-l^+,y^-,y_\perp) 
+\mathcal{H}_{kl}(\bar z)\,
{\bb{D}_{kl}^\dagger}(\,-l^+,x^-,x_\perp)\,\overline{F}_{-i}(y^-,\, y_\perp)
\Big]_{l^+ = 0} , 
\end{align}
where we  defined
\begin{align}\label{eq: H_kl}
\mathcal{H}_{kl}(z)
\equiv
\delta_{kl}-\frac{4P^kP^l}{P_\perp^2+\epsilon_f^2}+\frac{2P^k P^l }{ z q^-}\frac{\partial}{\partial l^+}.
\end{align} 
 Note that the gluon momentum component $k_g^+$ entering the Fourier phase should also be expanded; its explicit expression is given in Eq.~\eqref{eq: kg_btb}.
Substituting the squared amplitude into Eq.~\eqref{eq:diff_xsec_def},
we obtain the differential cross section for longitudinally polarized
virtual photons. 

In Sec.~\ref{sec:cmp_smlx}, we compare our  general-$x$ factorization framework result with the one obtained in the the small-x approach.

\subsubsection{Transversely polarized photon}
For transverse polarization, the squared amplitude is more involved due to additional tensor structures; in particular, the $\sigma^{-i}$ term now contributes. The result can nevertheless be organized systematically as a sum of operator structures multiplied by kinematic coefficients:
\begin{align}
\label{eq:trans_gen}
|i\mathcal{M}|_T^2 = &\,4 \,e^2   \int  \frac{dx^- d^2x_\perp dy^- d^2y_\perp}{(2q^-)^2}\,{e^{i {k}_g^+  (x^- - y^-)-i\Delta_{\perp}  (x_\perp - y_\perp) }}\,{\rm{Tr}}\,
\bigg[
 A^{ij}\, \overline{F}_{-i}(x^-,x_\perp)\, \overline{F}_{-j}(y^-,y_\perp) \na
&
+ B^i\,\Big(\overline{F}_{-i}(x^-,x_\perp)\,\overline{F}_{+-}(y^-,y_\perp)
        +\overline{F}_{+-}(x^-,x_\perp)\,\overline{F}_{-i}(y^-,y_\perp)\Big) \na [4pt]
& 
+ C^i \Big(\overline{F}_{ij}(x^-,x_\perp)\overline{F}_{-j}(y^-,y_\perp)
        \! +\!\overline{F}_{-j}(x^-,x_\perp)\overline{F}_{ij}(y^-,y_\perp)\Big)   \na[4pt]
        &+ \left(  E^{ikl} \,\overline{F}_{-i}(x^-,x_\perp)\,{\bb{D}_{kl}^\dagger}(\,l^+,y^-,y_\perp)+ E^{ikl}\,{\bb{D}_{kl}}(\,l^+,y^-,y_\perp)\,\overline{F}_{-i}(y^-,y_\perp)\right)_{l^+=0} \na[4pt]
& 
+ \left( G^{ikl} \, \overline{F}_{-i}(x^-,x_\perp)\,{\bb{D}_{kl} }(\,-l^+,y^-,y_\perp)
+G^{ikl}\, {\bb{D}_{kl}^\dagger}(\,-l^+,x^-,x_\perp)\,\overline{F}_{-i}(y^-,y_\perp) \right)_{l^+=0} \bigg]\,,
\end{align}
with
\begin{align}
A^{ij} &=  (k_1^+ k_2^- + k_1^- k_2^+) \,\bb{C}_1^i \bb{C}_1^j
         - 8 g^{ij} k_1^- k_2^- (\bb{S}_1^2 + \bb{S}_2^2)
        - 2\Big[
             k_1^- \bb{S}_2 (\bb{C}_1^i k_2^j + \bb{C}_1^j k_2^i)
           - k_2^- \bb{S}_1 (\bb{C}_1^i k_1^j + \bb{C}_1^j k_1^i)
           \Big]  ,\na
 [4pt]
B^i &= -k_2^- \big(2(\bb{S}_1+\bb{S}_2)-\bb{C}_2^+\big)
          (2\bb{S}_1 k_1^i + k_1^+ \bb{C}_1^i)
   - k_1^- \big(2(\bb{S}_1+\bb{S}_2)+\bb{C}_2^+\big)
          (2\bb{S}_2 k_2^i - k_2^+ \bb{C}_1^i)  ,\na
 [4pt]
C^i &= 8\big(\bb{S}_1 k_1^i k_2^- - \bb{S}_2 k_2^i k_1^-\big)
       (\bb{S}_1- \bb{S}_2)   ,\na
 [4pt]
D^i &= 2\,(\bb{S}_1 k_1^i k_2^- - \bb{S}_2 k_2^i k_1^-)
      + \bb{C}_1^i (k_1^+ k_2^- + k_2^+ k_1^-)
  ,\na
 [4pt]
E^{ikl} &= -\Big[
4\, k_1^-k_2^-\,(  g^{ik}\,\bb{S}_1 \bb{C}_3^l  + g^{il} \,\bb{S}_1 \bb{C}_4^k
          + g^{il}\, \bb{S}_2 \bb{C}_5^k )
 - \bb{C}_1^i\big( k_2^- (k_1^k\,\bb{C}_3^l+k_1^l\,\bb{C}_4^k)
                     - k_1^- k_2^l \bb{C}_5^k\big)
\Big]+D^i\, \bb{T}_1^{kl}  ,\na
 [4pt]
G^{ikl}&= -\Big[
 4\, k_1^-k_2^-\,(g^{ik} \,\bb{S}_1 \bb{C}_6^l +g^{ik} \,\bb{S}_2 \bb{C}_7^l 
          + g^{il} \,\bb{S}_2 \bb{C}_8^k)
 + \bb{C}_1^i\big( k_1^- (k_2^k\,\bb{C}_7^l+k_2^l\,\bb{C}_8^k)
                     - k_2^- k_1^k \bb{C}_6^l\big)
\Big]+D^i \,\bb{T}_2^{kl}  .
\end{align}
In the back--to--back limit, the expansion in $\Delta_\perp/P_\perp$, the coefficient functions reduce to
\begin{align}\label{eq:coeff_btb}
  A^{ij}_{\rm btb}&=\frac{2z\bar{z}(z^2+\bar{z}^2)}{({P}_\perp^2 +\epsilon_f^2)^2 } \Bigg[\delta^{ij} \left(1+\frac{2(\bar{z}-z)(\Delta\cdot P)}{({P}_\perp^2  +\epsilon_f^2 )}\right)
  -\frac{4P^iP^j \epsilon_f^2}{({P}_\perp^2  +\epsilon_f^2 )^2 }\left(1 -\frac{4(z-\bar z)(\Delta\!\cdot\! P) }{(P_\perp^2+\epsilon_f^2)}\right) -\frac{2(z-\bar z) \epsilon_f^2 }{({P}_\perp^2  +\epsilon_f^2 )^2  } (P^i \Delta^j+\Delta^i P^j)\na[4pt]
  &\qquad+\frac{(z-\bar{z}) }{({P}_\perp^2 +\epsilon_f^2)(z^2+\bar z^2) } \Big( (P^i \Delta^j+\Delta^i P^j) -2 (\Delta\cdot P)\delta^{ij}\Big) \Bigg](q^-)^2 \na
 [4pt]
  B^i_{\rm btb}&=-2z\bar{z}(z-\bar{z}) \frac{(P_\perp^2+\epsilon_f^2+(z^2+\bar{z}^2)Q^2)({P}_\perp^2 -\epsilon_f^2)}{({P}_\perp^2 +\epsilon_f^2)^4} P^i q^-,\qquad
  C^i_{\rm btb}=-\frac{2(z-\bar{z})(z^2+\bar{z}^2)}{({P}_\perp^2 +\epsilon_f^2)^2} P^i q^-,\na
 [4pt]
 \mathcal{K}^{ikl}&=\frac{2\,(q^-)^2\, z\,\bar z}{(P_\perp^2 +\epsilon_f^2)^3}
\left(
-2
\bigl(
 z^{\,2} \delta^{il} P^{k} + \bar z^{\,2}\delta^{ik} P^{l}
\bigr)
+
\frac{
8\,\epsilon_f^{2}\,(\bar z^{\,2}+z^{2})\,P^i P^k P^l
}{
\bigl(P^{2}+\epsilon_f^{2}\bigr)^{2}
}
+
\frac{
\delta^{kl}\,(\bar z^{\,2}+z^{2})\,P^{i}\,
\bigl(P^{2}-\epsilon_f^{2}\bigr)
}{
P^{2}+\epsilon_f^{2}
}
\right)\na
 [4pt]
 E^{ikl}_{\rm btb}&= \mathcal{K}^{ikl}
  -\frac{2 \, q^-\, \bar z\,(\bar z^{2}+z^{2})}{ \bigl(P^{2}+\epsilon_f^{2}\bigr)^2 }
\left[
{ -\delta^{il} P^k } 
+
\frac{ 4\,\epsilon_f^{2}\, P^i P^k P^l }{ \bigl(P^{2}+\epsilon_f^{2}\bigr)^2 }
\right]\frac{\partial}{\partial l^+},
\na
 [4pt]
 G^{ikl}_{\rm btb}&= \mathcal{K}^{ikl}-
\frac{2 \,q^- z\,(\bar z^{2}+z^{2}) }{\bigl(P^{2}+\epsilon_f^{2}\bigr)^{2}}
\left[
{- \delta^{ik} P^l } 
+
\frac{ 4\,\epsilon_f^{2}\, P^i P^k P^l }{ \bigl(P^{2}+\epsilon_f^{2}\bigr)^2 }
\right] \frac{\partial}{\partial l^+}
.
\end{align}

According to App.~\ref{sec:appendix_trnsfr}, the twist-3 operator
\(F_{ij}F_{-j}\) can be traded for \(F_{-i}F_{-j}\) multiplying a  kinematic higher--twist correction:  
\begin{equation}
\left(P^{i}\Delta^{j}-(\Delta\!\cdot\! P)\,\delta^{ij}\right)
F_{-i}F_{-j},
\end{equation}
and an additional genuine twist-3 contributions involving the three--gluon
operator \(F_{-i}F_{-j}F_{-k}\). This type of relationship between kinematic and dynamic twist operators is well established for quarks (see Refs.~\cite{Gamberg:2022lju,Boer:1997mf}). It follows from equations of motion (which includes the Bianchi identity) and was used to reduce the number of operators in the cross section. 
Therefore, the above expression can be rewritten as
\begin{align}
\label{eq:trans_kin_dyn}
|i\mathcal{M}|_T^2 =&\,4 \,e^2   \int  \frac{dx^- d^2x_\perp dy^- d^2y_\perp}{(2q^-)^2}\,{e^{i {k}_g^+  (x^- - y^-)-i\Delta_{\perp}  (x_\perp - y_\perp) }}\,{\rm{Tr}}\,
\bigg[
 \bar{A}^{ij}\, \overline{F}_{-i}(x^-,x_\perp)\, \overline{F}_{-j}(y^-,y_\perp) \na
&
+ \bar{B}^i\,\Big(\overline{F}_{-i}(x^-,x_\perp)\,\overline{F}_{+-}(y^-,y_\perp)
        +\overline{F}_{+-}(x^-,x_\perp)\,\overline{F}_{-i}(y^-,y_\perp)\Big) \na [4pt]
        &+ \left(  \bar{E}^{ikl} \,\overline{F}_{-i}(x^-,x_\perp)\,{\bb{D}_{kl}^\dagger}(\,l^+,y^-,y_\perp)+  \bar{E}^{ikl}\,{\bb{D}_{kl}}(\,l^+,x^-,x_\perp)\,\overline{F}_{-i}(y^-,y_\perp)\right)_{l^+=0} \na[4pt]
& 
+ \left(  \bar G^{ikl} \, \overline{F}_{-i}(x^-,x_\perp)\,{\bb{D}_{kl}}(\,-l^+,y^-,y_\perp)
+\bar G^{ikl}\, {\bb{D}_{kl}^\dagger}(\,-l^+,x^-,x_\perp)\,\overline{F}_{-i}(y^-,y_\perp) \right)_{l^+=0} \bigg]\,,
\end{align}
with
\begin{align}
\label{eq:trans_kin_dyn_coeffs}
\bar{A}^{ij}=&{A}^{ij}-\bar{C}(P^i\Delta^j-(\Delta\!\cdot\!P)\delta^{ij}),
\qquad
\bar{C} = { 8}\,\big(\bb{S}_1 k_2^- - \bb{S}_2 k_1^-\big)
       (\bb{S}_1- \bb{S}_2) /k_g^+  ,\na
 [4pt]
\bar{E}^{ikl} =& {E}^{ikl}
+\bar{C}\,(P^k\delta^{il}-P^l\delta^{ik}),
\qquad
\bar{G}^{ikl}={G}^{ikl}+\bar{C}\,(P^k\delta^{il}-P^l\delta^{ik}).
\end{align}
In the back-to-back limit, $\bar{C}$ simplifies to 
\begin{align}
\label{eq:trans_btb_coeff2}
\bar{C}_{\rm btb}
= \,\frac{4\,z\bar z(z-\bar{z})^2}
       {\bigl(P_\perp^2+\epsilon_f^2\bigr)^3}\,
(q^-)^2 .
\end{align}
The obtained Eqs.~\eqref{eq:trans_kin_dyn}--\eqref{eq:trans_btb_coeff2}  provide the factorized expression for dijet cross section for the transversely polarized photon in the back--to--back limit up to twist-3 accuracy.

{
Our TMD-factorized results are not limited to DIS kinematics with large $Q^2$. In the back-to-back limit, the hard scale can be set by the relative transverse momentum $P_\perp$ of the dijet system rather than by the photon virtuality. Consequently, the TMD-factorized cross section and twist expansion remain well defined even as $Q^2 \to 0$, as relevant for photoproduction or ultra-peripheral collisions. In this limit, one sets $\epsilon_f^2 = z\bar{z} Q^2 \to 0$ in the expressions above, and the longitudinal photon contribution becomes negligible, leaving only non-zero transverse polarization.
}

{The structure of the cross section further exhibits a useful simplifying pattern. Near the symmetric point $z \simeq 1/2$, the cross section simplifies (see Eqs.~\eqref{eq:lg_btb}--\eqref{eq: H_kl} and \eqref{eq:trans_gen}--\eqref{eq:trans_btb_coeff2}), leaving only contributions involving $F_{-i}$. This provides a clean starting point for isolating that component. Moving away from $z=1/2$ activates additional operator structures, allowing, in principle, for a systematic separation of the remaining gluon TMDs.}

\section{Small-x limit}\label{sec:cmp_smlx}

In Ref.~\cite{Altinoluk:2024zom} back-to-back quark-antiquark dijet production was studied at leading order in strong coupling within the high-energy CGC framework, valid in the eikonal $x\to0$ limit, and results for the cross section was obtained up to kinematic and dynamic twist-3 through leading sub-eikonal expansion around $x=0$. In contrast, our results obtained through the MSTT TMD framework are valid for any general values of $x$. We, thus, can establish a direct connection between the two approaches by taking the small-$x$ limit of our general expressions. In this section, we demonstrate that our cross section expressions for both longitudinally and transversely polarized virtual photons in the small-$x$ limit reproduce the known CGC results of Ref.~\cite{Altinoluk:2024zom}.

\subsection{Longitudinally polarized photon}
For concreteness, we first consider the cross section for a longitudinally polarized virtual photon, given in Eq.~\eqref{eq:lg_btb}.
To connect with the CGC formulation, we expand the phase factor $e^{ik_g^+(x^- - y^-)}$ around $k_g^+ = xP^+ \to 0$:
\begin{align}\label{eq:lg_btb_eik}
|i\mathcal{M}|^2_L
=\;&
e^2 \int  {dx^- d^2x_\perp dy^- d^2y_\perp} \,{e^{-i \Delta_{\perp}  (x_\perp - y_\perp) }}\big(1 + i {k}_g^+  (x^- - y^-)\big)
\frac{16\,z^2\bar z^2\,\epsilon_f^2}{(P_\perp^2+\epsilon_f^2)^4}
\,{\rm Tr}\Big[
 {O}_1
+ {O}_2
+ {O}_3
+ {O}_4
\Big]
+\mathcal{O}\left(\frac{\Delta_\perp^2}{P_\perp^2}\right),
\end{align}
where the leading term contributes at both eikonal and sub-eikonal order, while the linear term contributes only at sub-eikonal order. In the small-$x$ limit, we find exact agreement with the results of Ref.~\cite{Altinoluk:2024zom}. Specifically, the two-body structures in ${O}_1$ and ${O}_2$ reproduce the corresponding terms at sub-eikonal order: the leading-twist contribution proportional to $\overline{F}_{-i}(x^-, x_\perp) \overline{F}_{-j}(y^-, y_\perp)$ matches Eq.~(115), while the twist-3 term involving $\overline{F}_{-i}(x^-, x_\perp) \overline{F}_{+-}(y^-, y_\perp)$ agrees with Eq.~(117).

The correspondence between our result in the small-$x$ limit and Ref.~\cite{Altinoluk:2024zom} for the three-body operators in $ {O}_3$ and ${O}_4$ is less direct. Considering first the eikonal limit,
\begin{align}
|i\mathcal{M}|^2_{L, \rm eik}\:=\;&
e^2 \int  {dx^- d^2x_\perp dy^- d^2y_\perp} \,{e^{-i \Delta_{\perp}  (x_\perp - y_\perp) }}
\frac{16\,z^2\bar z^2\,\epsilon_f^2}{(P_\perp^2+\epsilon_f^2)^4}
\,{\rm Tr}\Big[
 {O}_1
+ {O}_3
+ {O}_4
\Big]
+\mathcal{O}\left(\frac{\Delta_\perp^2}{P_\perp^2}\right).
\end{align}
In the eikonal limit, all longitudinal phase factors vanish, including those in the three-body operators. The three-body contribution ${O}_3 + {O}_4$ then becomes
\begin{align}
\label{eq:theta_remove}
 {O}_3
+ {O}_4\:
=&\,
P^i\,\tilde{\mathcal{H}}_{kl} \Big[
\overline{F}_{-i}(x^-,x_\perp)\,{\bb{D}_{kl}^\dagger}(0,y^-,y_\perp) 
+
{\bb{D}_{kl}}(\,0,x^-,x_\perp) \,\overline{F}_{-i}(y^-,\, y_\perp)
+\,
\overline{F}_{-i}(x^-,x_\perp)\,{\bb{D}_{kl}}(\,0,y^-,y_\perp) 
\na &+\,
{\bb{D}_{kl}^\dagger}(\,0,x^-,x_\perp)\,\overline{F}_{-i}(y^-,\, y_\perp)\Big]\na
=& \,
2\,P^i\,\tilde{\mathcal{H}}_{kl} \,{\rm Re}\Big[\overline{F}_{-i}(x^-,x_\perp)\,{\bb{D}_{kl}}(0,y^-,y_\perp)  
+
\overline{F}_{-i}(x^-,x_\perp)\,{\bb{D}_{kl}^\dagger}(\,0,y^-,y_\perp) \Big]
\end{align}
with 
\begin{align}
\tilde{\mathcal{H}}_{kl}(z)
\equiv
\delta_{kl}-\frac{4P^kP^l}{P_\perp^2+\epsilon_f^2}.
\end{align}
In Eq.~\eqref{eq:theta_remove}, the $\theta$-function contributions
from $\bb{D}_{kl}$ and $\bb{D}_{kl}^\dagger$
[see Eq.~\eqref{eq:def:theta_Dkl}] cancel pairwise.

The relevant (involving three body TMDs) sub-eikonal contribution to longitudinal polarization is given by 
\begin{align}
|i\mathcal{M}|^2_{L, \rm \,Seik}
=\;&
e^2 \int  {dx^- d^2x_\perp dy^- d^2y_\perp} \,{e^{-i \Delta_{\perp}  (x_\perp - y_\perp) }}\big( i {k}_g^+  (x^- - y^-)\big)
\frac{16\,z^2\bar z^2\,\epsilon_f^2}{(P_\perp^2+\epsilon_f^2)^4}
\,{\rm Tr}\Big[
 {O}_3
+ {O}_4
\Big]
+\ldots\,   
\end{align}
At sub-eikonal order, $ {O}_3$ and $ {O}_4$ decompose into two independent tensor structures proportional to $\delta_{kl}$ and $P^k P^l$. We treat these separately, beginning with the $\delta_{kl}$ term:\footnote{For clarity, we suppress the ``$2\,\mathrm{Re}$" projection, the transverse convolution (the associated $x_\perp,y_\perp$ integrations), and the overall kinematic coefficients in the intermediate steps.}

\begin{align}
\mathcal{I}_1\: = &
\int  {dx^-  dy^- }  P^i\,\delta_{kl} \:  (x^- - y^-)\,
\Big(
\overline{F}_{-i}(x^-,x_\perp)\,{\bb{D}_{kl}}(0,y^-,y_\perp) 
+
\overline{F}_{-i}(x^-,x_\perp)\,{\bb{D}_{kl}^\dagger}(\,0,y^-,y_\perp)  \Big) .
\end{align}
Expanding the definitions of ${\bb D}_{kl}(0,y^-,y_\perp)$ and
${\bb D}_{kl}^\dagger(0,y^-,y_\perp)$, and performing a change of integration variables in the latter, we obtain
\begin{align}
\mathcal{I}_1\:
=&  \int dx^-  dy^-   dy'^- (x^- - {\rm min}(y^-,y'^-))\:P^i\,\delta_{kl}\:  \overline{F}_{-i}(x^-,x_\perp)\:\overline{F}_{-k}(y'^-,y_\perp) \overline{F}_{-l}(y^-,y_\perp).
\end{align}
We now turn to the second tensor structure, proportional to $P^k P^l$.
Its explicit form can be written as
\begin{align}
\mathcal{I}_2\:
=&- \int dx^- dy^-\:P^i\,\frac{4 P^k P^l}{P_\perp^2 + \epsilon_f^2} \,\bigg(\overline{F}_{-i}(x^-,x_\perp) \int_{y^-}^{\infty}dy'^-\:(x^- - y^-)\:\overline{F}_{-k}(y'^-,y_\perp) \overline{F}_{-l}(y^-,y_\perp) \na &+ \overline{F}_{-i}(x^-,x_\perp)   \int_{y^-}^{\infty}dy'^-\:(x^- - y^-)\:\overline{F}_{-k}(y^-,y_\perp) \overline{F}_{-l}(y'^-,y_\perp) \bigg)\na &
- \int dx^- dy^-\:P^i\,\frac{4 P^k P^l}{P_\perp^2 + \epsilon_f^2}\bigg(\:\bar{z}\,\overline{F}_{-i}(x^-,x_\perp) \int_{y^-}^{\infty}dy'^-\:(y^- - y'^-)\:\overline{F}_{-k}(y'^-,y_\perp) \overline{F}_{-l}(y^-,y_\perp) \na &+ z \, \overline{F}_{-i}(x^-,x_\perp)   \int_{y^-}^{\infty}dy'^-\:(y^- - y'^-)\:\overline{F}_{-k}(y^-,y_\perp) \overline{F}_{-l}(y'^-,y_\perp)\bigg)
\end{align}
The first contribution originates from the expansion of the Fourier phase associated with $k_g^+$, while the second one arises from taking the derivative of the exponential factor involving $l^+$. Adding these two contributions, we obtain
\begin{align}
\mathcal{I}_2\: 
=&- \int dx^- dy^-\:P^i\,\:\frac{4 P^k P^l}{P_\perp^2 + \epsilon_f^2} \bigg( \overline{F}_{-i}(x^-,x_\perp) \int_{y^-}^{\infty}dy'^-\:(x^- - y^- + \bar{z} (y^- - y'^-))\:\overline{F}_{-k}(y'^-,y_\perp) \overline{F}_{-l}(y^-,y_\perp) \na &+ \overline{F}_{-i}(x^-,x_\perp)   \int_{y^-}^{\infty}dy'^-\:(x^- - y^- + z(y^- - y'^-))\:\overline{F}_{-k}(y^-,y_\perp) \overline{F}_{-l}(y'^-,y_\perp) \bigg)
\end{align}
In the second term, we perform a change of integration variables, which allows us to rewrite the integration range in a symmetric form. 
Combining the two terms and rearranging the integration variables, we arrive at
\begin{align}
\mathcal{I}_2\:&
=- \int dx^- dy^- dy'^-\:(x^- - z y^- - \bar{z} y'^-)\:P^i\,\frac{4 P^k P^l}{P_\perp^2 + \epsilon_f^2} \overline{F}_{-i}(x^-,x_\perp)\:\overline{F}_{-k}(y'^-,y_\perp) \overline{F}_{-l}(y^-,y_\perp).  
\end{align}

Collecting the two contributions, the complete eikonal and subeikonal three-body result for longitudinal polarization is given by
\begin{align}
&|i\mathcal{M}|^2_{L, \rm \,eik+sub}\bigg{|}_{FFF}\na
& =2e^2 \, {\rm Re}\int  {dx^- dy^- dy'^- d^2x_\perp d^2y_\perp} {e^{-i \Delta_{\perp}  (x_\perp - y_\perp) }}
\frac{16 z^2\bar z^2 \epsilon_f^2 }{(P_\perp^2+\epsilon_f^2)^4}  \,P^i \Bigg\{ \delta_{kl} \left[1+i\frac{(P_\perp^2+\epsilon_f^2)}{2z\bar z q^-} (x^- \!- {\rm min}(y^-,y'^-)) \right]  \na 
& \quad -\frac{4 P^k P^l}{P_\perp^2 + \epsilon_f^2} \left[1+i \frac{(P_\perp^2+\epsilon_f^2)}{2z\bar z q^-}(x^- - z y^- - \bar{z} y'^-)  \right]\Bigg\}
\,{\rm Tr}\Big[
\overline{F}_{-i}(x^-,x_\perp)\:\overline{F}_{-k}(y'^-,y_\perp) \overline{F}_{-l}(y^-,y_\perp)
\Big]
\end{align}
This expression is in exact agreement with the result obtained
in Ref.~\cite{Altinoluk:2024zom}.

\subsection{Transversely polarized photon}

For transverse photon polarization, the comparison proceeds similarly. At the level of individual operator structures, only the $F_{+-}F_{-i}$ contribution in Eq.~\eqref{eq:trans_gen} reproduces   Ref.~\cite{Altinoluk:2024zom} directly. In that work, part of the kinematic twist-3 contributions are reexpressed in terms of dynamic twist-3 operators, whereas in our formulation these kinematic effects remain embedded in the leading-twist sector. To make the correspondence explicit, we reorganize the kinematic twist-3 terms into $F_{ij}F_{-i}$ and $F_{-i}F_{-k}F_{-l}$ structures using Eq.~\eqref{eq: knmt_trsf_dynm}. Specifically, the second line of $A^{ij}_{\mathrm{btb}}$ can be recast in terms of the coefficient functions $C^i_{\mathrm{btb}}$ and $\mathcal{K}^{ikl}$. The coefficients in Eq.~\eqref{eq:trans_gen} then become
\begin{align}\label{eq:coeff_btb_trans}
 \tilde{A}^{ij}_{\rm btb}&=\!\frac{2 (q^-)^2z\bar{z}(z^2+\bar{z}^2)}{({P}_\perp^2 +\epsilon_f^2)^2 } \Bigg[\delta^{ij} \!\left(\!1+\frac{2(\bar{z}-z)(\Delta\cdot P)}{({P}_\perp^2  +\epsilon_f^2 )}\!\right)\!
  -\frac{4P^iP^j \epsilon_f^2}{({P}_\perp^2  +\epsilon_f^2 )^2 }\!\left(\!1 -\frac{4(z-\bar z)(\Delta\!\cdot\! P) }{(P_\perp^2+\epsilon_f^2)}\!\right)\! -\!\frac{2(z-\bar z) \epsilon_f^2 }{({P}_\perp^2  +\epsilon_f^2 )^2  } (P^i \Delta^j+\Delta^i P^j)  \Bigg], \na
 [4pt]
  \tilde{C}^i_{\rm btb}&=-\frac{2(z-\bar{z})z\bar{z}}{({P}_\perp^2 +\epsilon_f^2)^2} P^i q^-,\qquad
 \tilde{\mathcal{K}}^{ikl}=\frac{2\,(q^-)^2\, z\,\bar z(\bar z^{\,2}+z^{2})}{(P_\perp^2 +\epsilon_f^2)^3}
\left(
-\bigl(
  \delta^{il} P^{k} + \delta^{ik} P^{l}
\bigr)
+
\frac{
8\,\epsilon_f^{2} \,P^i P^k P^l
}{
\bigl(P^{2}+\epsilon_f^{2}\bigr)^{2}
}
+
\frac{
\delta^{kl}\,P^{i}\,
\bigl(P^{2}-\epsilon_f^{2}\bigr)
}{
P^{2}+\epsilon_f^{2}
}
\right)\na
\end{align}
It is straight forward to reproduce  $\tilde{A}^{ij}_{\rm btb}$ and
$\tilde{C}^i_{\rm btb}$.
We therefore concentrate on the triple-$F$ contribution, which is not as trivial. 

At the eikonal level, the triple-$F$ term reads
\begin{align}
\label{eq:trans_gen_eik}
|i\mathcal{M}|_{T,{\rm eik}}^2 = &\,8 \,e^2\, {\rm Re}  
\int  \frac{dx^- d^2x_\perp dy^- d^2y_\perp}{(2q^-)^2}\,
e^{-i\Delta_{\perp}  (x_\perp - y_\perp)}\,
\tilde{\mathcal{K}}^{ikl}\,
{\rm Tr}\Big[
\overline{F}_{-i}(x^-,x_\perp)\,
\overline{F}_{-k}(y'^-,y_\perp)
\overline{F}_{-l}(y^-,y_\perp)
\Big]
+ \cdots .
\end{align}
Beyond the eikonal approximation, the sub-eikonal contribution decomposes into four independent tensor structures proportional to $\delta^{il}P^k$, $\delta^{ik}P^l$, $P^i P^k P^l$, and $\delta^{kl}P^i$. The contributions proportional to $\delta^{kl}P^i$ and  $P^i P^k P^l$ coincide with those in the longitudinal case, differing only in their coefficient functions, and therefore require no further discussion. The remaining nontrivial contributions are those proportional to $\delta^{il}P^k$ and $\delta^{ik}P^l$.
We first consider the $\delta^{il}P^k$ term,
 \begin{align}\label{eq:P3_step1}
  \mathcal{I}_3 =& -\int dx^- dy^-\frac{   (\bar z^{\,2}+z^{2})}{(P_\perp^2 +\epsilon_f^2)^2} \,\bigg(\overline{F}_{-i}(x^-,x_\perp) \int_{y^-}^{\infty}dy'^-\:(x^- - y^-)\:\overline{F}_{-k}(y'^-,y_\perp) \overline{F}_{-l}(y^-,y_\perp) \na &+ \overline{F}_{-i}(x^-,x_\perp)   \int_{-\infty}^{y^-}dy'^-\:(x^- - y'^-)\:\overline{F}_{-k}(y'^-,y_\perp) \overline{F}_{-l}(y^-,y_\perp) \bigg)\na &
-\int dx^- dy^-\frac{  2\,(\bar z^{\,2}+z^{2})}{(P_\perp^2 +\epsilon_f^2)^2}  \bigg(\:\bar{z}\, \overline{F}_{-i}(x^-,x_\perp) \int_{y^-}^{\infty}dy'^-\:(y^- - y'^-)\:\overline{F}_{-k}(y'^-,y_\perp) \overline{F}_{-l}(y^-,y_\perp) \bigg)
\end{align}
The former contribution follows from the $k_g^+$ expansion of the Fourier phase, whereas the latter is produced by differentiating the exponential with respect to $l^+$. The associated $k_g^+$ factors are absorbed into the coefficients. Using $(x^- - y'^-)=(x^- - y^-)+(y^- - y'^-)$ in the first term of
Eq.~\eqref{eq:P3_step1}, and combining the two pieces proportional to $(x^- - y^-)$, we can rewrite
\begin{align}
  \mathcal{I}_3 =& -\int dx^- dy^-\frac{   (\bar z^{\,2}+z^{2})}{(P_\perp^2 +\epsilon_f^2)^2} \,\bigg(\overline{F}_{-i}(x^-,x_\perp) \int_{-\infty}^{\infty}dy'^-\:(x^- - y^-)\:\overline{F}_{-k}(y'^-,y_\perp) \overline{F}_{-l}(y^-,y_\perp) \na &
  + \overline{F}_{-i}(x^-,x_\perp)   \int_{-\infty}^{y^-}dy'^-\:(y^- - y'^-)\:\overline{F}_{-k}(y'^-,y_\perp) \overline{F}_{-l}(y^-,y_\perp) \bigg)\na &
-\int dx^- dy^-\frac{  2\,(\bar z^{\,2}+z^{2})}{(P_\perp^2 +\epsilon_f^2)^2}  \bigg(\:\bar{z}\,\overline{F}_{-i}(x^-,x_\perp) \int_{y^-}^{\infty}dy'^-\:(y^- - y'^-)\:\overline{F}_{-k}(y'^-,y_\perp) \overline{F}_{-l}(y^-,y_\perp) \bigg) 
\end{align}
Repeating the above manipulations, i.e., \ recombining the integration regions and reorganizing the kernels, we obtain
\begin{align}
  \mathcal{I}_3 =& -\int dx^- dy^-dy'^-\frac{   (\bar z^{\,2}+z^{2})}{(P_\perp^2 +\epsilon_f^2)^2}   \Big[x^- -y'^- - (\bar{z}-z)(y'^- - y^-)\theta(y'^- - y^-)\Big] \overline{F}_{-i}(x^-,x_\perp) \overline{F}_{-k}(y'^-,y_\perp) \overline{F}_{-l}(y^-,y_\perp) 
\end{align}
One proceeds with the $\delta^{ik}P^l$ along the same lines. 

Collecting the eikonal and sub-eikonal pieces, we obtain the full three-body contribution to the transverse photon polarization,
\begin{align}
&|i\mathcal{M}|^2_{T, \rm \,eik+sub}\bigg{|}_{FFF}\na
& =2e^2{\rm Re}\int  {dx^- dy^- dy'^- d^2x_\perp d^2y_\perp} {e^{-i \Delta_{\perp}  (x_\perp - y_\perp) }}
\frac{2z\bar z (z^2 +\bar z^2) }{(P_\perp^2+\epsilon_f^2)^3}  \Bigg\{ 
\frac{
P^i\delta^{kl}
\bigl(P^{2}-\epsilon_f^{2}\bigr)
}{
P^{2}+\epsilon_f^{2}
} \left[1+\!i\frac{(P_\perp^2+\epsilon_f^2)}{2z\bar z q^-} (x^- \!- {\rm min}(y^-,y'^-)) \right]  \na 
& \quad \! +\!
\frac{
8\epsilon_f^{2} P^i P^k P^l
}{
\bigl(P^{2}+\epsilon_f^{2}\bigr)^{2}
} \left[1\!+i \frac{(P_\perp^2\!+\!\epsilon_f^2)}{2z\bar z q^-}(x^- \!- \!z y^- \!-\! \bar{z} y'^-)  \right]\!-\delta^{il}P^k\left[ 1  + i\frac{(P_\perp^2\!+\!\epsilon_f^2)}{2z\bar z q^-}  \big(x^- \!-\!y'^- \!-\! (\bar{z}-z)(y'^- - y^-)\theta(y'^- \!-\! y^-)\big) \right]\na
&\quad-\delta^{ik}P^l\left[ 1 + i\frac{(P_\perp^2+\epsilon_f^2)}{2z\bar z q^-}  \big(x^- - y^- \!+\! (\bar{z}-z)(y^- - y'^-)\theta(y^- \!-\! y'^-)\big) \right]\Bigg\}
\,{\rm Tr}\Big[
\overline{F}_{-i}(x^-,x_\perp)\:\overline{F}_{-k}(y'^-,y_\perp) \overline{F}_{-l}(y^-,y_\perp)
\Big]\,.
\end{align}
This result agrees with the corresponding expression of Ref.~\cite{Altinoluk:2024zom}.

This comparison clarifies the relationship between our back-to-back expansion and the back-to-back limit of the CGC sub-eikonal expansion. In Ref.~\cite{Altinoluk:2024zom}, results are obtained via an eikonal expansion. At leading eikonal order, one captures twist-2 and part of the twist-3 operator structures, but without the plus-momentum phase. Sub-eikonal terms can be used to reconstruct the phase for twist-2 operators; however, the expansion does not allow systematic reconstruction of the phases for twist-3 operators, including both two-body and three-body contributions.

In contrast, within our back-to-back gradient expansion, the leading term generates the complete set of two-body operators together with the full plus-momentum phase to all orders in the eikonal approximation. Subleading terms then produce the three-body operators with their complete phase structure. This organization makes the formalism well-suited for extensions to twist-4 and four-body operators.

We emphasize, however, that our approach assumes back-to-back kinematics from the outset; extensions beyond this limit would require a genuine eikonal expansion along the lines of Ref.~\cite{Altinoluk:2022jkk} at small $x$ or resummation of higher twists at arbitrary $x$. The latter can be done for specific TMD operators to arbitrary kinematic twist order, as demonstrated by the iTMD formalism~\cite{Altinoluk:2019fui,Altinoluk:2019wyu,Kotko:2015ura,vanHameren:2016ftb}. 

\section{Conclusion}
\label{sec:outl}

{We computed the dijet cross section in deep inelastic scattering at leading order in $\alpha_s$ using the background field method, working in the back-to-back limit $\Delta_\perp \ll P_\perp$. Restricting to gluon background fields, we obtain a TMD-factorized expression valid through twist-3 accuracy in both dynamic twists and kinematic twists.  While our main focus in this work is dijet production in DIS, the factorized structure derived here relies only on the back-to-back kinematics and does not assume any ordering in Bjorken-$x$. It can therefore be applied across the broad kinematic range accessible at the EIC, from moderate to small $x$, provided the hard transverse scale $P_\perp$ remains perturbative. In particular, in the small-$x$ limit our expressions systematically reproduce the eikonal and sub-eikonal results of the CGC framework, while at moderate $x$ they retain the full longitudinal phase structure characteristic of standard TMD factorization. The same formalism can also be extended to photoproduction and ultra-peripheral collisions, where the virtuality $Q^2$ may become small, but the hard scale is set by $P_\perp$, allowing a consistent description of back-to-back dijet production beyond the strict DIS regime.}

Our approach employs a gradient expansion of the quark propagator in the transverse covariant momentum ${P}_\perp$ while retaining all powers of the longitudinal momentum ${ P}^+$. At leading order, the expansion yields longitudinal Wilson lines, with field-strength insertions appearing at subleading order {in $1/P_\perp^2$}. This expansion is naturally motivated by back-to-back kinematics: the small transverse momentum imbalance implies that the quark traverses the hadron at nearly fixed transverse position, with transverse displacement occurring only at light-cone infinity. This structure is consistent with the operator definition of TMDs, where the transverse gauge link resides at positive light-cone infinity (our gauge choice permits zero $A_\mu$ at asymptotic infinity, $x^-\to \infty$). 

A key advantage of this framework is that it retains the full longitudinal phase $e^{ixP^+z^-}$ in the TMD matrix elements without requiring an eikonal expansion. Our results agree with the back-to-back limit of CGC calculations \(x\to0\)~\cite{Dominguez:2011wm,Altinoluk:2024zom}, while providing a systematic determination of the Fourier phase  for twist-3 TMDs—information that was not accessible in the sub-eikonal approach of Ref.~\cite{Altinoluk:2024zom}.

We further reduce the operator basis using equations of motion. As shown in App.~\ref{sec:appendix_trnsfr}, matrix elements involving $F_{ij}$ can be rewritten in terms of kinematic and dynamic twist contributions, consistent with earlier observations for quark TMDs~\cite{Gamberg:2022lju,Boer:1997mf}. This reduction of possible operator structures might be crucial for the extraction of TMDs via global analyses~\cite{Boussarie:2023izj}.

The methods developed here provide a foundation for extending MSTT framework ~\cite{Mukherjee:2023snp,Mukherjee:2025aiw} through systematic inclusion of higher-twist contributions, such as $\bar{\psi}F\psi$, $FFF$ etc.  This is particularly relevant at small $x$, where a correspondence between CGC and higher-twist formalisms is expected~\cite{Fu:2024sba}.

Several extensions remain for future work. Inclusion of quark background fields will allow treatment of quark–gluon dijet production (this was considered at small $x$ in Ref.~\cite{Altinoluk:2023qfr}), involving operators such as $\bar{\psi}\gamma^\mu\psi$ and $\bar{\psi}F\psi$ at twist-3 accuracy. This requires careful treatment of quark equations of motion, with nontrivial implications for gauge invariance~\cite{Mukherjee:2025aiw}.

{ A key phenomenological extension of this work involves comparing e+p and e+A dijet cross sections. By analyzing these processes in near back-to-back kinematics, one can extract specific twist contributions and examine how nuclear shadowing manifests across the $P_\perp$ and $\Delta_\perp$ phase space~\cite{Nikolaev:1990ja}. In the small-$x$ regime, saturation effects and the associated nonlinear QCD dynamics in nuclei may provide an important direction for future study once contributions beyond tree level are incorporated~\cite{Mueller:1989st, Golec-Biernat:1998zce, Golec-Biernat:1999qor,Nikolaev:1994ce}.  }

Finally, this work sets the stage for an NLO calculation of dijet production at arbitrary $x$. At NLO, one encounters numerous diagrams with UV, collinear, soft, and rapidity divergences.  Known cancellation mechanisms should eliminate UV and soft divergences, while collinear and rapidity divergences are expected to match those in NLO TMD matching~\cite{Mukherjee:2023snp,Mukherjee:2025aiw}. Under this framework, a comparison with small-$x$ CGC calculations at one loop~\cite{Caucal:2021ent,Caucal:2023fsf,Caucal:2023nci} would provide a valuable consistency check. Relating the two factorizations will be essential to understanding the range of validity for the CGC effective theory. At least for leading twist operator $F_{i-} F_{i-}$ such a comparison can be made immediately. Judging by the agreement/disagreement with the cross sections, we can then set the stage for a comparison beyond leading twist at NLO.

\acknowledgements

We thank Y. Hatta, E. Iancu, T. Kar, Y. Mehtar-Tani and X. B. Tong for illuminating discussions and comments. 

This work is supported by the U.S. Department of Energy, Office of Science, Office of Nuclear Physics through Contract Nos.~DE-SC0012704 and DE-SC0020081, within the framework of Saturated Glue (SURGE) Topical Collaboration in Nuclear Theory, and the United States-Israel Binational Science Foundation grant \#2022132. 

\appendix

\section{Gauge Field Conventions}

For completeness, we collect here the conventions and definitions for the
most common objects appearing in the main text:
\begin{align}
  X Y &= X^\mu Y_\mu
      = X^+ Y_+ + X^- Y_- - X_\perp Y_\perp \,, \\
  X_\perp Y_\perp &= X_i Y_i \,, \\
  D_\mu &= \partial_\mu - i \,[A_\mu,\, \cdot\,]\,, \\
  [D_\mu, D_\nu]
  &= -i [F_{\mu\nu},\, \cdot\,] \,,
\end{align}
where the field--strength tensor is defined as
\begin{equation}
F_{\mu\nu}
= \partial_\mu A_\nu - \partial_\nu A_\mu - i [A_\mu, A_\nu] \,.
\end{equation}
Throughout this work, the background gauge field was rescaled to absorb
the coupling constant $g$.

\section{Schwinger notation}
\label{app:schwinger}

In this paper, we use the Schwinger notation  (see Ref.~\cite{Schwinger:1951nm} and application to the background field method in Refs.~\cite{Balitsky:1987bk,Balitsky:2015qba,Balitsky:2016dgz,Balitsky:2017gis,Balitsky:2017flc,Mukherjee:2023snp,Balitsky:2023hmh,Mukherjee:2025aiw,Balitsky:2025bup}),
in which the coherent states $|x)$ and $|p)$ are eigenstates of the position and momentum operators, respectively:
\begin{eqnarray}
&&\hat{x}_\mu|x) = x_\mu|x);\ \ \ \hat{p}_\mu|p) = p_\mu|p)\,.
\end{eqnarray}
Both sets of states form a complete and orthonormal basis in the corresponding Hilbert spaces,
yielding two resolutions of identity:
\begin{eqnarray}
& \int d^4x\, |x)( x| = 1\, ,
\\ & \int \dhd^4p \, |p) (p| = 1,
\label{eq:res_of_iden}
\end{eqnarray}
where  
\begin{align}
\dhd^d p   = d^d p/(2\pi)^d\,.
\label{eq:BalitNot}
\end{align}

The scalar product between position and momentum eigenstates is given by
$$( x|p) = e^{-ipx}.$$
Using the resolution of identity, we can express matrix elements of functions of the momentum operator as
\begin{eqnarray}
&(x|f(\hat{p})|y) = \int \dhd^4p  (x|f(\hat{p}) |p) \, (p| y) =  \int \dhd^4p e^{-ip(x-y)}f(p)\,.
\label{eq:f_of_p}
\end{eqnarray}
for an arbitrary function $f(\hat{p})$

Additionally, we have the usual canonical commutation relation  
\begin{equation}
[p_\nu, f] = i \partial_\nu f\,. 
\label{eq:canonical}
\end{equation}
For brevity, we will typically omit the hat notation for operators and use the letter $p$ to denote the momentum operator.
Capital letters are reserved for the covariant momentum operator:
\begin{align}
    \mathrm{P}_\mu = p_\mu + A_\mu\,. 
\end{align}

\section{Back-to-back kinematics}
\label{App:Kinematic_setup}

We introduce the total and relative transverse momenta of the outgoing
quark--antiquark pair via
\begin{equation}
\left\{
\begin{aligned}
   {\Delta}_\perp &= {k}_{1\perp} + {k}_{2\perp}, \\
    {P}_\perp &= \bar{z}\, {k}_{1\perp} - z\, {k}_{2\perp},
\end{aligned}
\right.
\qquad \Longrightarrow \qquad
\left\{
\begin{aligned}
    {k}_{1\perp} &= z\,{\Delta}_\perp + {P}_\perp, \\
    {k}_{2\perp} &= \bar{z}\,{\Delta}_\perp - {P}_\perp .
\end{aligned}
\right.
\end{equation}
Here, $\vec{\Delta}_\perp$ denotes the transverse momentum imbalance in the quark--antiquark system, while ${P}_\perp$ represents
the relative transverse momentum. The longitudinal momentum fractions are defined as
$z \equiv k_1^-/q^-$ and $\bar z \equiv k_2^-/q^- = 1 - z$.

With these definitions, the momenta of the photon, background gluon, and the
produced quark and antiquark can be written as
\begin{align}
q^\mu &= (q^+,\, q^-,\, \mathbf{0}_\perp), \\[4pt]\label{eq: kg_btb}
k_g^\mu &= (x P^+ = k_1^+ + k_2^+ - q^+,\, 0,\,{\Delta}_\perp)\approx \left( \frac{P_\perp^2 + \epsilon_f^2}{2z\bar{z}q^-}+\frac{ \Delta_\perp^2}{2 q^-},\,0,\,{\Delta}_\perp \right), \\[4pt] 
k_1^\mu &= \left(
\frac{({P}_\perp + z\,{\Delta}_\perp)^{2}}{2 z\, q^-},
\ z\, q^-,
\ {z}\,{\Delta}_\perp + {P}_\perp
\right) \approx \left(
\frac{{P}_\perp^2 }{2 z q^-}+\frac{ (\Delta \cdot P) }{ q^-}+\frac{ z\Delta^2  }{ 2q^-},
\ z\, q^-,
\ { z}\,{\Delta}_\perp + {P}_\perp
\right) , \\[4pt]
k_2^\mu &= \left(
\frac{({P}_\perp - \bar z\,{\Delta}_\perp)^{2}}{2 \bar z\, q^-},
\  \bar z\, q^-,
\ { \bar z}\,{\Delta}_\perp - {P}_\perp
\right) \approx \left(
\frac{{P}_\perp^2 }{2 \bar{z} q^-}- \frac{ (\Delta \cdot P) }{ q^-}+\frac{ \bar z\Delta^2  }{ 2q^-},\, \bar z \,q^-, \,
 { \bar z}\,{\Delta}_\perp - {P}_\perp
\right).
\end{align}
Here the approximate equalities correspond to expanding the kinematics in the back-to-back limit
where the transverse momentum imbalance $\Delta_\perp$ is parametrically small compared to the hard transverse
scale $P_\perp$, i.e.  $|\Delta_\perp| \ll |P_\perp|$. 
The equations listed above follow from the on-shell conditions for the outgoing
quark and antiquark, together with the light-cone momentum parametrization
appropriate for high-energy scattering in the background-field framework.
 
We want to warn the reader, that  $P^+$ denotes the total light-cone momentum of the hadron and ${P}_\perp$ denotes the total transverse momentum of the dijet; both $P^+$ and ${P}_\perp$ are  {\it c-numbers}. They should be distinguished from the covariant momentum {\it operator} $\mathrm{P}_\mu = p_\mu + A_\mu$ defined in App.~\ref{app:schwinger}.

\section{Useful integrals}
\label{App:integrals}

The following plus momentum integral is ubiquitous  in our calculations
\begin{align}
\label{Eq:pplusint}
\int \dhd k^+ e^{-i k^+ x^-} \frac{1}{k^+ - a^+ + i \epsilon }  = -i \theta(x^-) e^{-i x^- a^+}\, .
\end{align}
where $\epsilon > 0$ is an infinitesimal regulator.

To evaluate the dijet production amplitude, we must perform nested integrations over the light-cone minus coordinates. For double integrations with the ordering $y^- < z^-$, we find:
\begin{align}
\label{Eq:nested1}
 \int dy^- \int_{y^-}^{\infty} dz^- e^{ik_g^+ y^-} 
&= -i \int dz^- \frac{e^{ik_g^+ z^-}}{k_g^+} \,, \\
 \int dy^- \int_{y^-}^{\infty} dz^- e^{ik_g^+ y^-} (z^- - y^-) 
    &= -\int dz^- \frac{e^{ik_g^+ z^-}}{(k_g^+)^2} \,, \\
 \int dy^- \int_{y^-}^{\infty} dz^- e^{ik_g^+ y^-} (z^- - y^-)^2 
    &= \int dz^- e^{ik_g^+ z^-} \frac{2i}{(k_g^+)^3} \,.
\end{align}
These formulas are obtained by applying the derivative identity $(z^- - y^-)^n \to (-i\partial/\partial k_g^+)^n$ acting on the exponential factor.
For triple integrations with the ordering $y^- < z_2^- < z_1^-$, the reduction procedure yields:
\begin{align}
\label{Eq:nested2}
\int dy^- \int_{y^-}^{\infty} dz_1^- \int_{y^-}^{z_1^-} dz_2^- e^{ik_g^+ y^-} (z_2^- - y^-) 
    &= -\int dz_2^- \int_{z_2^-}^{\infty} dz_1^- \frac{e^{ik_g^+ z_2^-}}{(k_g^+)^2} \,, \\
\int dy^- \int_{y^-}^{\infty} dz_1^- \int_{y^-}^{z_1^-} dz_2^- e^{ik_g^+ y^-} (z_1^- - y^-) 
    &= \int dz_2^- \int_{z_2^-}^{\infty} dz_1^- e^{ik_g^+ z_2^-} \left( \frac{i(z_2^- - z_1^-)}{k_g^+} - \frac{1}{(k_g^+)^2} \right) \,, \\
\int dy^- \int_{y^-}^{\infty} dz_1^- \int_{y^-}^{z_1^-} dz_2^- e^{ik_g^+ y^-} (z_1^- - y^-)(z_2^- - y^-) 
    &= \int dz_2^- \int_{z_2^-}^{\infty} dz_1^- e^{ik_g^+ z_2^-} \left( \frac{z_2^- - z_1^-}{(k_g^+)^2} + \frac{2i}{(k_g^+)^3} \right) \,.
\end{align}
In the last two equations, the integration over $y^-$ was performed, but explicit dependence on the remaining coordinates $z_1^-$ and $z_2^-$ persists. These residual double integrals appear naturally in the structure of the amplitude and must be evaluated in conjunction with the specific color and spinor traces of the diagrams.

\section{Useful identities}

\subsection{Wilson line propagator}
\label{sec:WL}
Our goal here is to derive the elementary propagator from $y^-$ to $x^-$, i.e., $(x^- | \frac{1}{{\rm P}^+ + i \epsilon} |y^-)$. There are numerous ways to proceed; we follow the most straightforward approach by expanding in powers of $A_-$ insertions:
\begin{align}
\label{eq:WLprop}
  (x^- | \frac{1}{{\rm P}^+ + i \epsilon} |y^-)  =
  (x^- | \frac{1}{p^+ + A_- + i\epsilon} |y^-)   =
  (x^- | \frac{1}{p^+ + i\epsilon}  -
    \frac{1}{p^+ + i\epsilon} A_-  \frac{1}{p^+ + i\epsilon}
  + ... |y^-)
\end{align}
Each term in Eq.~\eqref{eq:WLprop} can be readily evaluated using Eq.~\eqref{Eq:pplusint}. We explicitly compute the first two terms. The zeroth-order term gives
\begin{align}
  (x^-| \frac{1}{p^+ + i\epsilon}  |y^-) =
  \int \dhd q^+ (x^-| q^+)  (q^+ |y^-) \frac{1}{q^+ + i\epsilon} =
  \int \dhd q^+ e^{-iq^+ (x^--y^-)}  \frac{1}{q^+ + i\epsilon} =
  - i \theta(x^--y^-)
\end{align}
while the first-order term yields
\begin{align}
  & (x^-| \frac{1}{p^+ + i\epsilon} A_-   \frac{1}{p^+ + i\epsilon}  |y^-) =
  \int d z^- \int \dhd q_1^+ \dhd q_2^+
  e^{-i q_1^+ (x^--z^-) - i q_2^+ (z^--y^-)}
  \frac{1}{q_1^+ + i\epsilon}
  \frac{1}{q_2^+ + i\epsilon}
  A_-(z)
  \na & = (i)^2 \int dz^-  \theta(x^--z^-) \theta(z^--y^-) A_-(z^-)
  = (i)^2 \theta(x^- - y^-) \int_{y^-}^{x^-} dz^-  A_-(z^-)\,.
\end{align}
Higher-order terms follow the same way and can be evaluated analogously. 
Summing all contributions, we obtain the full propagator in terms of a Wilson line:
\begin{align}
\label{Eq:PropToWL}
  (x^- | \frac{1}{{\rm P}^+ + i \epsilon} |y^-)  = - i \theta(x^- - y^-)  \left(1 + i \int_{y^-}^{x^-} dz^- A_-(z^-) + ... \right)
  = - i \theta(x^- - y^-) [x^-, y^-]\,.
\end{align}
where $[x^-, y^-]$ is the Wilson line defined in Eq.~\eqref{eq:def-Wl-noxp}.

\subsection{Transverse derivative of a Wilson line}
From the definition, it is straightforward  to show that the derivative of the Wilson line is given by
\begin{align}
  &\partial_k [x^-, y^-]_{z_\perp}  = i A_k(x^-, z_\perp)  [x^-, y^-]_{z_\perp}  -  i [x^-, y^-]_{z_\perp} A_k(y^-, z_\perp) - i \int_{y^-}^{x^-} d\zeta^- [x^-, \zeta^-]_{z_\perp} F_{-k} (\zeta^-,z_\perp) [\zeta^-, y^-]_{z_\perp}\,.
  \label{eq:derWL}
\end{align}

\subsection{Commutators and anticommutators}
In this appendix, we collect several basic and advanced operator identities that are repeatedly used in the main text. For completeness, we list them here and briefly comment on their role in our derivations.

A frequently used special case of the Leibniz rule reads
\begin{align}
  \label{eq:Leibniz}
  [{\rm P}^2_\perp, {\cal O}] = \{{\rm P}_k, [{\rm P}_k, {\cal O}]\}\,.
\end{align}
The commutator of two covariant momenta yields the field-strength tensor,
\begin{align}
  \label{eq:PP_DP_com}
  [{\rm P}_\mu, {\rm P}_\nu] = i F_{\mu\nu}\,,
\end{align}
while commuting a covariant momentum with the field-strength tensor gives
\begin{align}
  \label{eq:comm_DF}
  [ {\rm P}_k, F_{-i} ] = i D_k F_{-i}\,.
\end{align}
Finally, we specify the identity used to commute covariant momenta across
Wilson lines. For a straight Wilson line extending along the minus
direction, one finds
\begin{align}
  \label{eq:wilson_comm}
  [x^-,y^-]\,{\rm P}_i(y^-)
  = {\rm P}_i(x^-)\,[x^-,y^-]
  - \int_{y^-}^{x^-} dz^-\,
    [x^-,z^-]\,F_{-i}(z^-)\,[z^-,y^-] \,.
\end{align}
This relation allows us to systematically move all covariant momenta to a
common coordinate, and introduces gauge-invariant insertions of the field-strength tensor along the Wilson line.

\section{Derivation of the reduction formula}
\subsection{Derivation of the Light-Cone LSZ Formula}\label{sec:LC_LSZ_simplify}
In this appendix, we derive the on-shell limit of the quark propagator given in the main text. The detailed calculation proceeds as follows:
\begin{align}
     \lim_{k^2\to 0} & k^2 (k| \frac{1}{{\rm P}^2 + \frac{1}{2} \sigma F + i \epsilon } |y)
      = \lim_{k^2\to 0}  \frac{k^2}{2k^-}\int d^4 x (k|x) (x|  \frac{1}{{\rm P}^+ - \frac{ {\rm P}_\perp^2 }{2 k^-} + {{\sigma} F \over 4 k^-} + i \epsilon }  |y)
     \na & = -i\lim_{k^2\to 0}  \left(k^+ - \frac{k_\perp^2}{2k^-} \right)\int d^4 x e^{i kx}  \delta(x^+-y^+) \theta(x^- - y^-)(x_\perp| e^{ -i (x^--y^-) \frac{ P_\perp^2 (x^-) }{2 k^-}} 
     {O}(x^-, y^-) |y_\perp)
     \na & = -i \lim_{k^2\to 0}  \int_{y^-}^\infty dx^- d^2 x_\perp \left[ \left( - i \frac{\partial}{\partial x^-}  - \frac{k_\perp^2}{2k^-} \right) e^{i kx} \right]_{x^+=y^+}  (x_\perp| e^{ -i (x^--y^-) \frac{ P_\perp^2 (x^-) }{2 k^-}} 
     {O}(x^-, y^-) |y_\perp)
      \na & = -i \lim_{k^2\to 0} \left( e^{iky} +  \int_{y^-}^\infty dx^- d^2 x_\perp   e^{i (k^- y^+ + k^+ x^- - k_\perp x_\perp)}  \left(  i \frac{\partial}{\partial x^-}  - \frac{k_\perp^2}{2k^-} \right)  (x_\perp| e^{ -i (x^--y^-) \frac{ P_\perp^2 (x^-) }{2 k^-}} 
         {O}(x^-, y^-) |y_\perp) \right)
      \na & = -i \lim_{k^2\to 0} \left( e^{iky} +  \int_{y^-}^\infty dx^- d^2 x_\perp   e^{i (k^- y^+ + k^+ y^- - k_\perp x_\perp) }   i  \frac{\partial}{\partial x^-}   (x_\perp| e^{i \frac{k_\perp^2}{2k^-} (x^--y^-)  -i (x^--y^-) \frac{ P_\perp^2 (x^-) }{2 k^-}} 
         {O}(x^-, y^-) |y_\perp) \right)
      \na & = -i \lim_{k^2\to 0} \left( e^{iky} +  \int_{y^-}^\infty dx^-   e^{i (k^- y^+ + k^+ y^-)}   i \frac{\partial}{\partial x^-}   (k_\perp| e^{i \frac{k_\perp^2}{2k^-} (x^- -y^-
      ) -i (x^--y^-) \frac{ P_\perp^2 (x^-) }{2 k^-}} 
         {O}(x^-, y^-) |y_\perp) \right)
     \na & =  \left.  e^{i (k^+ y^- + k^- y^+ )}   (k_\perp|  { O}(\infty, y^-) |y_\perp) \right|_{k^+ = \frac{k^2_\perp}{2k^-}}
     \,.
     \label{eq:lsz}
\end{align}
Here, the on-shell condition $k^2 \to 0$ implies $k^+ = k_\perp^2/(2k^-)$. In the final step, we evaluate the $x^-$ derivative and take the limit $x^- \to \infty$, using the asymptotic behavior ${\rm P}_\perp^2(x^- \to \infty) \to k_\perp^2$. The exponential factor becomes stationary in this limit, and the first term $e^{iky}$ vanishes due to the on-shell condition. The result demonstrates that the on-shell propagator is completely determined by the Wilson line operator ${\cal O}(\infty, y^-)$ evaluated at light-cone infinity.

\subsection{Simplification of the $DF$ Term}\label{sec:DF_simplify}

In the amplitude, operator insertions involving the covariant derivative 
$D_i F_{-i}$ can be reduced to combinations of single-- and double--field--strength 
structures. Using integration by parts together with Eq.~\eqref{eq:derWL}, we find
\begin{align}
& \int dz\, e^{i k \cdot  z } \, 
     [\infty,z^-]_{z_\perp} D_j F_{-i}(z^-, z_\perp)\,[z^-,\infty]_{z_\perp} \notag\\&= \int dz\, e^{i k \cdot  z } 
     \Bigl\{
       -i  k_j \,[\infty,z^-]_{z_\perp} F_{-i}(z^-,z_\perp)[z^-,\infty]_{z_\perp}
       +   i   \int_{z^-}^\infty dz'^- \,[\infty,z'^-]_{z_\perp} F_{-j} (z'^-,z_\perp) [z'^-,z^-] F_{-i} (z^-,z_\perp) [z^-,\infty]_{z_\perp}  \notag\\
&~~~~~~~~~~~~~~~~~-i   \int_{z^-}^{\infty} dz'^- \,[\infty,z^-]_{z_\perp}  F_{-i}(z^-,z_\perp)  [z^-, z'^-] F_{-j}(z'^-,z_\perp) [z'^-,\infty]_{z_\perp} 
     \Bigr\}\,.  
\label{eq:DF-simplify}
\end{align}

This identity shows explicitly that the covariant-derivative insertion produces  
(i) a momentum--weighted single $F$ insertion and  
(ii) two distinct $FF$ structures originating from the derivative acting on the Wilson line.  
The same structure applies to $D_i(\sigma F)$ with the replacement 
$F_{-i}\rightarrow (\sigma F)$, and we do not repeat the expression.

\section{Dynamics in the target}
\label{sec:dynintar}

In this appendix, we present a detailed derivation of the scalar propagator in a background field with explicit $z^+$ dependence, generalizing the discussion of the main text.
Firstly, we have
\begin{equation}
\begin{aligned}
    D(x,y)\:&=\:(x|\frac{1}{{\rm P}^2}|y)
    \:\\&=\:(x| \frac{1}{2p^-p^+ + \{p^-, A_-\}-P_\perp^2}|y)
    \:\\&=\:(x| \frac{1}{2p^-p^+ + 2p^- A_- -i \partial^- A_- - P_\perp^2}|y),
\end{aligned}
\end{equation}
where $A_-$ acts on a Schwinger ket $|z^+)$ according to
\begin{equation}
A_-|z^+)=A_-(z^+)\,|z^+)\,.
\end{equation}
This definition follows the same convention as that introduced in App.~\ref{app:schwinger}.

Let us now consider an operator expansion around $y^+$ 
\begin{equation}
\begin{aligned}
A_-\: \approx \:A_-(y^+) + (\hat{z}^+ - y^+)\:\partial^- A_-(y^+),
\end{aligned}
\end{equation}
where $\hat{z}^+$ is the position operator while $A_-(y^+)$ and $\partial^- A_-(y^+)$ are c-numbers on the  Hilbert space spanned by $|z^+)$, i.e. $[\hat{z}^+,A_-(y^+)] = 0$. 
In all expressions below, $\hat z^+$ is the only
operator acting on the $|z^+)$ space. We thus obtain
\begin{equation}
\begin{aligned}
  D(x,y)\:&=\:(x| \frac{1}{ 2p^- {\rm P}^+ - {\rm P}_\perp^2 + 2p^- (\hat{z}^+ - y^+)\:\partial^- A_- -i \partial^- A_- }|y).
\end{aligned}
\end{equation}
We now focus on the terms that are not included in the main text:
\begin{equation}
\label{eq:expandzplusprop}
\begin{aligned}
D_{+-}(x,y)\:=&\:-(x|{1 \over 2p^- {\rm P}^+ - {\rm P}_\perp^2 + i \epsilon}(2p^- (\hat{z}^+ - y^+)\:\partial^- A_- -i \partial^- A_-){1 \over  2p^- {\rm P}^+ - {\rm P}_\perp^2 + i \epsilon}|y)\: .
\end{aligned}
\end{equation}
We consider the individual contributions to this expression. Focusing on the first term and inserting a resolution of the identity in the $z^+$ space, we obtain
\begin{align}
D_{+-}^{\rm with\; z^+}(x,y)
=&
-\!\int dz^+\,
(x|\frac{2p^-}{2p^- {\rm P}^+ -  {\rm P}_\perp^2 + i\epsilon}|z^+)\,
(z^+ - y^+)\,\partial^- A_-(y^+)\,
(z^+|\frac{1}{2p^-  {\rm P}^+ -  {\rm P}_\perp^2 + i\epsilon}|y).
\end{align}
Writing the longitudinal propagators in momentum space and subsequently introducing a resolution of the identity in $z^-$, this expression can be cast into the form
\begin{align}
D_{+-}^{\rm with\; z^+}(x,y)
=&
-\!\int dz^+ dz^-\!
\int \dhd p_1^- \frac{\dhd p_2^-}{2p_2^-}\,
e^{-i(x^+-z^+)p_1^-} e^{-i(z^+-y^+)p_2^-}\,(z^+-y^+)
\na [4pt]
&\times
(x^-,x_\perp|
\frac{1}{ {\rm P}^+ - \frac{ {\rm P}_\perp^2}{2p_1^-} + i\epsilon}
|z^-)\,
\partial^- A_-(z^-,y^+)\,
(z^-|
\frac{1}{ {\rm P}^+ - \frac{ {\rm P}_\perp^2}{2p_2^-} + i\epsilon}
|y^-,y_\perp)\,.
\end{align}
The remaining Schwinger matrix element can be evaluated at leading order in ${\rm P}_\perp^2$. After resumming the external transverse--momentum insertions,
we have
\begin{equation}
\label{eq:Pplus_kernel}
(x|  \frac{1}{{\rm P}^+ - \frac{ {\rm P}_\perp^2 }{2 k^-} + i \epsilon}  |y)
=
\delta(x^+-y^+)\,\theta(x^- - y^-)\,
(x_\perp|\,
\bigl[-i\,e^{-i (x^- - y^-)\frac{{\rm P}_\perp^2(x^-)}{2k^-}}\bigr]\,
[x^-, y^-]\,
|y_\perp)\, .
\end{equation}
Substituting Eq.~\eqref{eq:Pplus_kernel} into the $z^+$--dependent contribution, we obtain
\begin{align}
\label{eq:Dpm_zplus_xy}
D_{+-}^{{\rm with}\;z^+}(x,y)
=&\int dz^+\!\int_{y^-}^{x^-}\!dz^-\!
\int \dhd p_1^-\,\frac{\dhd p_2^-}{2p_2^-}\,
e^{-i (x^+ - z^+) p_1^-}\,e^{-i (z^+ - y^+) p_2^-}\,(z^+ - y^+)
\nonumber\\
&\times
(x_\perp|\,
e^{-i (x^- - z^-)\frac{{\rm P}_\perp^2(x^-)}{2p_1^-}}\,[x^-,z^-]\,
\partial^- A_-(z^-,y^+)\,
e^{-i (z^- - y^-)\frac{{\rm P}_\perp^2(z^-)}{2p_2^-}}\,[z^-,y^-]
|y_\perp)\, .
\end{align}
Applying the LSZ prescription, we define
\begin{align}
\label{eq:Dpm_zplus_ky_def}
D_{+-}^{{\rm with}\;z^+}(k,y)
\equiv
\lim_{k^2\to0}k^2\int d^4x\,e^{ik\cdot x}\,
D_{+-}^{{\rm with}\;z^+}(x,y)\, .
\end{align}
Carrying out the $x^+$ integral and performing the on--shell reduction, the result can be brought to the form
\begin{align}
\label{eq:Dpm_zplus_ky_simpl}
D_{+-}^{{\rm with}\;z^+}(k,y)
=&\, i\,(2k^-)
\int dz^+\!\int_{y^-}^{\infty}\!dz^-\!
\int\frac{\dhd p_2^-}{2p_2^-}\,
e^{i z^+ k^-}\,e^{-i (z^+ - y^+) p_2^-}\,(z^+ - y^+)
\nonumber\\
&\times
(k_\perp|\,
e^{\,i z^- \frac{k_\perp^2}{2k^-}}\,[\infty^-,z^-]\,
\partial^- A_-(z^-,y^+)\,
e^{-i (z^- - y^-)\frac{{\rm P}_\perp^2(z^-)}{2p_2^-}}\,[z^-,y^-]
|y_\perp)\, .
\end{align}
To facilitate the $z^+$ integration, we rewrite the factor $(z^+ - y^+)$ as a derivative with respect to the conjugate momentum $p_2^-$, using
\begin{equation}
(z^+ - y^+)\,e^{-i (z^+ - y^+) p_2^-}
= i\,\frac{\partial}{\partial p_2^-}
e^{-i (z^+ - y^+) p_2^-}\, .
\end{equation}
In this way, the $z^+$ integration can now be performed straightforwardly, producing a delta function $\delta(k^- - p_2^-)$.
Subsequently, we carry out an integration by parts with respect to $p_2^-$, which fixes $p_2^- = k^-$ and generates two contributions.
As a result, we obtain
\begin{equation}
\label{eq:D+-_with_z}
\begin{aligned}
D_{+-}^{{\rm with}\;z^+}(k,y)
=&-\int_{y^-}^{\infty}\!dz^-\,\frac{1}{k^-}\,
e^{i y^+ k^-}\,
(k_\perp|\,
e^{\,i z^- \frac{k_\perp^2}{2k^-}}\,[\infty^-,z^-]\,
\partial^- A_-(z^-,y^+)\,
e^{-i (z^- - y^-)\frac{{\rm P}_\perp^2(z^-)}{2k^-}}\,[z^-,y^-]
|y_\perp)
\\[0.3em]
&
+i\int_{y^-}^{\infty}\!dz^-(z^-\! -\! y^-)
e^{i y^+ k^-}
(k_\perp|
e^{i z^- \frac{k_\perp^2}{2k^-}}[\infty^-,z^-] 
\partial^- A_-(z^-,y^+) 
\frac{{\rm P}_\perp^2(z^-)}{2(k^-)^2} 
e^{-i (z^- - y^-)\frac{{\rm P}_\perp^2(z^-)}{2k^-}} [z^-,y^-]
|y_\perp)\, .
\end{aligned}
\end{equation}
Following the same strategy as above, we now consider the contribution
without an explicit $(z^+ - y^+)$ dependence in Eq.~\eqref{eq:expandzplusprop}.
The corresponding term can be written as
\begin{align}\label{eq:D+-_no_z}
D_{+-}^{\rm no\; z^+}
=&
-\!\int dz^+\,
(x|\frac{1}{2p^- {\rm P}^+ - {\rm P}_\perp^2 + i\epsilon}|z^+)\,
\bigl(-i\partial^- A_-(y^+)\bigr)\,
(z^+|\frac{1}{2p^- {\rm P}^+ - {\rm P}_\perp^2 + i\epsilon}|y)\, 
\na =&
\int_{y^-}^{\infty}\!dz^-\,\frac{1}{2k^-}\,
e^{i y^+ k^-}\,
(k_\perp|\,
e^{\,i z^- \frac{k_\perp^2}{2k^-}}\,[\infty^-,z^-]\,
\partial^- A_-(z^-,y^+)\,
e^{-i (z^- - y^-)\frac{{\rm P}_\perp^2(z^-)}{2k^-}}\,[z^-,y^-]
|y_\perp)\,.
\end{align}
Finally, we combine the contributions from Eq.\eqref{eq:D+-_with_z} and Eq. \eqref{eq:D+-_no_z}. We may now commute the transverse exponential evaluated at $z^-$ to the surface at infinity across $\partial^- A_-(z^-,y^+)$ and the Wilson line $[\infty^-,z^-]$. This procedure generates additional transverse derivative terms, which are suppressed by higher powers of transverse momenta and therefore correspond to higher--twist contributions. Neglecting such terms, we obtain
\begin{equation}
\begin{aligned}
D_{+-}(k,y)
=&\,e^{i (y^+ k^- + y^- k^+)}
\int_{y^-}^{\infty}\!dz^-\,
(k_\perp|[\infty^-,z^-]\,
\partial^- A_-(z^-,y^+)\,[z^-,y^-]
|y_\perp)
\\
&\times\left[
-\frac{1}{2k^-}
+i (z^- - y^-)\frac{k_\perp^2}{2(k^-)^2}
\right] .
\end{aligned}
\end{equation}
Using the light--cone gauge condition $A_+=0$, we may rewrite
$\partial^- A_-(z^-,y^+)$ in terms of the field strength
$F_{+-}(z^-,y^+)$. The final result then reads
\begin{equation}
\begin{aligned}
D_{+-}(k,y)
=&\,e^{i (y^+ k^- + y^- k^+)}
\int_{y^-}^{\infty}\!dz^-\,
(k_\perp|[\infty^-,z^-]\,
F_{+-}(z^-,y^+)\,[z^-,y^-]
|y_\perp)
\\
&\times\left[
-\frac{1}{2k^-}
+i (z^- - y^-)\frac{k_\perp^2}{2(k^-)^2}
\right] .
\end{aligned}
\end{equation}

\section{Relation between kinematic and dynamic twists from the Bianchi identity}
\label{sec:appendix_trnsfr} 
In this appendix, following Appendix~C of Ref.~\cite{Altinoluk:2024zom}, 
we show that a specific higher-order kinematic contribution arising from the
expansion of leading-twist operators can be rewritten in terms of genuine
twist-3 operator structures.
In particular, we demonstrate that the linear combination
$(P^i \Delta^j - (\Delta\!\cdot\! P)\delta^{ij})\,F_{-i}$
can be expressed as a sum of operators involving $F_{ij}$ and
the form $F_{-i}F_{-j}$.
To this end, we start from the following kinematic contribution to the amplitude,
\begin{align}
\mathcal{M}_{\rm KT}^j\:&=\:(P^i \Delta^j - (\Delta\cdot P) \delta^{ij})\:\int dz^-\:e^{-i k^+ z^-}\int d^2 z_\perp\:e^{-i \Delta_\perp  z_\perp}\:[\infty, z^-]_{z_\perp}\:F_{-i}(z^-,z_\perp)[z^-,\infty]_{z_\perp}
\na&
=\:- \epsilon^{im}\:\epsilon^{jn} P^n \Delta^m\:\int dz^-\:e^{-i k^+ z^-}\int d^2 z_\perp\:e^{-i \Delta_\perp  z_\perp}\:[\infty, z^-]_{z_\perp}\:F_{-i}(z^-,z_\perp)[z^-,\infty]_{z_\perp} \na&
=\:-i\:\epsilon^{im}\:\epsilon^{jn} P^n\:\int dz^-\:e^{-i k^+ z^-}\int d^2 z_\perp\:\Big({\partial \over \partial z_\perp^m}e^{-i \Delta_\perp  z_\perp}\Big)\:[\infty, z^-]_{z_\perp}\:F_{-i}(z^-,z_\perp)[z^-,\infty]_{z_\perp} \na&
=i\:\epsilon^{im}\:\epsilon^{jn} P^n\:\int dz^-\:e^{-i k^+ z^-}\int d^2 z_\perp\:e^{-i \Delta_\perp  z_\perp}\:{\partial \over \partial z_\perp^m}\Big([\infty, z^-]_{z_\perp}\:F_{-i}(z^-,z_\perp)[z^-,\infty]_{z_\perp}\Big), 
\end{align}
where $\delta^{im}\delta^{jn}-\delta^{in}\delta^{jm}=\epsilon^{ij}\epsilon^{mn}$, and the subscript ``${\rm KT}$'' denotes a higher--order kinematic contribution originating from the expansion of leading--twist operators.
Using Eq.~\eqref{eq:derWL}, the above expression can be reorganized as
\begin{align}\label{eq:M_kt_2}
\mathcal{M}_{\rm KT}^j\:=&\:i\:\epsilon^{im}\:\epsilon^{jn} P^n\:\int dz^-\:e^{-i k^+ z^-}\int d^2 z_\perp\:e^{-i \Delta_\perp  z_\perp}\:\Bigg([\infty, z^-]_{z_\perp}\:D_mF_{-i}(z^-,z_\perp)[z^-,\infty]_{z_\perp} \na &-i \int_{z^-}^{\infty}\:dz'^-\: [\infty,z'^-]_{z_\perp} F_{-m}(z'^-,z_\perp)[z'^-, z^-]_{z_\perp} F_{-i}(z^-,z_\perp) [z^-,\infty]_{z_\perp} \na &+i \int_{z^-}^{\infty}\:dz'^-\: [\infty,z^-]_{z_\perp}F_{-i}(z^-,z_\perp)[z^-,z'^-]_{z_\perp}F_{-m}(z'^-,z_\perp)[z'^-,\infty]_{z_\perp}\Bigg).
\end{align}
Using the antisymmetry of $\epsilon^{im}$ together with the Jacobi identity
for the adjoint covariant derivative,
\begin{equation}
D_i F_{jk} + D_j F_{ki} + D_k F_{ij} = 0 \, ,
\end{equation}
we can simplify the first term appearing in Eq.~\eqref{eq:M_kt_2}. Specifically, we have
\begin{align}
\epsilon^{im} D_m F_{-i}
&= \frac{\epsilon^{im}}{2} \left( D_m F_{-i} + D_i F_{m-} \right)
= -\frac{\epsilon^{im}}{2}\, D_- F_{im} \, .
\label{eq:beforetrick}
\end{align}
Substituting Eq.~\eqref{eq:beforetrick} into Eq.~\eqref{eq:M_kt_2} yields
\begin{align}
\mathcal{M}_{\rm KT}^j\:=&\:i\:\epsilon^{im}\:\epsilon^{jn} P^n\:\int dz^-\:e^{-i k^+ z^-}\int d^2 z_\perp\:e^{-i \Delta_\perp  z_\perp}\:\Bigg(-{ \frac12}[\infty, z^-]_{z_\perp}\:D_-F_{im}(z^-,z_\perp)[z^-,\infty]_{z_\perp} \na&-i \int_{z^-}^{\infty}\:dz'^-\: [\infty,z'^-]_{z_\perp} F_{-m}(z'^-,z_\perp)[z'^-, z^-]_{z_\perp} F_{-i}(z^-,z_\perp) [z^-,\infty]_{z_\perp]} \na &+i \int_{z^-}^{\infty}\:dz'^-\: [\infty,z^-]_{z_\perp}F_{-i}(z^-,z_\perp)[z^-,z'^-]_{z_\perp}F_{-m}(z'^-,z_\perp)[z'^-,\infty]_{z_\perp}\Bigg).
\end{align}
Using the relation derived above, the first term can be rewritten as a total
derivative with respect to $z^-$,
\begin{align}
&\mathcal{M}_{\rm KT}^j\:=\:i\:\epsilon^{im}\:\epsilon^{jn} P^n\:\int dz^-\:e^{-i k^+ z^-}\int d^2 z_\perp\:e^{-i \Delta_\perp  z_\perp}\:\Bigg(-{ \frac12}\partial_-([\infty, z^-]_{z_\perp}\:F_{im}(z^-,z_\perp)[z^-,\infty]_{z_\perp}) \na &-i \int_{z^-}^{\infty}\:dz'^-\: [\infty,z'^-]_{z_\perp} F_{-m}(z'^-,z_\perp)[z'^-, z^-]_{z_\perp} F_{-i}(z^-,z_\perp) [z^-,\infty]_{z_\perp} \na &+i \int_{z^-}^{\infty}\:dz'^-\: [\infty,z^-]_{z_\perp}F_{-i}(z^-,z_\perp)[z^-,z'^-]_{z_\perp}F_{-m}(z'^-,z_\perp)[z'^-,\infty]_{z_\perp]}\Bigg).
\end{align}
Performing an integration by parts with respect to $z^-$ and acting on the
Fourier phase $e^{-ik^+ z^-}$, we obtain
\begin{align}
\mathcal{M}_{\rm KT}^j\:=&\:i\:\epsilon^{im}\:\epsilon^{jn} P^n\:\int dz^-\:e^{-i k^+ z^-}\int d^2 z_\perp\:e^{-i \Delta_\perp  z_\perp}\:\Bigg(-{ \frac12}ik^+\:([\infty, z^-]_{z_\perp}\:F_{im}(z^-,z_\perp)[z^-,\infty]_{z_\perp}) \na &-i \int_{z^-}^{\infty}\:dz'^-\: [\infty,z'^-]_{z_\perp} F_{-m}(z'^-,z_\perp)[z'^-, z^-]_{z_\perp} F_{-i}(z^-,z_\perp) [z^-,\infty]_{z_\perp} \na &+i \int_{z^-}^{\infty}\:dz'^-\: [\infty,z^-]_{z_\perp}F_{-i}(z^-,z_\perp)[z^-,z'^-]_{z_\perp}F_{-m}(z'^-,z_\perp)[z'^-,\infty]_{z_\perp}\Bigg)
\end{align}
Note that the final equation has a different form compared to \cite{Altinoluk:2024zom} as we retain the full exponential $e^{i k^+ z^-}$ in our calculations. We can invert the above expression and write an equation that relates $F_{im}$ to $F_{-i}$ and $F_{-i} F_{-m}$ as
\begin{align}\label{eq: knmt_trsf_dynm}
&\int dz^-\:e^{-i k^+ z^-}\int d^2 z_\perp\:e^{-i \Delta_\perp  z_\perp}\Big(- P^i\:k^+\:[\infty, z^-]_{z_\perp}\:F_{ij}(z^-,z_\perp)[z^-,\infty]_{z_\perp} \na &- (P^i \Delta^j - (\Delta\cdot P) \delta^{ij})\:[\infty, z^-]_{z_\perp}\:F_{-i}(z^-,z_\perp)[z^-,\infty]_{z_\perp} \na &
+(P^i-P^j) \int_{z^-}^{\infty}\:dz'^-\: [\infty,z'^-]_{z_\perp} F_{-i}(z'^-,z_\perp)[z'^-, z^-]_{z_\perp} F_{-j}(z^-,z_\perp) [z^-,\infty]_{z_\perp} \na &+(P^i-P^j)\:\int_{z^-}^{\infty}\:dz'^-\: [\infty,z^-]_{z_\perp}F_{-i}(z^-,z_\perp)[z^-,z'^-]_{z_\perp}F_{-j
}(z'^-,z_\perp)[z'^-,\infty]_{z_\perp}
\Big)\:=\:0\, .
\end{align}

\section{Dirac traces}
\label{app:Dirac_trace}

In this appendix, we summarize the Dirac traces required for evaluating the dijet cross section presented in Sec.~\ref{x_section}. Throughout this work, we consider massless quarks and employ the completeness relations
\begin{align}
\sum_s u_s(k_1)\bar u_s(k_1)&=\slashed{k_1},
&
\sum_s v_s(k_2)\bar v_s(k_2)&=\slashed{k_2},
\label{eq:spin_sum}
\end{align}
which allow us to convert products of amplitudes into traces over Dirac matrices. The tables also include the twist-4 structures, although they do not contribute to the twist-3 cross section considered in the main text.

For a longitudinally polarized virtual photon,
$\slashed{\epsilon}^{\,L}\propto\gamma^-$, so both photon vertices reduce to $\gamma^-$. Consequently, all single $\sigma^{-i}F_{-i}$ insertions vanish because $\gamma^-\sigma^{-i}\gamma^-=0$, leaving fewer independent operator structures than in the transverse case. The corresponding traces are
listed in Table~\ref{tab: trace_summary}; Items~7--10 are twist-4 and listed for completeness; $i,j,k,l,m,n$ denote transverse indices.

 \begin{table*}[h]
  \centering
  \footnotesize  
  \renewcommand\arraystretch{1.25} 
\setlist[enumerate]{leftmargin=1.4em,itemsep=4pt,topsep=3pt}
  \setlength{\tabcolsep}{4pt}
  \noindent\begin{minipage}{\textwidth}
    \hrule\vspace{4pt}
    \textbf{Longitudinal}\par\smallskip
    \begin{minipage}[t]{\linewidth}
      \begin{enumerate}[]
        \item $\Tr{\slashed{k_1}\gamma^-\slashed{k_2}\gamma^-}{F}_{-i}F_{-j}
              \;=\;{8\,k_1^- k_2^-}{F}_{-i}F_{-j}$.

        \item \text{All the terms proportional to } $ \sigma^{-i} F_{-i}$ \text{ vanish.}

        \item $\Tr{\slashed{k}_1 \gamma^{-} \slashed{k}_2 \sigma^{kl}\gamma^{-}}\,{F}_{-i}F_{kl}
              \;=\;\Tr{\slashed{k}_1 \gamma^{-} \slashed{k}_2 \gamma^{-} \sigma^{kl}}\,{F}_{-i}F_{kl}
              \;=\;0$.
        \item $\Tr{\slashed{k_1}\sigma^{mn}\gamma^- \slashed{k_2}\gamma^-}\,F_{mn}{F}_{-j}
              \;=\; \Tr{\slashed{k_1}\gamma^- \sigma^{mn}\slashed{k_2}\gamma^-}\,F_{mn}{F}_{-j}
              \;=\; 0$.
              \item $\Tr{\slashed{k}_1 \gamma^{-} \slashed{k}_2 \sigma^{+-}\gamma^{-}}\,{F}_{-i}F_{+-}
        \;=\;\Tr{\slashed{k_1}\sigma^{+-}\gamma^- \slashed{k_2}\gamma^-}\,F_{+-}{F}_{-i} 
              \;=\;-8ik_1^-k_2^-F_{+-}{F}_{-i}$.
        \item $
             \Tr{\slashed{k}_1 \gamma^{-} \slashed{k}_2 \gamma^{-} \sigma^{+-}}\,{F}_{-i}F_{+-} \;=\; \Tr{\slashed{k_1}\gamma^- \sigma^{+-}\slashed{k_2}\gamma^-}\,F_{+-}{F}_{-i}
              \;=\; 8ik_1^-k_2^-F_{+-}{F}_{-i}$.
       \item $
             \Tr{\slashed{k}_1(\sigma_{+-}\gamma^-)\slashed{k}_2(\sigma_{+-}\gamma^-)}F_{+-}{F}_{+-}
          =
         \Tr{\slashed{k}_1(\gamma^-\sigma_{+-})\slashed{k}_2(\gamma^-\sigma_{+-})}F_{+-}{F}_{+-}
       = -8\,k_1^- k_2^- F_{+-}{F}_{+-}$.
      \item $
        \Tr{\slashed{k}_1(\sigma_{+-}\gamma^-)\slashed{k}_2(\gamma^-\sigma_{+-})} F_{+-}{F}_{+-}
      =
     \Tr{\slashed{k}_1(\gamma^-\sigma_{+-})\slashed{k}_2(\sigma_{+-}\gamma^-)}  F_{+-}{F}_{+-}
    = 8\,k_1^- k_2^- F_{+-}{F}_{+-} $.
    \item $
      \Tr{\slashed{k}_1(\sigma_{mn}\gamma^-)\slashed{k}_2(\sigma_{\mu\nu}\gamma^-)}F_{mn}{F}_{\mu\nu}
=
\Tr{\slashed{k}_1(\sigma_{mn}\gamma^-)\slashed{k}_2(\gamma^-\sigma_{\mu\nu})}F_{mn}{F}_{\mu\nu} =16\,k_1^- k_2^- F_{mn} F^{mn}.
$
\item $
\Tr{\slashed{k}_1(\gamma^-\sigma_{mn})\slashed{k}_2(\sigma_{\mu\nu}\gamma^-)}F_{mn}{F}_{\mu\nu}
=
\Tr{\slashed{k}_1(\gamma^-\sigma_{mn})\slashed{k}_2(\gamma^-\sigma_{\mu\nu})}F_{mn}{F}_{\mu\nu}
=16\,k_1^- k_2^- F_{mn} F^{mn} $.
\item $
\Tr{\slashed{k}_1(\sigma_{+-}\gamma^-)\slashed{k}_2(\sigma_{mn}\gamma^-)}
=
\Tr{\slashed{k}_1(\sigma_{+-}\gamma^-)\slashed{k}_2(\gamma^-\sigma_{mn})}
 =
\Tr{\slashed{k}_1(\gamma^-\sigma_{+-})\slashed{k}_2(\sigma_{mn}\gamma^-)}
=
... 
=0 $.
      \end{enumerate}
    \end{minipage}
    \vspace{4pt}\hrule
  \end{minipage}
  \caption{Dirac traces for a longitudinally polarized virtual photon.
The twist-4 structures are included for completeness.
}
  \label{tab: trace_summary}
\end{table*}

For a transversely polarized virtual photon, the photon vertices $\gamma^m$ and $\gamma^n$ ($m,n=1,2$) are contracted with the polarization
projector\(
\frac12\sum_{\lambda=\pm1}
\epsilon_m^\lambda
\epsilon_n^{\lambda *}
=
-\frac12 g_{mn}.\)
Unlike the longitudinal case, the single $\sigma^{-i}F_{-i}$ insertions survive and generate the additional operator structures $C^i$ and the $\sigma^{-k}$ components of
$E^{ikl}$ and $G^{ikl}$ appearing in Eqs.~\eqref{eq:trans_gen}--\eqref{eq:coeff_btb}. Combining the traces
listed in Table~\ref{tab:trace_summary_T} with the coefficients of
Eq.~\eqref{eq:coeff_expand} reproduces the coefficient functions $A^{ij}$, $B^i$, $C^i$, $E^{ikl}$, and $G^{ikl}$. Throughout this appendix, $i,j,k,l$ denote transverse indices and $k_{1,2}^i$ are the transverse components of the jet momenta. Items~8--10 are again listed only for completeness.

 \begin{table*}[h]
  \centering
  \footnotesize
  \renewcommand\arraystretch{1.25}
\setlist[enumerate]{leftmargin=1.4em,itemsep=4pt,topsep=3pt}
  \setlength{\tabcolsep}{4pt}
  \noindent\begin{minipage}{\textwidth}
    \hrule\vspace{4pt}
    \textbf{Transverse}\par\smallskip
    \begin{minipage}[t]{\linewidth}
      \begin{enumerate}[]
        \item $-\tfrac12 g_{mn}\Tr{\slashed{k}_1\gamma^m\slashed{k}_2\gamma^n}\,F_{-i}F_{-j}
              \;=\; 4\,(k_1^+ k_2^- + k_1^- k_2^+)\,F_{-i}F_{-j}$.

        \item Unlike the longitudinal case, the single $\sigma^{-i}F_{-i}$ terms are nonvanishing:\\[2pt]
              $-\tfrac12 g_{mn}\Tr{\slashed{k}_1\,\sigma^{-i}\gamma^m\,\slashed{k}_2\,\gamma^n}\,F_{-i}F_{-j}
                \;=\; -4i\,k_2^-\,k_1^i\,F_{-i}F_{-j}$,\\[2pt]
              $-\tfrac12 g_{mn}\Tr{\slashed{k}_1\,\gamma^m\sigma^{-i}\,\slashed{k}_2\,\gamma^n}\,F_{-i}F_{-j}
                \;=\; +4i\,k_1^-\,k_2^i\,F_{-i}F_{-j}$,\\[2pt]
              $-\tfrac12 g_{mn}\Tr{\slashed{k}_1\,\gamma^m\,\slashed{k}_2\,\gamma^n\sigma^{-j}}\,F_{-i}F_{-j}
                \;=\; +4i\,k_2^-\,k_1^j\,F_{-i}F_{-j}$,\\[2pt]
              $-\tfrac12 g_{mn}\Tr{\slashed{k}_1\,\gamma^m\,\slashed{k}_2\,\sigma^{-j}\gamma^n}\,F_{-i}F_{-j}
                \;=\; -4i\,k_1^-\,k_2^j\,F_{-i}F_{-j}$.

        \item Double $\sigma^{-i}$ insertions (entering the $\bb{S}_1^2,\bb{S}_2^2$ part of $A^{ij}$):\\[2pt]
              $-\tfrac12 g_{mn}\Tr{\slashed{k}_1\,\sigma^{-i}\gamma^m\,\slashed{k}_2\,\gamma^n\sigma^{-j}}\,F_{-i}F_{-j}
              \;=\;
              -\tfrac12 g_{mn}\Tr{\slashed{k}_1\,\gamma^m\sigma^{-i}\,\slashed{k}_2\,\sigma^{-j}\gamma^n}\,F_{-i}F_{-j}
                \;=\; -8\,k_1^- k_2^-\,g^{ij}\,F_{-i}F_{-j}$,\\[2pt]
              $-\tfrac12 g_{mn}\Tr{\slashed{k}_1\,\sigma^{-i}\gamma^m\,\slashed{k}_2\,\sigma^{-j}\gamma^n}\,F_{-i}F_{-j}
              \;=\;
              -\tfrac12 g_{mn}\Tr{\slashed{k}_1\,\gamma^m\sigma^{-i}\,\slashed{k}_2\,\gamma^n\sigma^{-j}}\,F_{-i}F_{-j}
                \;=\; 0 $.

        \item $-\tfrac12 g_{mn}\Tr{\slashed{k}_1\,\gamma^m\,\slashed{k}_2\,\sigma^{+-}\gamma^n}\,F_{-i}F_{+-}
              \;=\;
              -\tfrac12 g_{mn}\Tr{\slashed{k}_1\,\gamma^m\,\slashed{k}_2\,\gamma^n\sigma^{+-}}\,F_{-i}F_{+-}
              \;=\; 4i\,(k_1^- k_2^+ - k_1^+ k_2^-)\,F_{-i}F_{+-}$.

        \item $-\tfrac12 g_{mn}\Tr{\slashed{k}_1\,\sigma^{-i}\gamma^m\,\slashed{k}_2\,\sigma^{+-}\gamma^n}\,F_{-i}F_{+-}
              \;=\;
              -\tfrac12 g_{mn}\Tr{\slashed{k}_1\,\sigma^{-i}\gamma^m\,\slashed{k}_2\,\gamma^n\sigma^{+-}}\,F_{-i}F_{+-}
              \;=\; -4\,k_2^-\,k_1^i\,F_{-i}F_{+-}$,\\[2pt]
              $-\tfrac12 g_{mn}\Tr{\slashed{k}_1\,\gamma^m\sigma^{-i}\,\slashed{k}_2\,\sigma^{+-}\gamma^n}\,F_{-i}F_{+-}
              \;=\;
              -\tfrac12 g_{mn}\Tr{\slashed{k}_1\,\gamma^m\sigma^{-i}\,\slashed{k}_2\,\gamma^n\sigma^{+-}}\,F_{-i}F_{+-}
              \;=\; -4\,k_1^-\,k_2^i\,F_{-i}F_{+-}$.

        \item $-\tfrac12 g_{mn}\Tr{\slashed{k}_1\,\sigma^{ij}\gamma^m\,\slashed{k}_2\,\gamma^n}\,F_{ij}F_{-l}
              \;=\;
              -\tfrac12 g_{mn}\Tr{\slashed{k}_1\,\gamma^m\sigma^{ij}\,\slashed{k}_2\,\gamma^n}\,F_{ij}F_{-l}
              \;=\; 0$.

        \item Mixed $\sigma^{ij}$--$\sigma^{-l}$ insertions:\\[2pt]
              $-\tfrac12 g_{mn}\Tr{\slashed{k}_1\,\sigma^{ij}\gamma^m\,\slashed{k}_2\,\gamma^n\sigma^{-l}}\,F_{ij}F_{-l}
                \;=\; 4\,k_2^-\,\big(k_1^j g^{il}-k_1^i g^{jl}\big)\,F_{ij}F_{-l}$,\\[2pt]
              $-\tfrac12 g_{mn}\Tr{\slashed{k}_1\,\gamma^m\sigma^{ij}\,\slashed{k}_2\,\gamma^n\sigma^{-l}}\,F_{ij}F_{-l}
                \;=\; 4\,k_2^-\,\big(k_1^i g^{jl}-k_1^j g^{il}\big)\,F_{ij}F_{-l}$,\\[2pt]
              $-\tfrac12 g_{mn}\Tr{\slashed{k}_1\,\sigma^{ij}\gamma^m\,\slashed{k}_2\,\sigma^{-l}\gamma^n}\,F_{ij}F_{-l}
                \;=\; 4\,k_1^-\,\big(k_2^i g^{jl}-k_2^j g^{il}\big)\,F_{ij}F_{-l}$,\\[2pt]
              $-\tfrac12 g_{mn}\Tr{\slashed{k}_1\,\gamma^m\sigma^{ij}\,\slashed{k}_2\,\sigma^{-l}\gamma^n}\,F_{ij}F_{-l}
                \;=\; 4\,k_1^-\,\big(k_2^j g^{il}-k_2^i g^{jl}\big)\,F_{ij}F_{-l}$.
        \item $F_{+-}F_{+-}$: all four orderings coincide,\\[2pt]
              $-\tfrac12 g_{mn}\Tr{\slashed{k}_1(\sigma^{+-}\gamma^m)\slashed{k}_2(\sigma^{+-}\gamma^n)}\,F_{+-}F_{+-}
              \;=\; 4\,(k_1^+k_2^-+k_1^-k_2^+)\,F_{+-}F_{+-}$.

        \item $F_{+-}F_{ij}$: the mixed $\sigma^{+-}$--$\sigma^{ij}$ traces vanish,
              $-\tfrac12 g_{mn}\Tr{\slashed{k}_1(\sigma^{+-}\gamma^m)\slashed{k}_2(\sigma^{ij}\gamma^n)}\,F_{+-}F_{ij}=0$.

       \item  $F_{ij}F_{kl}$: writing $K\equiv k_1^+k_2^-+k_1^-k_2^+$, the crossed orderings
              reduce to the single invariant $F_{ij}F^{ij}$,\\[2pt]
              $-\tfrac12 g_{mn}\Tr{\slashed{k}_1\,\sigma^{ij}\gamma^m\,\slashed{k}_2\,\gamma^n\sigma^{kl}}\,F_{ij}F_{kl}
              \;=\;
              -\tfrac12 g_{mn}\Tr{\slashed{k}_1\,\gamma^m\sigma^{ij}\,\slashed{k}_2\,\sigma^{kl}\gamma^n}\,F_{ij}F_{kl}
              \;=\; 8\,K\,F_{ij}(x)F^{ij}(y)$,\\[2pt]
              while the same-side orderings reduce, after using the two-dimensional
              transverse antisymmetry of $F_{ij}$, to\\[2pt]
              $-\tfrac12 g_{mn}\Tr{\slashed{k}_1\,\sigma^{ij}\gamma^m\,\slashed{k}_2\,\sigma^{kl}\gamma^n}\,F_{ij}F_{kl}
              \;=\;
              -\tfrac12 g_{mn}\Tr{\slashed{k}_1\,\gamma^m\sigma^{ij}\,\slashed{k}_2\,\gamma^n\sigma^{kl}}\,F_{ij}F_{kl}
              \;=\; -8\,K\,F_{ij}(x)F^{ij}(y)$.

      \end{enumerate}
    \end{minipage}
    \vspace{4pt}\hrule
  \end{minipage}
  \caption{Dirac traces for a transversely polarized virtual photon. The photon vertices are contracted with the transverse polarization projector $-\tfrac12 g_{mn}$.
The twist-4 structures are included for completeness and do not contribute to the twist-3 cross section. }
  \label{tab:trace_summary_T}
\end{table*}

\bibliographystyle{apsrev4-2}
\bibliography{dijet}

\begin{thebibliography}{57}%
\makeatletter
\providecommand \@ifxundefined [1]{%
 \@ifx{#1\undefined}
}%
\providecommand \@ifnum [1]{%
 \ifnum #1\expandafter \@firstoftwo
 \else \expandafter \@secondoftwo
 \fi
}%
\providecommand \@ifx [1]{%
 \ifx #1\expandafter \@firstoftwo
 \else \expandafter \@secondoftwo
 \fi
}%
\providecommand \natexlab [1]{#1}%
\providecommand \enquote  [1]{``#1''}%
\providecommand \bibnamefont  [1]{#1}%
\providecommand \bibfnamefont [1]{#1}%
\providecommand \citenamefont [1]{#1}%
\providecommand \href@noop [0]{\@secondoftwo}%
\providecommand \href [0]{\begingroup \@sanitize@url \@href}%
\providecommand \@href[1]{\@@startlink{#1}\@@href}%
\providecommand \@@href[1]{\endgroup#1\@@endlink}%
\providecommand \@sanitize@url [0]{\catcode `\\12\catcode `\$12\catcode
  `\&12\catcode `\#12\catcode `\^12\catcode `\_12\catcode `\%12\relax}%
\providecommand \@@startlink[1]{}%
\providecommand \@@endlink[0]{}%
\providecommand \url  [0]{\begingroup\@sanitize@url \@url }%
\providecommand \@url [1]{\endgroup\@href {#1}{\urlprefix }}%
\providecommand \urlprefix  [0]{URL }%
\providecommand \Eprint [0]{\href }%
\providecommand \doibase [0]{https://doi.org/}%
\providecommand \selectlanguage [0]{\@gobble}%
\providecommand \bibinfo  [0]{\@secondoftwo}%
\providecommand \bibfield  [0]{\@secondoftwo}%
\providecommand \translation [1]{[#1]}%
\providecommand \BibitemOpen [0]{}%
\providecommand \bibitemStop [0]{}%
\providecommand \bibitemNoStop [0]{.\EOS\space}%
\providecommand \EOS [0]{\spacefactor3000\relax}%
\providecommand \BibitemShut  [1]{\csname bibitem#1\endcsname}%
\let\auto@bib@innerbib\@empty
\bibitem [{\citenamefont {Page}\ \emph {et~al.}(2020)\citenamefont {Page},
  \citenamefont {Chu},\ and\ \citenamefont {Aschenauer}}]{Page:2019gbf}%
  \BibitemOpen
  \bibfield  {author} {\bibinfo {author} {\bibfnamefont {B.~S.}\ \bibnamefont
  {Page}}, \bibinfo {author} {\bibfnamefont {X.}~\bibnamefont {Chu}},\ and\
  \bibinfo {author} {\bibfnamefont {E.~C.}\ \bibnamefont {Aschenauer}},\ }\href
  {https://doi.org/10.1103/PhysRevD.101.072003} {\bibfield  {journal} {\bibinfo
   {journal} {Phys. Rev. D}\ }\textbf {\bibinfo {volume} {101}},\ \bibinfo
  {pages} {072003} (\bibinfo {year} {2020})},\ \Eprint
  {https://arxiv.org/abs/1911.00657} {arXiv:1911.00657 [hep-ph]} \BibitemShut
  {NoStop}%
\bibitem [{\citenamefont {Metz}\ and\ \citenamefont
  {Zhou}(2011)}]{Metz:2011wb}%
  \BibitemOpen
  \bibfield  {author} {\bibinfo {author} {\bibfnamefont {A.}~\bibnamefont
  {Metz}}\ and\ \bibinfo {author} {\bibfnamefont {J.}~\bibnamefont {Zhou}},\
  }\href {https://doi.org/10.1103/PhysRevD.84.051503} {\bibfield  {journal}
  {\bibinfo  {journal} {Phys. Rev. D}\ }\textbf {\bibinfo {volume} {84}},\
  \bibinfo {pages} {051503} (\bibinfo {year} {2011})},\ \Eprint
  {https://arxiv.org/abs/1105.1991} {arXiv:1105.1991 [hep-ph]} \BibitemShut
  {NoStop}%
\bibitem [{\citenamefont {Dumitru}\ \emph {et~al.}(2015)\citenamefont
  {Dumitru}, \citenamefont {Lappi},\ and\ \citenamefont
  {Skokov}}]{Dumitru:2015gaa}%
  \BibitemOpen
  \bibfield  {author} {\bibinfo {author} {\bibfnamefont {A.}~\bibnamefont
  {Dumitru}}, \bibinfo {author} {\bibfnamefont {T.}~\bibnamefont {Lappi}},\
  and\ \bibinfo {author} {\bibfnamefont {V.}~\bibnamefont {Skokov}},\ }\href
  {https://doi.org/10.1103/PhysRevLett.115.252301} {\bibfield  {journal}
  {\bibinfo  {journal} {Phys. Rev. Lett.}\ }\textbf {\bibinfo {volume} {115}},\
  \bibinfo {pages} {252301} (\bibinfo {year} {2015})},\ \Eprint
  {https://arxiv.org/abs/1508.04438} {arXiv:1508.04438 [hep-ph]} \BibitemShut
  {NoStop}%
\bibitem [{\citenamefont {Dumitru}\ \emph {et~al.}(2019)\citenamefont
  {Dumitru}, \citenamefont {Skokov},\ and\ \citenamefont
  {Ullrich}}]{Dumitru:2018kuw}%
  \BibitemOpen
  \bibfield  {author} {\bibinfo {author} {\bibfnamefont {A.}~\bibnamefont
  {Dumitru}}, \bibinfo {author} {\bibfnamefont {V.}~\bibnamefont {Skokov}},\
  and\ \bibinfo {author} {\bibfnamefont {T.}~\bibnamefont {Ullrich}},\ }\href
  {https://doi.org/10.1103/PhysRevC.99.015204} {\bibfield  {journal} {\bibinfo
  {journal} {Phys. Rev. C}\ }\textbf {\bibinfo {volume} {99}},\ \bibinfo
  {pages} {015204} (\bibinfo {year} {2019})},\ \Eprint
  {https://arxiv.org/abs/1809.02615} {arXiv:1809.02615 [hep-ph]} \BibitemShut
  {NoStop}%
\bibitem [{\citenamefont {M{\"a}ntysaari}\ \emph {et~al.}(2020)\citenamefont
  {M{\"a}ntysaari}, \citenamefont {Mueller}, \citenamefont {Salazar},\ and\
  \citenamefont {Schenke}}]{Mantysaari:2019hkq}%
  \BibitemOpen
  \bibfield  {author} {\bibinfo {author} {\bibfnamefont {H.}~\bibnamefont
  {M{\"a}ntysaari}}, \bibinfo {author} {\bibfnamefont {N.}~\bibnamefont
  {Mueller}}, \bibinfo {author} {\bibfnamefont {F.}~\bibnamefont {Salazar}},\
  and\ \bibinfo {author} {\bibfnamefont {B.}~\bibnamefont {Schenke}},\ }\href
  {https://doi.org/10.1103/PhysRevLett.124.112301} {\bibfield  {journal}
  {\bibinfo  {journal} {Phys. Rev. Lett.}\ }\textbf {\bibinfo {volume} {124}},\
  \bibinfo {pages} {112301} (\bibinfo {year} {2020})},\ \Eprint
  {https://arxiv.org/abs/1912.05586} {arXiv:1912.05586 [nucl-th]} \BibitemShut
  {NoStop}%
\bibitem [{\citenamefont {Caucal}\ \emph {et~al.}(2024)\citenamefont {Caucal},
  \citenamefont {Salazar}, \citenamefont {Schenke}, \citenamefont {Stebel},\
  and\ \citenamefont {Venugopalan}}]{Caucal:2023fsf}%
  \BibitemOpen
  \bibfield  {author} {\bibinfo {author} {\bibfnamefont {P.}~\bibnamefont
  {Caucal}}, \bibinfo {author} {\bibfnamefont {F.}~\bibnamefont {Salazar}},
  \bibinfo {author} {\bibfnamefont {B.}~\bibnamefont {Schenke}}, \bibinfo
  {author} {\bibfnamefont {T.}~\bibnamefont {Stebel}},\ and\ \bibinfo {author}
  {\bibfnamefont {R.}~\bibnamefont {Venugopalan}},\ }\href
  {https://doi.org/10.1103/PhysRevLett.132.081902} {\bibfield  {journal}
  {\bibinfo  {journal} {Phys. Rev. Lett.}\ }\textbf {\bibinfo {volume} {132}},\
  \bibinfo {pages} {081902} (\bibinfo {year} {2024})},\ \Eprint
  {https://arxiv.org/abs/2308.00022} {arXiv:2308.00022 [hep-ph]} \BibitemShut
  {NoStop}%
\bibitem [{\citenamefont {Boer}\ \emph {et~al.}(2011)\citenamefont {Boer} \emph
  {et~al.}}]{Boer:2011fh}%
  \BibitemOpen
  \bibfield  {author} {\bibinfo {author} {\bibfnamefont {D.}~\bibnamefont
  {Boer}} \emph {et~al.},\ }\href@noop {} {\  (\bibinfo {year} {2011})},\
  \Eprint {https://arxiv.org/abs/1108.1713} {arXiv:1108.1713 [nucl-th]}
  \BibitemShut {NoStop}%
\bibitem [{\citenamefont {Meissner}\ \emph {et~al.}(2009)\citenamefont
  {Meissner}, \citenamefont {Metz},\ and\ \citenamefont
  {Schlegel}}]{Meissner:2009ww}%
  \BibitemOpen
  \bibfield  {author} {\bibinfo {author} {\bibfnamefont {S.}~\bibnamefont
  {Meissner}}, \bibinfo {author} {\bibfnamefont {A.}~\bibnamefont {Metz}},\
  and\ \bibinfo {author} {\bibfnamefont {M.}~\bibnamefont {Schlegel}},\ }\href
  {https://doi.org/10.1088/1126-6708/2009/08/056} {\bibfield  {journal}
  {\bibinfo  {journal} {JHEP}\ }\textbf {\bibinfo {volume} {08}},\ \bibinfo
  {pages} {056}},\ \Eprint {https://arxiv.org/abs/0906.5323} {arXiv:0906.5323
  [hep-ph]} \BibitemShut {NoStop}%
\bibitem [{\citenamefont {Lorc{\'e}}\ and\ \citenamefont
  {Pasquini}(2013)}]{Lorce:2013pza}%
  \BibitemOpen
  \bibfield  {author} {\bibinfo {author} {\bibfnamefont {C.}~\bibnamefont
  {Lorc{\'e}}}\ and\ \bibinfo {author} {\bibfnamefont {B.}~\bibnamefont
  {Pasquini}},\ }\href {https://doi.org/10.1007/JHEP09(2013)138} {\bibfield
  {journal} {\bibinfo  {journal} {JHEP}\ }\textbf {\bibinfo {volume} {09}},\
  \bibinfo {pages} {138}},\ \Eprint {https://arxiv.org/abs/1307.4497}
  {arXiv:1307.4497 [hep-ph]} \BibitemShut {NoStop}%
\bibitem [{\citenamefont {Mulders}\ and\ \citenamefont
  {Rodrigues}(2001)}]{Mulders:2000sh}%
  \BibitemOpen
  \bibfield  {author} {\bibinfo {author} {\bibfnamefont {P.~J.}\ \bibnamefont
  {Mulders}}\ and\ \bibinfo {author} {\bibfnamefont {J.}~\bibnamefont
  {Rodrigues}},\ }\href {https://doi.org/10.1103/PhysRevD.63.094021} {\bibfield
   {journal} {\bibinfo  {journal} {Phys. Rev. D}\ }\textbf {\bibinfo {volume}
  {63}},\ \bibinfo {pages} {094021} (\bibinfo {year} {2001})},\ \Eprint
  {https://arxiv.org/abs/hep-ph/0009343} {arXiv:hep-ph/0009343} \BibitemShut
  {NoStop}%
\bibitem [{\citenamefont {Dominguez}\ \emph {et~al.}(2011)\citenamefont
  {Dominguez}, \citenamefont {Marquet}, \citenamefont {Xiao},\ and\
  \citenamefont {Yuan}}]{Dominguez:2011wm}%
  \BibitemOpen
  \bibfield  {author} {\bibinfo {author} {\bibfnamefont {F.}~\bibnamefont
  {Dominguez}}, \bibinfo {author} {\bibfnamefont {C.}~\bibnamefont {Marquet}},
  \bibinfo {author} {\bibfnamefont {B.-W.}\ \bibnamefont {Xiao}},\ and\
  \bibinfo {author} {\bibfnamefont {F.}~\bibnamefont {Yuan}},\ }\href
  {https://doi.org/10.1103/PhysRevD.83.105005} {\bibfield  {journal} {\bibinfo
  {journal} {Phys. Rev. D}\ }\textbf {\bibinfo {volume} {83}},\ \bibinfo
  {pages} {105005} (\bibinfo {year} {2011})},\ \Eprint
  {https://arxiv.org/abs/1101.0715} {arXiv:1101.0715 [hep-ph]} \BibitemShut
  {NoStop}%
\bibitem [{\citenamefont {del Castillo}\ \emph {et~al.}(2021)\citenamefont {del
  Castillo}, \citenamefont {Echevarria}, \citenamefont {Makris},\ and\
  \citenamefont {Scimemi}}]{delCastillo:2020omr}%
  \BibitemOpen
  \bibfield  {author} {\bibinfo {author} {\bibfnamefont {R.~F.}\ \bibnamefont
  {del Castillo}}, \bibinfo {author} {\bibfnamefont {M.~G.}\ \bibnamefont
  {Echevarria}}, \bibinfo {author} {\bibfnamefont {Y.}~\bibnamefont {Makris}},\
  and\ \bibinfo {author} {\bibfnamefont {I.}~\bibnamefont {Scimemi}},\ }\href
  {https://doi.org/10.1007/JHEP01(2021)088} {\bibfield  {journal} {\bibinfo
  {journal} {JHEP}\ }\textbf {\bibinfo {volume} {01}},\ \bibinfo {pages}
  {088}},\ \Eprint {https://arxiv.org/abs/2008.07531} {arXiv:2008.07531
  [hep-ph]} \BibitemShut {NoStop}%
\bibitem [{\citenamefont {Dumitru}\ and\ \citenamefont
  {Skokov}(2016)}]{Dumitru:2016jku}%
  \BibitemOpen
  \bibfield  {author} {\bibinfo {author} {\bibfnamefont {A.}~\bibnamefont
  {Dumitru}}\ and\ \bibinfo {author} {\bibfnamefont {V.}~\bibnamefont
  {Skokov}},\ }\href {https://doi.org/10.1103/PhysRevD.94.014030} {\bibfield
  {journal} {\bibinfo  {journal} {Phys. Rev. D}\ }\textbf {\bibinfo {volume}
  {94}},\ \bibinfo {pages} {014030} (\bibinfo {year} {2016})},\ \Eprint
  {https://arxiv.org/abs/1605.02739} {arXiv:1605.02739 [hep-ph]} \BibitemShut
  {NoStop}%
\bibitem [{\citenamefont {Fleming}\ \emph {et~al.}(2003)\citenamefont
  {Fleming}, \citenamefont {Leibovich},\ and\ \citenamefont
  {Mehen}}]{Fleming:2003gt}%
  \BibitemOpen
  \bibfield  {author} {\bibinfo {author} {\bibfnamefont {S.}~\bibnamefont
  {Fleming}}, \bibinfo {author} {\bibfnamefont {A.~K.}\ \bibnamefont
  {Leibovich}},\ and\ \bibinfo {author} {\bibfnamefont {T.}~\bibnamefont
  {Mehen}},\ }\href {https://doi.org/10.1103/PhysRevD.68.094011} {\bibfield
  {journal} {\bibinfo  {journal} {Phys. Rev. D}\ }\textbf {\bibinfo {volume}
  {68}},\ \bibinfo {pages} {094011} (\bibinfo {year} {2003})},\ \Eprint
  {https://arxiv.org/abs/hep-ph/0306139} {arXiv:hep-ph/0306139} \BibitemShut
  {NoStop}%
\bibitem [{\citenamefont {Becher}\ and\ \citenamefont
  {Neubert}(2011)}]{Becher:2010tm}%
  \BibitemOpen
  \bibfield  {author} {\bibinfo {author} {\bibfnamefont {T.}~\bibnamefont
  {Becher}}\ and\ \bibinfo {author} {\bibfnamefont {M.}~\bibnamefont
  {Neubert}},\ }\href {https://doi.org/10.1140/epjc/s10052-011-1665-7}
  {\bibfield  {journal} {\bibinfo  {journal} {Eur. Phys. J. C}\ }\textbf
  {\bibinfo {volume} {71}},\ \bibinfo {pages} {1665} (\bibinfo {year}
  {2011})},\ \Eprint {https://arxiv.org/abs/1007.4005} {arXiv:1007.4005
  [hep-ph]} \BibitemShut {NoStop}%
\bibitem [{\citenamefont {McLerran}\ and\ \citenamefont
  {Venugopalan}(1994{\natexlab{a}})}]{McLerran:1993ni}%
  \BibitemOpen
  \bibfield  {author} {\bibinfo {author} {\bibfnamefont {L.~D.}\ \bibnamefont
  {McLerran}}\ and\ \bibinfo {author} {\bibfnamefont {R.}~\bibnamefont
  {Venugopalan}},\ }\href {https://doi.org/10.1103/PhysRevD.49.2233} {\bibfield
   {journal} {\bibinfo  {journal} {Phys. Rev. D}\ }\textbf {\bibinfo {volume}
  {49}},\ \bibinfo {pages} {2233} (\bibinfo {year} {1994}{\natexlab{a}})},\
  \Eprint {https://arxiv.org/abs/hep-ph/9309289} {arXiv:hep-ph/9309289}
  \BibitemShut {NoStop}%
\bibitem [{\citenamefont {McLerran}\ and\ \citenamefont
  {Venugopalan}(1994{\natexlab{b}})}]{McLerran:1993ka}%
  \BibitemOpen
  \bibfield  {author} {\bibinfo {author} {\bibfnamefont {L.~D.}\ \bibnamefont
  {McLerran}}\ and\ \bibinfo {author} {\bibfnamefont {R.}~\bibnamefont
  {Venugopalan}},\ }\href {https://doi.org/10.1103/PhysRevD.49.3352} {\bibfield
   {journal} {\bibinfo  {journal} {Phys. Rev. D}\ }\textbf {\bibinfo {volume}
  {49}},\ \bibinfo {pages} {3352} (\bibinfo {year} {1994}{\natexlab{b}})},\
  \Eprint {https://arxiv.org/abs/hep-ph/9311205} {arXiv:hep-ph/9311205}
  \BibitemShut {NoStop}%
\bibitem [{\citenamefont {Iancu}\ and\ \citenamefont
  {Venugopalan}(2003)}]{Iancu:2003xm}%
  \BibitemOpen
  \bibfield  {author} {\bibinfo {author} {\bibfnamefont {E.}~\bibnamefont
  {Iancu}}\ and\ \bibinfo {author} {\bibfnamefont {R.}~\bibnamefont
  {Venugopalan}},\ }\bibinfo {title} {{The Color glass condensate and
  high-energy scattering in QCD}},\ in\ \href
  {https://doi.org/10.1142/9789812795533_0005} {\emph {\bibinfo {booktitle}
  {{Quark-gluon plasma 4}}}},\ \bibinfo {editor} {edited by\ \bibinfo {editor}
  {\bibfnamefont {R.~C.}\ \bibnamefont {Hwa}}\ and\ \bibinfo {editor}
  {\bibfnamefont {X.-N.}\ \bibnamefont {Wang}}}\ (\bibinfo {year} {2003})\ pp.\
  \bibinfo {pages} {249--3363},\ \Eprint {https://arxiv.org/abs/hep-ph/0303204}
  {arXiv:hep-ph/0303204} \BibitemShut {NoStop}%
\bibitem [{\citenamefont {Gelis}\ \emph {et~al.}(2010)\citenamefont {Gelis},
  \citenamefont {Iancu}, \citenamefont {Jalilian-Marian},\ and\ \citenamefont
  {Venugopalan}}]{Gelis:2010nm}%
  \BibitemOpen
  \bibfield  {author} {\bibinfo {author} {\bibfnamefont {F.}~\bibnamefont
  {Gelis}}, \bibinfo {author} {\bibfnamefont {E.}~\bibnamefont {Iancu}},
  \bibinfo {author} {\bibfnamefont {J.}~\bibnamefont {Jalilian-Marian}},\ and\
  \bibinfo {author} {\bibfnamefont {R.}~\bibnamefont {Venugopalan}},\ }\href
  {https://doi.org/10.1146/annurev.nucl.010909.083629} {\bibfield  {journal}
  {\bibinfo  {journal} {Ann. Rev. Nucl. Part. Sci.}\ }\textbf {\bibinfo
  {volume} {60}},\ \bibinfo {pages} {463} (\bibinfo {year} {2010})},\ \Eprint
  {https://arxiv.org/abs/1002.0333} {arXiv:1002.0333 [hep-ph]} \BibitemShut
  {NoStop}%
\bibitem [{\citenamefont {Kovner}(2005)}]{Kovner:2005pe}%
  \BibitemOpen
  \bibfield  {author} {\bibinfo {author} {\bibfnamefont {A.}~\bibnamefont
  {Kovner}},\ }\href@noop {} {\bibfield  {journal} {\bibinfo  {journal} {Acta
  Phys. Polon. B}\ }\textbf {\bibinfo {volume} {36}},\ \bibinfo {pages} {3551}
  (\bibinfo {year} {2005})},\ \Eprint {https://arxiv.org/abs/hep-ph/0508232}
  {arXiv:hep-ph/0508232} \BibitemShut {NoStop}%
\bibitem [{\citenamefont {Boussarie}\ \emph {et~al.}(2021)\citenamefont
  {Boussarie}, \citenamefont {M{\"a}ntysaari}, \citenamefont {Salazar},\ and\
  \citenamefont {Schenke}}]{Boussarie:2021ybe}%
  \BibitemOpen
  \bibfield  {author} {\bibinfo {author} {\bibfnamefont {R.}~\bibnamefont
  {Boussarie}}, \bibinfo {author} {\bibfnamefont {H.}~\bibnamefont
  {M{\"a}ntysaari}}, \bibinfo {author} {\bibfnamefont {F.}~\bibnamefont
  {Salazar}},\ and\ \bibinfo {author} {\bibfnamefont {B.}~\bibnamefont
  {Schenke}},\ }\href {https://doi.org/10.1007/JHEP09(2021)178} {\bibfield
  {journal} {\bibinfo  {journal} {JHEP}\ }\textbf {\bibinfo {volume} {09}},\
  \bibinfo {pages} {178}},\ \Eprint {https://arxiv.org/abs/2106.11301}
  {arXiv:2106.11301 [hep-ph]} \BibitemShut {NoStop}%
\bibitem [{\citenamefont {Caucal}\ \emph {et~al.}(2021)\citenamefont {Caucal},
  \citenamefont {Salazar},\ and\ \citenamefont {Venugopalan}}]{Caucal:2021ent}%
  \BibitemOpen
  \bibfield  {author} {\bibinfo {author} {\bibfnamefont {P.}~\bibnamefont
  {Caucal}}, \bibinfo {author} {\bibfnamefont {F.}~\bibnamefont {Salazar}},\
  and\ \bibinfo {author} {\bibfnamefont {R.}~\bibnamefont {Venugopalan}},\
  }\href {https://doi.org/10.1007/JHEP11(2021)222} {\bibfield  {journal}
  {\bibinfo  {journal} {JHEP}\ }\textbf {\bibinfo {volume} {11}},\ \bibinfo
  {pages} {222}},\ \Eprint {https://arxiv.org/abs/2108.06347} {arXiv:2108.06347
  [hep-ph]} \BibitemShut {NoStop}%
\bibitem [{\citenamefont {Caucal}\ \emph {et~al.}(2023)\citenamefont {Caucal},
  \citenamefont {Salazar}, \citenamefont {Schenke}, \citenamefont {Stebel},\
  and\ \citenamefont {Venugopalan}}]{Caucal:2023nci}%
  \BibitemOpen
  \bibfield  {author} {\bibinfo {author} {\bibfnamefont {P.}~\bibnamefont
  {Caucal}}, \bibinfo {author} {\bibfnamefont {F.}~\bibnamefont {Salazar}},
  \bibinfo {author} {\bibfnamefont {B.}~\bibnamefont {Schenke}}, \bibinfo
  {author} {\bibfnamefont {T.}~\bibnamefont {Stebel}},\ and\ \bibinfo {author}
  {\bibfnamefont {R.}~\bibnamefont {Venugopalan}},\ }\href
  {https://doi.org/10.1007/JHEP08(2023)062} {\bibfield  {journal} {\bibinfo
  {journal} {JHEP}\ }\textbf {\bibinfo {volume} {08}},\ \bibinfo {pages}
  {062}},\ \Eprint {https://arxiv.org/abs/2304.03304} {arXiv:2304.03304
  [hep-ph]} \BibitemShut {NoStop}%
\bibitem [{\citenamefont {Altinoluk}\ \emph
  {et~al.}(2023{\natexlab{a}})\citenamefont {Altinoluk}, \citenamefont {Beuf},
  \citenamefont {Czajka},\ and\ \citenamefont {Tymowska}}]{Altinoluk:2022jkk}%
  \BibitemOpen
  \bibfield  {author} {\bibinfo {author} {\bibfnamefont {T.}~\bibnamefont
  {Altinoluk}}, \bibinfo {author} {\bibfnamefont {G.}~\bibnamefont {Beuf}},
  \bibinfo {author} {\bibfnamefont {A.}~\bibnamefont {Czajka}},\ and\ \bibinfo
  {author} {\bibfnamefont {A.}~\bibnamefont {Tymowska}},\ }\href
  {https://doi.org/10.1103/PhysRevD.107.074016} {\bibfield  {journal} {\bibinfo
   {journal} {Phys. Rev. D}\ }\textbf {\bibinfo {volume} {107}},\ \bibinfo
  {pages} {074016} (\bibinfo {year} {2023}{\natexlab{a}})},\ \Eprint
  {https://arxiv.org/abs/2212.10484} {arXiv:2212.10484 [hep-ph]} \BibitemShut
  {NoStop}%
\bibitem [{\citenamefont {Altinoluk}\ \emph {et~al.}(2025)\citenamefont
  {Altinoluk}, \citenamefont {Beuf}, \citenamefont {Czajka},\ and\
  \citenamefont {Marquet}}]{Altinoluk:2024zom}%
  \BibitemOpen
  \bibfield  {author} {\bibinfo {author} {\bibfnamefont {T.}~\bibnamefont
  {Altinoluk}}, \bibinfo {author} {\bibfnamefont {G.}~\bibnamefont {Beuf}},
  \bibinfo {author} {\bibfnamefont {A.}~\bibnamefont {Czajka}},\ and\ \bibinfo
  {author} {\bibfnamefont {C.}~\bibnamefont {Marquet}},\ }\href
  {https://doi.org/10.1103/PhysRevD.111.014010} {\bibfield  {journal} {\bibinfo
   {journal} {Phys. Rev. D}\ }\textbf {\bibinfo {volume} {111}},\ \bibinfo
  {pages} {014010} (\bibinfo {year} {2025})},\ \Eprint
  {https://arxiv.org/abs/2410.00612} {arXiv:2410.00612 [hep-ph]} \BibitemShut
  {NoStop}%
\bibitem [{\citenamefont {Mukherjee}\ \emph {et~al.}(2024)\citenamefont
  {Mukherjee}, \citenamefont {Skokov}, \citenamefont {Tarasov},\ and\
  \citenamefont {Tiwari}}]{Mukherjee:2023snp}%
  \BibitemOpen
  \bibfield  {author} {\bibinfo {author} {\bibfnamefont {S.}~\bibnamefont
  {Mukherjee}}, \bibinfo {author} {\bibfnamefont {V.~V.}\ \bibnamefont
  {Skokov}}, \bibinfo {author} {\bibfnamefont {A.}~\bibnamefont {Tarasov}},\
  and\ \bibinfo {author} {\bibfnamefont {S.}~\bibnamefont {Tiwari}},\ }\href
  {https://doi.org/10.1103/PhysRevD.109.034035} {\bibfield  {journal} {\bibinfo
   {journal} {Phys. Rev. D}\ }\textbf {\bibinfo {volume} {109}},\ \bibinfo
  {pages} {034035} (\bibinfo {year} {2024})},\ \Eprint
  {https://arxiv.org/abs/2311.16402} {arXiv:2311.16402 [hep-ph]} \BibitemShut
  {NoStop}%
\bibitem [{\citenamefont {Mukherjee}\ \emph {et~al.}(2025)\citenamefont
  {Mukherjee}, \citenamefont {Skokov}, \citenamefont {Tarasov},\ and\
  \citenamefont {Tiwari}}]{Mukherjee:2025aiw}%
  \BibitemOpen
  \bibfield  {author} {\bibinfo {author} {\bibfnamefont {S.}~\bibnamefont
  {Mukherjee}}, \bibinfo {author} {\bibfnamefont {V.~V.}\ \bibnamefont
  {Skokov}}, \bibinfo {author} {\bibfnamefont {A.}~\bibnamefont {Tarasov}},\
  and\ \bibinfo {author} {\bibfnamefont {S.}~\bibnamefont {Tiwari}},\ }\href
  {https://doi.org/10.1103/zrz3-3rbh} {\bibfield  {journal} {\bibinfo
  {journal} {Phys. Rev. D}\ }\textbf {\bibinfo {volume} {111}},\ \bibinfo
  {pages} {114034} (\bibinfo {year} {2025})},\ \Eprint
  {https://arxiv.org/abs/2502.15889} {arXiv:2502.15889 [hep-ph]} \BibitemShut
  {NoStop}%
\bibitem [{\citenamefont {Abbott}(1981)}]{Abbott:1980hw}%
  \BibitemOpen
  \bibfield  {author} {\bibinfo {author} {\bibfnamefont {L.~F.}\ \bibnamefont
  {Abbott}},\ }\href {https://doi.org/10.1016/0550-3213(81)90371-0} {\bibfield
  {journal} {\bibinfo  {journal} {Nucl. Phys. B}\ }\textbf {\bibinfo {volume}
  {185}},\ \bibinfo {pages} {189} (\bibinfo {year} {1981})}\BibitemShut
  {NoStop}%
\bibitem [{\citenamefont {Kovchegov}\ \emph {et~al.}(2016)\citenamefont
  {Kovchegov}, \citenamefont {Pitonyak},\ and\ \citenamefont
  {Sievert}}]{Kovchegov:2015pbl}%
  \BibitemOpen
  \bibfield  {author} {\bibinfo {author} {\bibfnamefont {Y.~V.}\ \bibnamefont
  {Kovchegov}}, \bibinfo {author} {\bibfnamefont {D.}~\bibnamefont
  {Pitonyak}},\ and\ \bibinfo {author} {\bibfnamefont {M.~D.}\ \bibnamefont
  {Sievert}},\ }\href {https://doi.org/10.1007/JHEP01(2016)072} {\bibfield
  {journal} {\bibinfo  {journal} {JHEP}\ }\textbf {\bibinfo {volume} {01}},\
  \bibinfo {pages} {072}},\ \bibinfo {note} {[Erratum: JHEP 10, 148 (2016)]},\
  \Eprint {https://arxiv.org/abs/1511.06737} {arXiv:1511.06737 [hep-ph]}
  \BibitemShut {NoStop}%
\bibitem [{\citenamefont {Kovchegov}\ \emph {et~al.}(2017)\citenamefont
  {Kovchegov}, \citenamefont {Pitonyak},\ and\ \citenamefont
  {Sievert}}]{Kovchegov:2016zex}%
  \BibitemOpen
  \bibfield  {author} {\bibinfo {author} {\bibfnamefont {Y.~V.}\ \bibnamefont
  {Kovchegov}}, \bibinfo {author} {\bibfnamefont {D.}~\bibnamefont
  {Pitonyak}},\ and\ \bibinfo {author} {\bibfnamefont {M.~D.}\ \bibnamefont
  {Sievert}},\ }\href {https://doi.org/10.1103/PhysRevD.95.014033} {\bibfield
  {journal} {\bibinfo  {journal} {Phys. Rev. D}\ }\textbf {\bibinfo {volume}
  {95}},\ \bibinfo {pages} {014033} (\bibinfo {year} {2017})},\ \Eprint
  {https://arxiv.org/abs/1610.06197} {arXiv:1610.06197 [hep-ph]} \BibitemShut
  {NoStop}%
\bibitem [{\citenamefont {Lehmann}\ \emph {et~al.}(1955)\citenamefont
  {Lehmann}, \citenamefont {Symanzik},\ and\ \citenamefont
  {Zimmermann}}]{Lehmann:1954rq}%
  \BibitemOpen
  \bibfield  {author} {\bibinfo {author} {\bibfnamefont {H.}~\bibnamefont
  {Lehmann}}, \bibinfo {author} {\bibfnamefont {K.}~\bibnamefont {Symanzik}},\
  and\ \bibinfo {author} {\bibfnamefont {W.}~\bibnamefont {Zimmermann}},\
  }\href {https://doi.org/10.1007/BF02731765} {\bibfield  {journal} {\bibinfo
  {journal} {Nuovo Cim.}\ }\textbf {\bibinfo {volume} {1}},\ \bibinfo {pages}
  {205} (\bibinfo {year} {1955})}\BibitemShut {NoStop}%
\bibitem [{\citenamefont {Schwinger}(1951)}]{Schwinger:1951nm}%
  \BibitemOpen
  \bibfield  {author} {\bibinfo {author} {\bibfnamefont {J.~S.}\ \bibnamefont
  {Schwinger}},\ }\href {https://doi.org/10.1103/PhysRev.82.664} {\bibfield
  {journal} {\bibinfo  {journal} {Phys. Rev.}\ }\textbf {\bibinfo {volume}
  {82}},\ \bibinfo {pages} {664} (\bibinfo {year} {1951})}\BibitemShut
  {NoStop}%
\bibitem [{\citenamefont {Roy}\ and\ \citenamefont
  {Venugopalan}(2018)}]{Roy:2018jxq}%
  \BibitemOpen
  \bibfield  {author} {\bibinfo {author} {\bibfnamefont {K.}~\bibnamefont
  {Roy}}\ and\ \bibinfo {author} {\bibfnamefont {R.}~\bibnamefont
  {Venugopalan}},\ }\href {https://doi.org/10.1007/JHEP05(2018)013} {\bibfield
  {journal} {\bibinfo  {journal} {JHEP}\ }\textbf {\bibinfo {volume} {05}},\
  \bibinfo {pages} {013}},\ \Eprint {https://arxiv.org/abs/1802.09550}
  {arXiv:1802.09550 [hep-ph]} \BibitemShut {NoStop}%
\bibitem [{\citenamefont {Kar}\ \emph {et~al.}(2026)\citenamefont {Kar},
  \citenamefont {Tarasov},\ and\ \citenamefont {Skokov}}]{Kar:2026vzk}%
  \BibitemOpen
  \bibfield  {author} {\bibinfo {author} {\bibfnamefont {T.}~\bibnamefont
  {Kar}}, \bibinfo {author} {\bibfnamefont {A.}~\bibnamefont {Tarasov}},\ and\
  \bibinfo {author} {\bibfnamefont {V.}~\bibnamefont {Skokov}, \bibfnamefont
  {V.}},\ }\href {https://doi.org/10.1103/8ggz-1434} {\bibfield  {journal}
  {\bibinfo  {journal} {Phys. Rev. D}\ }\textbf {\bibinfo {volume} {114}},\
  \bibinfo {pages} {034010} (\bibinfo {year} {2026})},\ \Eprint
  {https://arxiv.org/abs/2603.08805} {arXiv:2603.08805 [hep-ph]} \BibitemShut
  {NoStop}%
\bibitem [{\citenamefont {Rodini}\ and\ \citenamefont
  {Vladimirov}(2022)}]{Rodini:2022wki}%
  \BibitemOpen
  \bibfield  {author} {\bibinfo {author} {\bibfnamefont {S.}~\bibnamefont
  {Rodini}}\ and\ \bibinfo {author} {\bibfnamefont {A.}~\bibnamefont
  {Vladimirov}},\ }\href {https://doi.org/10.1007/JHEP08(2022)031} {\bibfield
  {journal} {\bibinfo  {journal} {JHEP}\ }\textbf {\bibinfo {volume} {08}},\
  \bibinfo {pages} {031}},\ \bibinfo {note} {[Erratum: JHEP 12, 048
  (2022)]}\BibitemShut {NoStop}%
\bibitem [{\citenamefont {Beppu}\ \emph {et~al.}(2010)\citenamefont {Beppu},
  \citenamefont {Koike}, \citenamefont {Tanaka},\ and\ \citenamefont
  {Yoshida}}]{Beppu:2010qn}%
  \BibitemOpen
  \bibfield  {author} {\bibinfo {author} {\bibfnamefont {H.}~\bibnamefont
  {Beppu}}, \bibinfo {author} {\bibfnamefont {Y.}~\bibnamefont {Koike}},
  \bibinfo {author} {\bibfnamefont {K.}~\bibnamefont {Tanaka}},\ and\ \bibinfo
  {author} {\bibfnamefont {S.}~\bibnamefont {Yoshida}},\ }\href
  {https://doi.org/10.1103/PhysRevD.82.054005} {\bibfield  {journal} {\bibinfo
  {journal} {Phys. Rev. D}\ }\textbf {\bibinfo {volume} {82}},\ \bibinfo
  {pages} {054005} (\bibinfo {year} {2010})},\ \Eprint
  {https://arxiv.org/abs/1007.2034} {arXiv:1007.2034 [hep-ph]} \BibitemShut
  {NoStop}%
\bibitem [{\citenamefont {Gamberg}\ \emph {et~al.}(2022)\citenamefont
  {Gamberg}, \citenamefont {Kang}, \citenamefont {Shao}, \citenamefont
  {Terry},\ and\ \citenamefont {Zhao}}]{Gamberg:2022lju}%
  \BibitemOpen
  \bibfield  {author} {\bibinfo {author} {\bibfnamefont {L.}~\bibnamefont
  {Gamberg}}, \bibinfo {author} {\bibfnamefont {Z.-B.}\ \bibnamefont {Kang}},
  \bibinfo {author} {\bibfnamefont {D.~Y.}\ \bibnamefont {Shao}}, \bibinfo
  {author} {\bibfnamefont {J.}~\bibnamefont {Terry}},\ and\ \bibinfo {author}
  {\bibfnamefont {F.}~\bibnamefont {Zhao}},\ }\href@noop {} {\  (\bibinfo
  {year} {2022})},\ \Eprint {https://arxiv.org/abs/2211.13209}
  {arXiv:2211.13209 [hep-ph]} \BibitemShut {NoStop}%
\bibitem [{\citenamefont {Boer}\ \emph {et~al.}(1997)\citenamefont {Boer},
  \citenamefont {Jakob},\ and\ \citenamefont {Mulders}}]{Boer:1997mf}%
  \BibitemOpen
  \bibfield  {author} {\bibinfo {author} {\bibfnamefont {D.}~\bibnamefont
  {Boer}}, \bibinfo {author} {\bibfnamefont {R.}~\bibnamefont {Jakob}},\ and\
  \bibinfo {author} {\bibfnamefont {P.~J.}\ \bibnamefont {Mulders}},\ }\href
  {https://doi.org/10.1016/S0550-3213(97)00456-2} {\bibfield  {journal}
  {\bibinfo  {journal} {Nucl. Phys. B}\ }\textbf {\bibinfo {volume} {504}},\
  \bibinfo {pages} {345} (\bibinfo {year} {1997})},\ \Eprint
  {https://arxiv.org/abs/hep-ph/9702281} {arXiv:hep-ph/9702281} \BibitemShut
  {NoStop}%
\bibitem [{\citenamefont {Altinoluk}\ \emph {et~al.}(2019)\citenamefont
  {Altinoluk}, \citenamefont {Boussarie},\ and\ \citenamefont
  {Kotko}}]{Altinoluk:2019fui}%
  \BibitemOpen
  \bibfield  {author} {\bibinfo {author} {\bibfnamefont {T.}~\bibnamefont
  {Altinoluk}}, \bibinfo {author} {\bibfnamefont {R.}~\bibnamefont
  {Boussarie}},\ and\ \bibinfo {author} {\bibfnamefont {P.}~\bibnamefont
  {Kotko}},\ }\href {https://doi.org/10.1007/JHEP05(2019)156} {\bibfield
  {journal} {\bibinfo  {journal} {JHEP}\ }\textbf {\bibinfo {volume} {05}},\
  \bibinfo {pages} {156}},\ \Eprint {https://arxiv.org/abs/1901.01175}
  {arXiv:1901.01175 [hep-ph]} \BibitemShut {NoStop}%
\bibitem [{\citenamefont {Altinoluk}\ and\ \citenamefont
  {Boussarie}(2019)}]{Altinoluk:2019wyu}%
  \BibitemOpen
  \bibfield  {author} {\bibinfo {author} {\bibfnamefont {T.}~\bibnamefont
  {Altinoluk}}\ and\ \bibinfo {author} {\bibfnamefont {R.}~\bibnamefont
  {Boussarie}},\ }\href {https://doi.org/10.1007/JHEP10(2019)208} {\bibfield
  {journal} {\bibinfo  {journal} {JHEP}\ }\textbf {\bibinfo {volume} {10}},\
  \bibinfo {pages} {208}},\ \Eprint {https://arxiv.org/abs/1902.07930}
  {arXiv:1902.07930 [hep-ph]} \BibitemShut {NoStop}%
\bibitem [{\citenamefont {Kotko}\ \emph {et~al.}(2015)\citenamefont {Kotko},
  \citenamefont {Kutak}, \citenamefont {Marquet}, \citenamefont {Petreska},
  \citenamefont {Sapeta},\ and\ \citenamefont {van Hameren}}]{Kotko:2015ura}%
  \BibitemOpen
  \bibfield  {author} {\bibinfo {author} {\bibfnamefont {P.}~\bibnamefont
  {Kotko}}, \bibinfo {author} {\bibfnamefont {K.}~\bibnamefont {Kutak}},
  \bibinfo {author} {\bibfnamefont {C.}~\bibnamefont {Marquet}}, \bibinfo
  {author} {\bibfnamefont {E.}~\bibnamefont {Petreska}}, \bibinfo {author}
  {\bibfnamefont {S.}~\bibnamefont {Sapeta}},\ and\ \bibinfo {author}
  {\bibfnamefont {A.}~\bibnamefont {van Hameren}},\ }\href
  {https://doi.org/10.1007/JHEP09(2015)106} {\bibfield  {journal} {\bibinfo
  {journal} {JHEP}\ }\textbf {\bibinfo {volume} {09}},\ \bibinfo {pages}
  {106}},\ \Eprint {https://arxiv.org/abs/1503.03421} {arXiv:1503.03421
  [hep-ph]} \BibitemShut {NoStop}%
\bibitem [{\citenamefont {van Hameren}\ \emph {et~al.}(2016)\citenamefont {van
  Hameren}, \citenamefont {Kotko}, \citenamefont {Kutak}, \citenamefont
  {Marquet}, \citenamefont {Petreska},\ and\ \citenamefont
  {Sapeta}}]{vanHameren:2016ftb}%
  \BibitemOpen
  \bibfield  {author} {\bibinfo {author} {\bibfnamefont {A.}~\bibnamefont {van
  Hameren}}, \bibinfo {author} {\bibfnamefont {P.}~\bibnamefont {Kotko}},
  \bibinfo {author} {\bibfnamefont {K.}~\bibnamefont {Kutak}}, \bibinfo
  {author} {\bibfnamefont {C.}~\bibnamefont {Marquet}}, \bibinfo {author}
  {\bibfnamefont {E.}~\bibnamefont {Petreska}},\ and\ \bibinfo {author}
  {\bibfnamefont {S.}~\bibnamefont {Sapeta}},\ }\href
  {https://doi.org/10.1007/JHEP12(2016)034} {\bibfield  {journal} {\bibinfo
  {journal} {JHEP}\ }\textbf {\bibinfo {volume} {12}},\ \bibinfo {pages}
  {034}},\ \bibinfo {note} {[Erratum: JHEP 02, 158 (2019)]},\ \Eprint
  {https://arxiv.org/abs/1607.03121} {arXiv:1607.03121 [hep-ph]} \BibitemShut
  {NoStop}%
\bibitem [{\citenamefont {Boussarie}\ \emph {et~al.}(2023)\citenamefont
  {Boussarie} \emph {et~al.}}]{Boussarie:2023izj}%
  \BibitemOpen
  \bibfield  {author} {\bibinfo {author} {\bibfnamefont {R.}~\bibnamefont
  {Boussarie}} \emph {et~al.},\ }\href@noop {} {\  (\bibinfo {year} {2023})},\
  \Eprint {https://arxiv.org/abs/2304.03302} {arXiv:2304.03302 [hep-ph]}
  \BibitemShut {NoStop}%
\bibitem [{\citenamefont {Fu}\ \emph {et~al.}(2025)\citenamefont {Fu},
  \citenamefont {Kang}, \citenamefont {Salazar}, \citenamefont {Wang},\ and\
  \citenamefont {Xing}}]{Fu:2024sba}%
  \BibitemOpen
  \bibfield  {author} {\bibinfo {author} {\bibfnamefont {Y.}~\bibnamefont
  {Fu}}, \bibinfo {author} {\bibfnamefont {Z.-B.}\ \bibnamefont {Kang}},
  \bibinfo {author} {\bibfnamefont {F.}~\bibnamefont {Salazar}}, \bibinfo
  {author} {\bibfnamefont {X.-N.}\ \bibnamefont {Wang}},\ and\ \bibinfo
  {author} {\bibfnamefont {H.}~\bibnamefont {Xing}},\ }\href
  {https://doi.org/10.1103/ckhv-5213} {\bibfield  {journal} {\bibinfo
  {journal} {Phys. Rev. D}\ }\textbf {\bibinfo {volume} {112}},\ \bibinfo
  {pages} {014029} (\bibinfo {year} {2025})},\ \Eprint
  {https://arxiv.org/abs/2406.01684} {arXiv:2406.01684 [hep-ph]} \BibitemShut
  {NoStop}%
\bibitem [{\citenamefont {Altinoluk}\ \emph
  {et~al.}(2023{\natexlab{b}})\citenamefont {Altinoluk}, \citenamefont
  {Armesto},\ and\ \citenamefont {Beuf}}]{Altinoluk:2023qfr}%
  \BibitemOpen
  \bibfield  {author} {\bibinfo {author} {\bibfnamefont {T.}~\bibnamefont
  {Altinoluk}}, \bibinfo {author} {\bibfnamefont {N.}~\bibnamefont {Armesto}},\
  and\ \bibinfo {author} {\bibfnamefont {G.}~\bibnamefont {Beuf}},\ }\href
  {https://doi.org/10.1103/PhysRevD.108.074023} {\bibfield  {journal} {\bibinfo
   {journal} {Phys. Rev. D}\ }\textbf {\bibinfo {volume} {108}},\ \bibinfo
  {pages} {074023} (\bibinfo {year} {2023}{\natexlab{b}})},\ \Eprint
  {https://arxiv.org/abs/2303.12691} {arXiv:2303.12691 [hep-ph]} \BibitemShut
  {NoStop}%
\bibitem [{\citenamefont {Nikolaev}\ and\ \citenamefont
  {Zakharov}(1991)}]{Nikolaev:1990ja}%
  \BibitemOpen
  \bibfield  {author} {\bibinfo {author} {\bibfnamefont {N.~N.}\ \bibnamefont
  {Nikolaev}}\ and\ \bibinfo {author} {\bibfnamefont {B.~G.}\ \bibnamefont
  {Zakharov}},\ }\href {https://doi.org/10.1007/BF01483577} {\bibfield
  {journal} {\bibinfo  {journal} {Z. Phys. C}\ }\textbf {\bibinfo {volume}
  {49}},\ \bibinfo {pages} {607} (\bibinfo {year} {1991})}\BibitemShut
  {NoStop}%
\bibitem [{\citenamefont {Mueller}(1990)}]{Mueller:1989st}%
  \BibitemOpen
  \bibfield  {author} {\bibinfo {author} {\bibfnamefont {A.~H.}\ \bibnamefont
  {Mueller}},\ }\href {https://doi.org/10.1016/0550-3213(90)90173-B} {\bibfield
   {journal} {\bibinfo  {journal} {Nucl. Phys. B}\ }\textbf {\bibinfo {volume}
  {335}},\ \bibinfo {pages} {115} (\bibinfo {year} {1990})}\BibitemShut
  {NoStop}%
\bibitem [{\citenamefont {Golec-Biernat}\ and\ \citenamefont
  {Wusthoff}(1998)}]{Golec-Biernat:1998zce}%
  \BibitemOpen
  \bibfield  {author} {\bibinfo {author} {\bibfnamefont {K.~J.}\ \bibnamefont
  {Golec-Biernat}}\ and\ \bibinfo {author} {\bibfnamefont {M.}~\bibnamefont
  {Wusthoff}},\ }\href {https://doi.org/10.1103/PhysRevD.59.014017} {\bibfield
  {journal} {\bibinfo  {journal} {Phys. Rev. D}\ }\textbf {\bibinfo {volume}
  {59}},\ \bibinfo {pages} {014017} (\bibinfo {year} {1998})},\ \Eprint
  {https://arxiv.org/abs/hep-ph/9807513} {arXiv:hep-ph/9807513} \BibitemShut
  {NoStop}%
\bibitem [{\citenamefont {Golec-Biernat}\ and\ \citenamefont
  {Wusthoff}(1999)}]{Golec-Biernat:1999qor}%
  \BibitemOpen
  \bibfield  {author} {\bibinfo {author} {\bibfnamefont {K.~J.}\ \bibnamefont
  {Golec-Biernat}}\ and\ \bibinfo {author} {\bibfnamefont {M.}~\bibnamefont
  {Wusthoff}},\ }\href {https://doi.org/10.1103/PhysRevD.60.114023} {\bibfield
  {journal} {\bibinfo  {journal} {Phys. Rev. D}\ }\textbf {\bibinfo {volume}
  {60}},\ \bibinfo {pages} {114023} (\bibinfo {year} {1999})},\ \Eprint
  {https://arxiv.org/abs/hep-ph/9903358} {arXiv:hep-ph/9903358} \BibitemShut
  {NoStop}%
\bibitem [{\citenamefont {Nikolaev}\ and\ \citenamefont
  {Zakharov}(1994)}]{Nikolaev:1994ce}%
  \BibitemOpen
  \bibfield  {author} {\bibinfo {author} {\bibfnamefont {N.~N.}\ \bibnamefont
  {Nikolaev}}\ and\ \bibinfo {author} {\bibfnamefont {B.~G.}\ \bibnamefont
  {Zakharov}},\ }\href {https://doi.org/10.1016/0370-2693(94)90877-X}
  {\bibfield  {journal} {\bibinfo  {journal} {Phys. Lett. B}\ }\textbf
  {\bibinfo {volume} {332}},\ \bibinfo {pages} {184} (\bibinfo {year}
  {1994})},\ \Eprint {https://arxiv.org/abs/hep-ph/9403243}
  {arXiv:hep-ph/9403243} \BibitemShut {NoStop}%
\bibitem [{\citenamefont {Balitsky}\ and\ \citenamefont
  {Braun}(1989)}]{Balitsky:1987bk}%
  \BibitemOpen
  \bibfield  {author} {\bibinfo {author} {\bibfnamefont {I.~I.}\ \bibnamefont
  {Balitsky}}\ and\ \bibinfo {author} {\bibfnamefont {V.~M.}\ \bibnamefont
  {Braun}},\ }\href {https://doi.org/10.1016/0550-3213(89)90168-5} {\bibfield
  {journal} {\bibinfo  {journal} {Nucl. Phys. B}\ }\textbf {\bibinfo {volume}
  {311}},\ \bibinfo {pages} {541} (\bibinfo {year} {1989})}\BibitemShut
  {NoStop}%
\bibitem [{\citenamefont {Balitsky}\ and\ \citenamefont
  {Tarasov}(2015)}]{Balitsky:2015qba}%
  \BibitemOpen
  \bibfield  {author} {\bibinfo {author} {\bibfnamefont {I.}~\bibnamefont
  {Balitsky}}\ and\ \bibinfo {author} {\bibfnamefont {A.}~\bibnamefont
  {Tarasov}},\ }\href {https://doi.org/10.1007/JHEP10(2015)017} {\bibfield
  {journal} {\bibinfo  {journal} {JHEP}\ }\textbf {\bibinfo {volume} {10}},\
  \bibinfo {pages} {017}},\ \Eprint {https://arxiv.org/abs/1505.02151}
  {arXiv:1505.02151 [hep-ph]} \BibitemShut {NoStop}%
\bibitem [{\citenamefont {Balitsky}\ and\ \citenamefont
  {Tarasov}(2016)}]{Balitsky:2016dgz}%
  \BibitemOpen
  \bibfield  {author} {\bibinfo {author} {\bibfnamefont {I.}~\bibnamefont
  {Balitsky}}\ and\ \bibinfo {author} {\bibfnamefont {A.}~\bibnamefont
  {Tarasov}},\ }\href {https://doi.org/10.1007/JHEP06(2016)164} {\bibfield
  {journal} {\bibinfo  {journal} {JHEP}\ }\textbf {\bibinfo {volume} {06}},\
  \bibinfo {pages} {164}},\ \Eprint {https://arxiv.org/abs/1603.06548}
  {arXiv:1603.06548 [hep-ph]} \BibitemShut {NoStop}%
\bibitem [{\citenamefont {Balitsky}\ and\ \citenamefont
  {Tarasov}(2018)}]{Balitsky:2017gis}%
  \BibitemOpen
  \bibfield  {author} {\bibinfo {author} {\bibfnamefont {I.}~\bibnamefont
  {Balitsky}}\ and\ \bibinfo {author} {\bibfnamefont {A.}~\bibnamefont
  {Tarasov}},\ }\href {https://doi.org/10.1007/JHEP05(2018)150} {\bibfield
  {journal} {\bibinfo  {journal} {JHEP}\ }\textbf {\bibinfo {volume} {05}},\
  \bibinfo {pages} {150}},\ \Eprint {https://arxiv.org/abs/1712.09389}
  {arXiv:1712.09389 [hep-ph]} \BibitemShut {NoStop}%
\bibitem [{\citenamefont {Balitsky}\ and\ \citenamefont
  {Tarasov}(2017)}]{Balitsky:2017flc}%
  \BibitemOpen
  \bibfield  {author} {\bibinfo {author} {\bibfnamefont {I.}~\bibnamefont
  {Balitsky}}\ and\ \bibinfo {author} {\bibfnamefont {A.}~\bibnamefont
  {Tarasov}},\ }\href {https://doi.org/10.1007/JHEP07(2017)095} {\bibfield
  {journal} {\bibinfo  {journal} {JHEP}\ }\textbf {\bibinfo {volume} {07}},\
  \bibinfo {pages} {095}},\ \Eprint {https://arxiv.org/abs/1706.01415}
  {arXiv:1706.01415 [hep-ph]} \BibitemShut {NoStop}%
\bibitem [{\citenamefont {Balitsky}(2023)}]{Balitsky:2023hmh}%
  \BibitemOpen
  \bibfield  {author} {\bibinfo {author} {\bibfnamefont {I.}~\bibnamefont
  {Balitsky}},\ }\href {https://doi.org/10.1007/JHEP03(2023)029} {\bibfield
  {journal} {\bibinfo  {journal} {JHEP}\ }\textbf {\bibinfo {volume} {03}},\
  \bibinfo {pages} {029}},\ \Eprint {https://arxiv.org/abs/2301.01717}
  {arXiv:2301.01717 [hep-ph]} \BibitemShut {NoStop}%
\bibitem [{\citenamefont {Balitsky}(2025)}]{Balitsky:2025bup}%
  \BibitemOpen
  \bibfield  {author} {\bibinfo {author} {\bibfnamefont {I.}~\bibnamefont
  {Balitsky}},\ }\href {https://doi.org/10.5506/APhysPolB.56.3-A20} {\bibfield
  {journal} {\bibinfo  {journal} {Acta Phys. Polon. B}\ }\textbf {\bibinfo
  {volume} {56}},\ \bibinfo {pages} {3} (\bibinfo {year} {2025})},\ \Eprint
  {https://arxiv.org/abs/2502.00986} {arXiv:2502.00986 [hep-ph]} \BibitemShut
  {NoStop}%
\end{thebibliography}%

\end{document}